\documentclass[notitlepage,aps,pra,twocolumn,superscriptaddress,amsmath,showpacs,tightenlines,longbibliography,10pt]{revtex4-1}
\usepackage{amssymb}
\usepackage{amsmath}
\usepackage{dcolumn}
\usepackage{graphicx}
\usepackage{mathrsfs}
\usepackage{subfigure}
\usepackage{booktabs}
\usepackage{color}
\usepackage{docmute}
\usepackage{hyphenat}
\usepackage{CJKutf8}

\newcommand{\ket}[1]{\mbox{$|#1\rangle$}}
\newcommand{\bra}[1]{\mbox{$\langle#1|$}}
\newcommand{\average}[1]{\mbox{$\langle#1\rangle$}}

\usepackage{url}
\usepackage[colorlinks]{hyperref}
\hypersetup{%
	plainpages=true,
	breaklinks=true,
	hypertexnames=false,
	pageanchor=true,
	bookmarksnumbered=true,
	colorlinks=true,
	linkcolor={blue},
	citecolor={red},
	urlcolor={blue},
	anchorcolor={black}
}

\begin{document}
	
\title{Strong spin squeezing induced by weak squeezing of light inside a cavity}
		
\author{Wei Qin}
\affiliation{Theoretical Quantum Physics Laboratory, RIKEN Cluster
for Pioneering Research, Wako-shi, Saitama 351-0198, Japan}
		
\author{Ye-Hong Chen}
\affiliation{Theoretical Quantum Physics Laboratory, RIKEN Cluster
for Pioneering Research, Wako-shi, Saitama 351-0198, Japan}
		
\author{Xin Wang}
\affiliation{Theoretical Quantum Physics Laboratory, RIKEN Cluster
for Pioneering Research, Wako-shi, Saitama 351-0198, Japan}
\affiliation{Institute of Quantum Optics and Quantum Information,
School of Science, Xi'an Jiaotong University, Xi'an 710049, China}
		
\author{Adam Miranowicz}
\affiliation{Theoretical Quantum Physics Laboratory, RIKEN Cluster
for Pioneering Research, Wako-shi, Saitama 351-0198, Japan}
\affiliation{Faculty of Physics, Adam Mickiewicz University,
61-614 Pozna\'n, Poland}
		
\author{Franco Nori}
\affiliation{Theoretical Quantum Physics Laboratory, RIKEN Cluster
for Pioneering Research, Wako-shi, Saitama 351-0198, Japan}
\affiliation{Department of Physics, The University of Michigan,
Ann Arbor, Michigan 48109-1040, USA}

\begin{abstract}
We propose a simple method for generating spin squeezing of atomic ensembles in a Floquet cavity subject to a weak, detuned two-photon driving. We demonstrate that {\it the weak squeezing of light inside the cavity can, counterintuitively, induce strong spin squeezing}. This is achieved by exploiting the anti-Stokes scattering process of a photon pair interacting with an atom. Specifically, {\it one photon of the photon pair is scattered into the cavity resonance by absorbing partially the energy of the other photon whose remaining energy excites the atom}. The scattering, combined with a Floquet sideband, provides an alternative mechanism to implement Heisenberg-limited spin squeezing. Our proposal does {\it not} need multiple classical and cavity-photon drivings applied to atoms in ensembles, and therefore its experimental feasibility is greatly improved compared to other cavity-based schemes. As an example, we demonstrate a possible implementation with a superconducting resonator coupled to a nitrogen-vacancy electronic-spin ensemble.
\end{abstract}

\date{\today}

\maketitle

\section{Introduction}
In analogy to squeezed states of light, spin squeezing in atomic ensembles~\cite{kitagawa1993squeezed,wineland1992spin,wineland1994squeezed,ma2011quantum} describes the reduction of quantum fluctuation noise in one component of a collective pseudospin, at the expense of increased quantum fluctuation noise in the other component. This property is an essential ingredient for high-precision quantum metrology and also enables various quantum-information applications~\cite{ma2011quantum,pezze2018quantum}. For this reason, significant effort has been devoted to generating spin squeezing; such effort includes exploiting atom-atom collisions in Bose-Einstein condensates~\cite{sorensen2001many,orzel2001squeezed,esteve2008squeezing,riedel2010atom,gross2010nonlinear,lucke2011twin,yu2014creating,luo2017deterministic,fadel2018spatial}, and atom-light interactions in atomic ensembles~\cite{kuzmich1997spin,hald1999spin,julsgaard2001experimental,kuzmich2000generation,koschorreck2010quantum,chalopin2018quantum,evrard2019enhanced}. In particular, cavity quantum electrodynamics~\cite{you2011atomic,gu2017microwave}, which can strongly couple atoms to cavity photons, is considered as an ideal platform for spin squeezing implementations~\cite{banerjee1996generation,sorensen2002entangling,leroux2010implementation,schleier2010states,bohnet2014reduced,hosten2016measurement,cox2016deterministic,zhang2017cavity,lewis2018robust,braverman2019near,song2019generation}.  Here, we propose a fundamentally different approach to prepare atomic spin-squeezed states in cavities, and demonstrate that the weak squeezing of the cavity field can induce strong spin squeezing.

One-axis twisting (OAT) and two-axis twisting (TAT) are two basic mechanisms to generate spin-squeezed states~\cite{kitagawa1993squeezed,ma2011quantum}. In high-precision measurements, TAT is considered to be superior to OAT~\cite{ma2011quantum}, because TAT can reduce quantum fluctuation noise to the fundamental Heisenberg limit $\propto N^{-1}$, lower than the OAT-allowed limit $\propto N^{-2/3}$. Here, $N$ refers to the number of atoms in an ensemble. Note that both mechanisms depend on controlled unitary dynamics, such that they are extremely fragile to dissipation and also require high-precision control for time evolution. Alternatively, dissipation, when treated as a resource~\cite{verstraete2009quantum,krauter2011entanglement,lin2013dissipative,qin2017heralded,qin2018exponentially}, has also been exploited to implement Heisenberg-limited squeezing~\cite{parkins2006unconditional,zheng2012generation,dalla2013dissipative,ma2013dissipation}. In dissipative protocols, atomic ensembles can be driven to a spin-squeezed steady state. However, these TAT and dissipative schemes have not been experimentally demonstrated because of their high complexity. This is partially attributed to the need for multiple classical and cavity-photon drivings applied to atoms. For example, various approaches for spin squeezing in cavities rely on a double off-resonant Raman transition (i.e., the double-$\Lambda$ transition)~\cite{sorensen2002entangling,zheng2012generation,dalla2013dissipative,ma2013dissipation,zhang2017cavity,borregaard2017one,parkins2006unconditional,liu2019spin}. It is generally difficult to realize such a transition for each atom in ensembles for spin squeezing. 

In this manuscript, we propose a simplification by introducing a weak and detuned two-photon driving for a Floquet cavity, and demonstrate the dissipative preparation of steady-state spin squeezing (SSSS), with Heisenberg scaling. Remarkably, light squeezing inside the cavity in our proposal is very weak and can be understood as {\it a seed for strong spin squeezing}. This is essentially different from the process that directly transfers squeezing from light to atomic ensembles~\cite{kuzmich1997spin,hald1999spin,julsgaard2001experimental,jensen2011quantum,yan2017establishing}. Such weak squeezing of light avoids two-photon correlation noise and thermal noise, which can give rise to the so-called $3$~dB limit in degenerate parametric amplification processes~\cite{milburn1981production} and can greatly limit spin squeezing. 

Furthermore, in contrast to other cavity-based proposals for Heisenberg-limited spin squeezing, our method does {\it not} require multiple classical and cavity-photon drivings on atoms, thus significantly reducing the experimental complexity. The key element underlying our method is the absorption of a detuned-driving photon pair: one of these photons is absorbed by the cavity and the other one by an atom. This process can be understood as {\it anti-Stokes scattering, of one photon of the driving photon pair, into the cavity resonance by absorbing part of the energy of the other photon, which excites the atom with its remaining energy}. As opposed to typical Raman scattering~\cite{boyd2003nonlinear}, {\it the scattered photon} in the description above {\it absorbs the energy of another photon}, rather than the excitation of matter, e.g., atoms, molecules, or mechanics.

\begin{figure}[t]
	\centering
	\includegraphics[width=6.5cm]{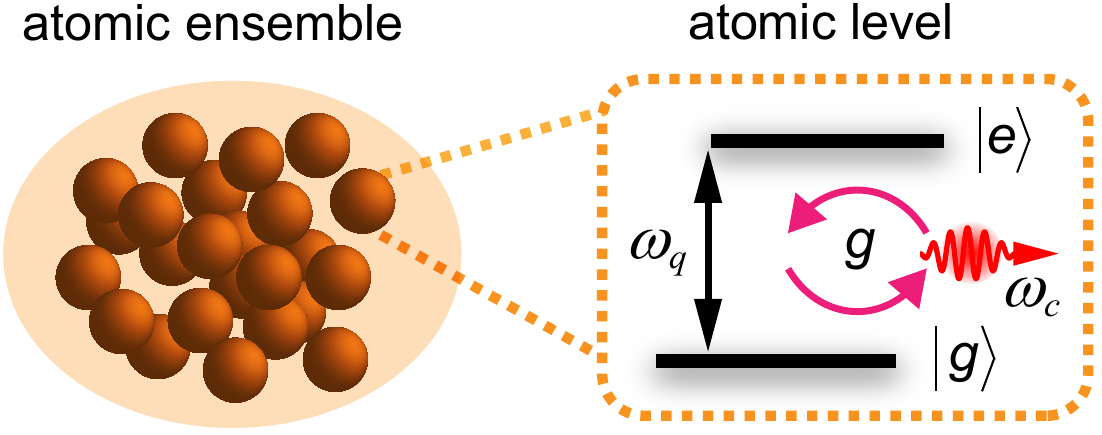}
	\caption{An atomic ensemble consisting of $N$ identical two-level atoms with the ground state $|g\rangle$ and the excited state $|e\rangle$. Here, $\omega_{q}$ is the atomic-transition frequency, $\omega_{c}$ the cavity frequency, and $g$ the single-atom coupling to the cavity mode.}\label{fig_schematics1}
\end{figure}

\section{Physical model}

\begin{figure}[t]
	\centering
	\includegraphics[width=8.3cm]{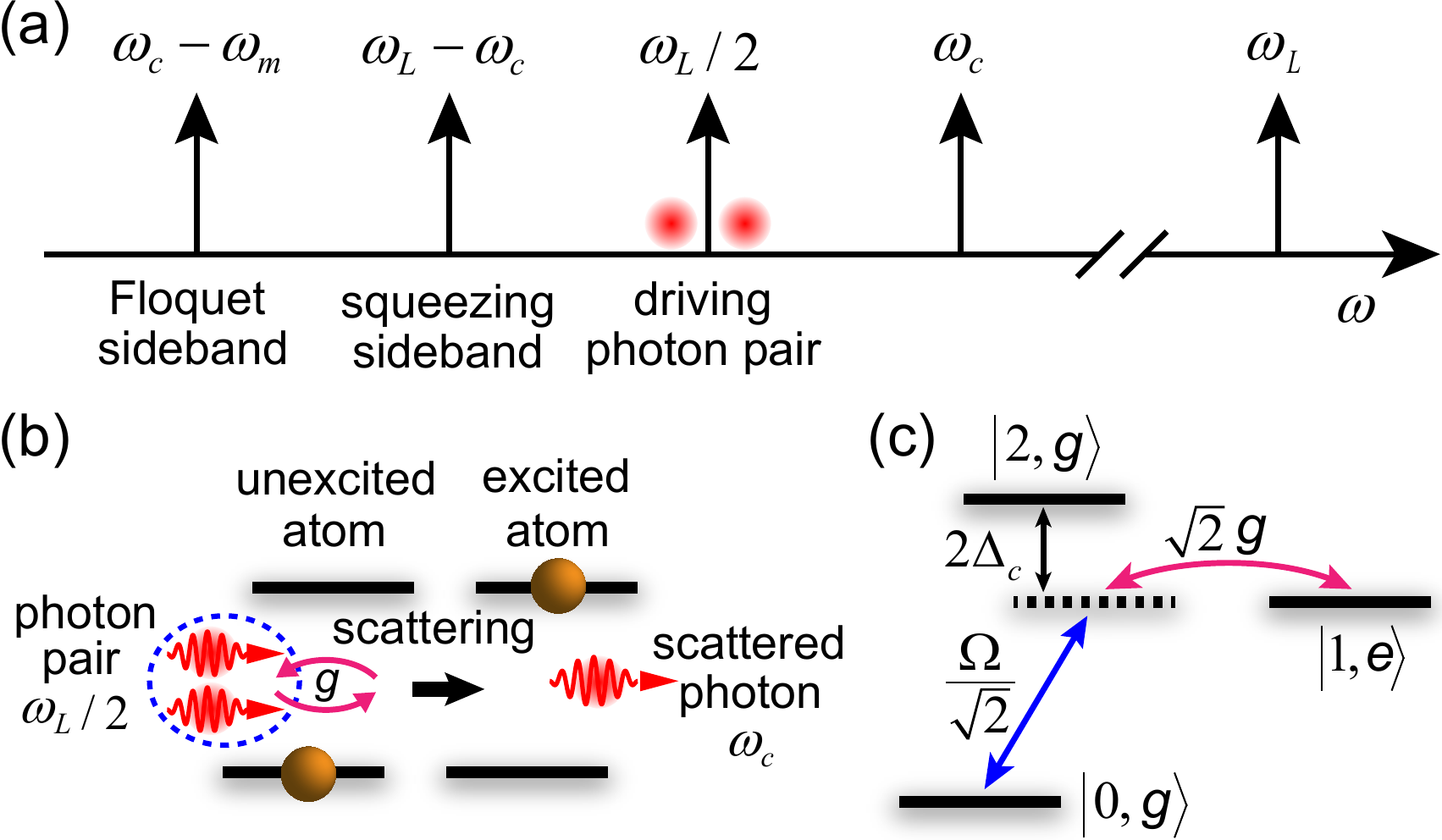}
	\caption{(a) Frequency-domain picture of a Floquet cavity driven by a weak and detuned parametric driving. The two-photon driving at frequency $\omega_{L}$, when driving the single-mode cavity of frequency $\omega_{c}$, can produce photon pairs at $\omega_{L}/2$, and induce a squeezing sideband at $\omega_{L}-\omega_{c}$. Owing to a cavity-frequency modulation with frequency $\omega_{m}$, there also exists a Floquet sideband at $\omega_{c}-\omega_{m}$. (b) Raman scattering of a driving photon pair interacting with an atom. If the squeezing sideband in (a) is tuned to the atomic resonance $\omega_{q}$, one photon of the photon pair at $\omega_{L}/2$ absorbs partially the energy of the other photon and is scattered into the cavity resonance $\omega_{c}$, and simultaneously the atom is excited by the remaining energy of the absorbed photon. (c) Transition mechanism responsible for Raman scattering described in (b). The weak, detuned two-photon driving ($\Omega$) and the cavity mode ($g$) couple the states $|0,g\rangle$ and $|1,e\rangle$ via a virtual intermediate state.}\label{fig_schematics2}
\end{figure}

We consider an ensemble consisting of $N$ two-level atoms in a single-mode cavity of frequency $\omega_{c}$, as shown in  Fig.~\ref{fig_schematics1}. For simplicity, these atoms are assumed to be identical, such that they have the same transition frequency $\omega_{q}$ and their transitions from the ground state $|g\rangle$ to the excited state $|e\rangle$ are driven by the same coupling $g$ to the cavity photon. This atomic ensemble can be described using collective spin operators $S_{\alpha}=\frac{1}{2}\sum_{j=1}^{N}\sigma_{j}^{\alpha}$, where $\sigma_{j}^{\alpha}$  ($\alpha=x,y,z$) are the Pauli matrices for the $j$th atom.  
The cavity mode is driven by a weak, detuned two-photon driving, e.g., with amplitude $\Omega$, frequency $\omega_{L}$, and phase $\theta_{L}$. Such a parametric driving can produce photon pairs at $\omega_{L}/2$ and induce a squeezing sideband at $\omega_{L}-\omega_{c}$ [see Fig.~\ref{fig_schematics2}(a)]. If this sideband is tuned to the atomic resonance $\omega_{q}$ (i.e., $\omega_{q}\approx\omega_{L}-\omega_{c}$), one photon of the driving photon pair is then scattered into the cavity resonance by absorbing a small part of the energy of the other photon; at the same time the main part of the absorbed-photon energy resonantly excites an atom [see Fig.~\ref{fig_schematics2}(b)]. We further assume that the cavity frequency $\omega_{c}$ is periodically modulated with amplitude $A_{m}$ and frequency $\omega_{m}$, and ensure that $\omega_{q}\approx\omega_{c}-\omega_{m}$. In this case, a detuned atom can emit a photon into the cavity resonance via a Floquet sideband at $\omega_{c}-\omega_{m}$ [see Fig.~\ref{fig_schematics2}(a)]. The above dynamics demonstrates that the cavity-photon creation gives rise to a competition between the atomic excitation and deexcitation. 

To be specific, we consider the Hamiltonian 
\begin{equation}
	H\left(t\right)=H_{0}+H_{1}\left(t\right),
\end{equation}
with $H_{0}=\Delta_{c}a^{\dag}a+\Delta_{q}S_{z}
+g\left(aS_{+}+a^{\dag}S_{-}\right)+\frac{1}{2}\Omega\left(e^{i\theta_{L}}a^{2}+{\rm H.c.}\right)$, and
$H_{1}\left(t\right)=A_{m}\sin\left(\omega_{m}t\right)a^{\dag}a+\frac{1}{2}\Omega_{1}\!\left(t\right)\left(e^{i\theta_{L}}a^{2}+{\rm H.c.}\right)$.
Here, $\Delta_{c/q}=\omega_{c/q}-\omega_{L}/2$ and $S_{\pm}=S_{x}\pm iS_{y}$. In addition to the driving $\Omega$, we have also assumed another two-photon driving, which has the same frequency and phase as the driving $\Omega$, but with a time-dependent amplitude $\Omega_{1}\left(t\right)\approx\Omega A_{m}\sin\left(\omega_{m}t\right)/\Delta_{c}$. The use of such a driving is to suppress an undesired two-photon driving of the cavity mode, which is induced by the periodic modulation of the cavity frequency and can destroy the dynamics of generating SSSS. 

To describe the dissipative dynamics, we use the Lindblad dissipator, given by
$\mathcal{L}\left(o\right)\rho=2o\rho o^{\dag}-o^{\dag}o\rho-\rho o^{\dag}o$. Thus,  $\frac{\kappa}{2}\mathcal{L}\left(a\right)\rho$ corresponds to cavity loss at a rate $\kappa$, and $\frac{\gamma}{2}\sum_{j=1}^{N}\mathcal{L}\left(\sigma_{j}^{-}\right)\rho$, where $\sigma_{j}^{-}=\frac{1}{2}\left(\sigma_{j}^{x}-i\sigma_{j}^{y}\right)$, describes atomic spontaneous emission at a rate $\gamma$. It follows, on taking the Fourier transformation $\widetilde{\sigma}_{k}^{-}=\frac{1}{\sqrt{N}}\sum_{j}\exp\left(-ikj\right)\sigma_{j}^{-}$, that $S_{-}=\sqrt{N}\widetilde{\sigma}_{k=0}^{-}$, indicating that the collective spin operators are related only to the zero momentum mode~\cite{gelhausen2017many,shammah2018open,macri2020spin}. Consequently, we have $\sum_{j=1}^{N}\mathcal{L}\left(\sigma_{j}^{-}\right)\rho=\frac{1}{N}\mathcal{L}\left(S_{-}\right)\rho$, because different momentum modes are uncoupled and nonzero momentum modes only decay. The full dynamics of the system is therefore determined by the master equation 
\begin{equation}
\dot{\rho}=i\left[\rho,H\left(t\right)\right]+\frac{\kappa}{2}\mathcal{L}\left(a\right)\rho+\frac{\gamma}{2N}\mathcal{L}\left(S_{-}\right)\rho.
\end{equation}

\begin{figure}[t]
	\centering
	\includegraphics[width=8.3cm]{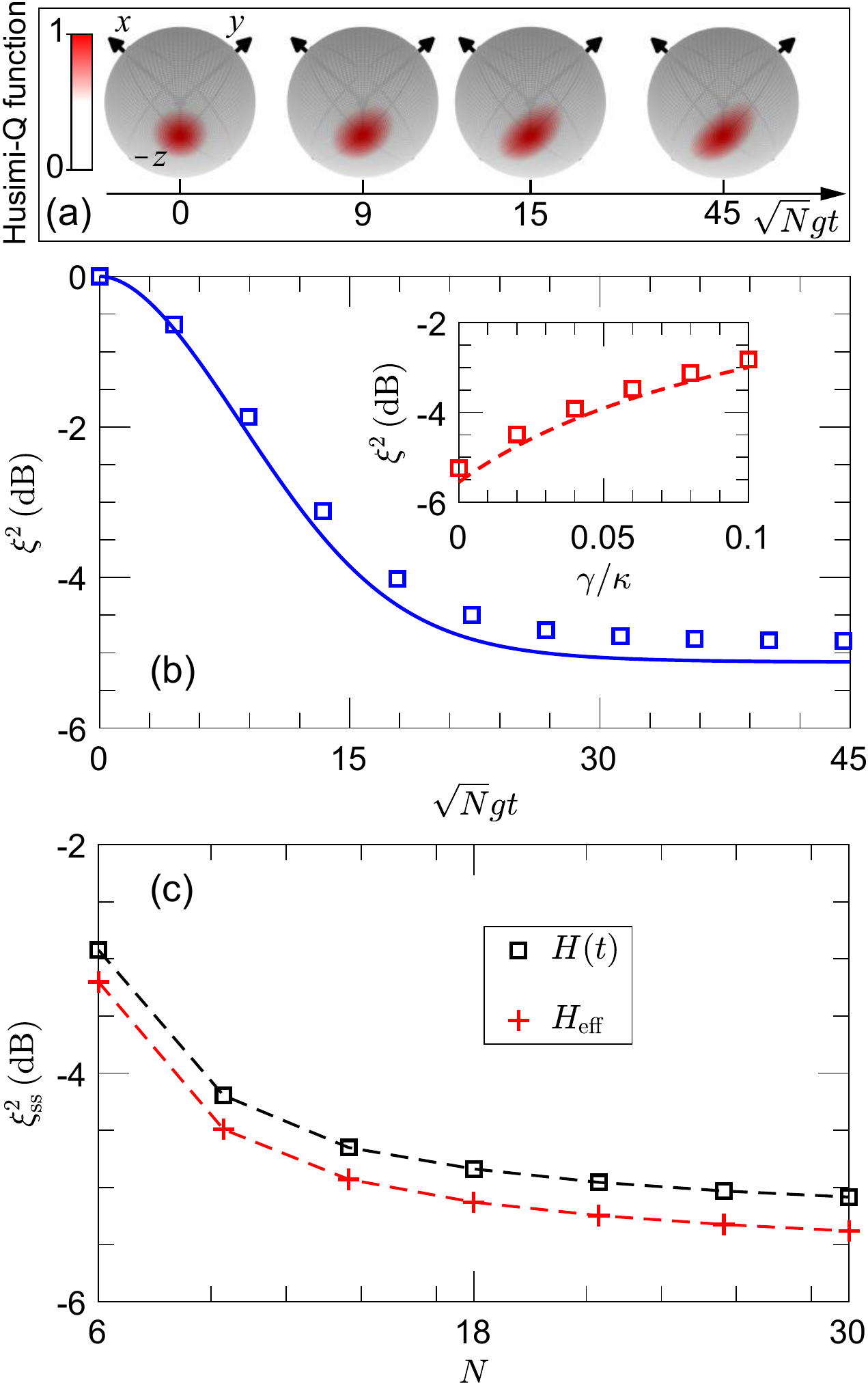}
	\caption{(a) Husimi distribution $Q\left(\theta, \phi\right)$ at different times. The distribution $Q\left(\theta,\phi\right)$ has been normalized to the range $\left[0,1\right]$. (b) Evolution of the squeezing parameter $\xi^{2}$. The inset shows an increase in $\xi^{2}$ with increasing $\gamma/\kappa$, at time $\sqrt{N}gt=45$. (c) Steady-state $\xi^{2}$ versus the number $N$ of atoms. Here, curves in (b) and crosses in (c) are predictions of  $H_{\rm eff}$, while all other plots are obtained from $H\left(t\right)$. This shows that $H_{\rm eff}$ can well describe the system dynamics. In (a) and (b), we assumed that $N=18$. In all plots, we assumed that $g=0.5\kappa$, $\Delta_{c}=200\kappa$, $\Omega=0.2\Delta_{c}$, $A_{m}=0.34\omega_{m}$, and that, except the inset in (b), $\gamma=0.01\kappa$. For time evolution, all atoms are initialized in the ground state and the cavity is in the vacuum.}\label{fig_spin}
\end{figure}

We begin by restricting our discussion to the limits $\left\{g, \Omega\right\}\ll\Delta_{c}$ and $A_{m}\ll\omega_{m}$. In such a case, the squeezing sideband resulting from the driving $\Omega$ enables a coupling in the form 
\begin{equation}
\exp\left(i\theta_{L}\right)aS_{-}+\exp\left(-i\theta_{L}\right)a^{\dag}S_{+},
\end{equation}
with strength $g\Omega/2\Delta_{c}$. The coupling becomes resonant when $\omega_{q}\approx\omega_{L}-\omega_{c}$. Such a coupling can be understood from the interaction between a driving photon pair and a single atom, as shown in Fig.~\ref{fig_schematics2}(c). The ground state $|0,g\rangle$ is driven to a virtual excited state via the two-photon driving $\Omega$ with detuning $\approx2\Delta_{c}$, and then is resonantly coupled to the state $|1,e\rangle$ via the atom-cavity coupling $g$. Here, the number in the ket refers to the cavity-photon number. This mechanism is responsible for anti-Stokes scattering of correlated photon pairs mentioned above. Furthermore,  for $\omega_{q}\approx\omega_{c}-\omega_{m}$, the coupling, 
\begin{equation}
a^{\dag}S_{-}+aS_{+},
\end{equation}
is also made resonant via a first-order Floquet sideband, but its strength becomes $gA_{m}/2\omega_{m}$. As we demonstrate in more detail in Appendix~\ref{Effective Hamiltonian and decay of the collective spin}, these two resonant couplings lead to an effective Hamiltonian  
\begin{equation}
H_{\rm eff}=ga^{\dagger}\left(G_{-}S_{-}+G_{+}S_{+}\right)+{\rm H.c.},
\end{equation}
where $G_{-}=A_{m}/2\omega_{m}$ and $G_{+}=\Omega/2\Delta_{c}$. Here, we have set $\theta_{L}=-\pi/2$ and a phase factor $i$ has been absorbed into $a$.  The dynamics driven by $H_{\rm eff}$ describes two distinct atomic transitions, which can cause the spin squeezed state to become a dark state~\cite{parkins2006unconditional,zheng2012generation,dalla2013dissipative,ma2013dissipation}. In particular, in the optimal case of $\gamma\rightarrow0$, assuming $G_{+}$ to be very close to $G_{-}$, it yields the maximally spin squeezed state corresponding to the Heisenberg-limited noise reduction $\propto 1/N$. In Fig.~\ref{fig_spin}(a) we plot the spin Husimi distribution $Q\left(\theta, \phi\right)$ using $H\left(t\right)$. Here, $Q\left(\theta, \phi\right)=\left(2N+1\right)/\left(4\pi\right)\langle{\rm CSS}|R^{\dag}\left(\theta, \phi\right)\rho R\left(\theta, \phi\right)|{\rm CSS}\rangle$, where $|{\rm CSS}\rangle$ refers to a coherent-spin state with all the atoms in the excited state, and $R\left(\theta, \phi\right)=\exp\left[i\theta\left(S_{x}\sin\phi-S_{y}\cos\phi\right)\right]$ is a rotation operator, which rotates $|{\rm CSS}\rangle$ by an angle $\theta$ about the axis $\left(-\sin\phi, \cos\phi, 0\right)$ of the collective Bloch sphere. We find, as predicted by $H_{\rm eff}$, that quantum noise is reduced along the $x$ direction, at the expense of increased quantum noise along the $y$ direction.

To quantify the degree of spin squeezing, we use the parameter defined as~\cite{wineland1992spin,wineland1994squeezed}:
\begin{equation}
\xi^{2}=N\frac{\average{\Delta S_{\perp}}^{2}_{\rm min}}{|\average{{\bf S}}|^{2}},
\end{equation}
where ${\bf S}=\left(S_{x},S_{y},S_{z}\right)$ is the total spin operator, and $\average{\Delta S_{\perp}}^{2}_{\rm min}=(\average{\left({\bf S}\cdot{\bf n_{\perp}}\right)^{2}}-\average{{\bf S}\cdot{\bf n_{\perp}}}^{2})_{\rm min}$ is the minimum spin fluctuation in the ${\bf n_{\perp}}$ direction perpendicular to the mean spin $\average{{\bf S}}$. Spin squeezed states, where quantum fluctuation in one quadrature is reduced below the standard quantum limit, exhibit $\xi^{2}<1$. We find from Fig.~\ref{fig_spin}(b) that a strong loss of a weakly and parametrically driven Floquet cavity can enable $\xi^{2}$ to be $\ll1$ in the steady state. In contrast, atomic spontaneous emission carries away information about spin-squeezed states, and hence limits spin squeezing, as plotted in the inset of Fig.~\ref{fig_spin}(b). In Fig.~\ref{fig_spin}(c), we plot the steady-state $\xi^{2}$, labeled $\xi^{2}_{\rm ss}$, versus the number $N$ of atoms. The enhancement of spin squeezing by increasing $N$ has a lower bound which, as demonstrated below, is determined by the ratio $G_{+}/G_{-}$ in the limit of $N\rightarrow\infty$. 

\section{Spin-wave approximation}
We now consider the case of $N\rightarrow\infty$, so that the dynamics of the collective spin can be mapped to a bosonic mode $b$, i.e., $S_{-}\approx\sqrt{N}b$. Here, we have assumed that the number of excited atoms is much smaller than the total number $N$, i.e., $\langle b^{\dag}b\rangle \ll N$, and have made the spin-wave approximation. The effective Hamiltonian is correspondingly transformed to 
\begin{equation}H_{\rm eff}^{\rm SWA}=G\sqrt{N}g\left(a^{\dagger}\beta+{\rm H.c.}\right),
\end{equation}
where $G^{2}=G_{-}^{2}-G_{+}^{2}$, and $\beta=\cosh\left(r\right)b+\sinh\left(r\right)b^{\dag}$, with $\tanh\left(r\right)=G_{+}/G_{-}$, describes a squeezed mode of the collective spin. The cavity loss thus can drive the mode $\beta$ to its vacuum, which corresponds to a squeezed vacuum state of the mode $b$. Under the spin-wave approximation, the parameter $\xi^{2}$ is likewise transformed to 
\begin{equation}
\xi^{2}_{\rm SWA}=1+2\left(\average{b^{\dag}b}-|\average{bb}|\right). 
\end{equation}
This implies that the two-atom correlation, $\langle bb\rangle$, characterizes a key signature of spin squeezing. 

\begin{figure}[t]
	\centering
	\includegraphics[width=8.3cm]{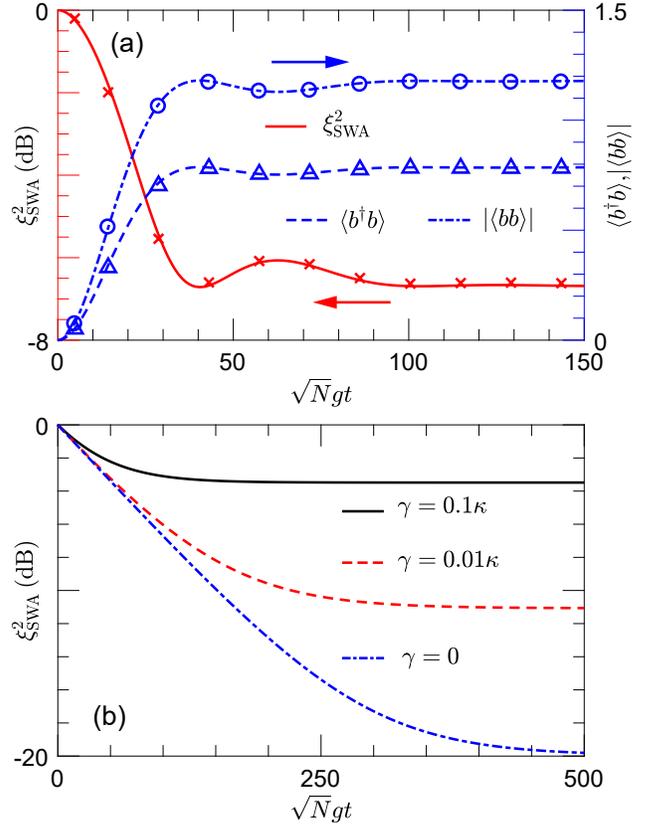}
	\caption{(a) Comparison between the effective (curves) and full (symbols) Hamiltonians under the spin-wave approximation. The spin-squeezing parameter ($\xi^{2}_{\rm SWA}$, left red axis) and the two-atom correlation ($\left|\langle bb\rangle\right|$, right blue axis) are shown. We have set $\omega_{m}\approx\omega_{c}-\omega_{q}+Ng^{2}/\Delta_{c}$. This yields an excellent agreement.  (b) Spin-squeezing parameter $\xi_{\rm SWA}^{2}$ given in Eq.~(\ref{sq_xi_SWA}) for $G_{+}/G_{-}=0.98$. In (a) we set: $\Delta_{c}=200\kappa$, $\Omega=0.1\Delta_{c}$, $A_{m}=0.15\omega_{m}$, $\gamma=0.01\kappa$; and in both plots: $\sqrt{N}g=10\kappa$.}\label{fig_SWA}
\end{figure}

In order to achieve $H_{\rm eff}^{\rm SWA}$, we have neglected the off-resonant coupling to the zero-order Floquet sideband, which lowers the degree of spin squeezing [see Figs.~\ref{fig_spin}(b) and~\ref{fig_spin}(c)]. Let us now consider this off-resonant coupling. In the limit $\sqrt{N}g\ll \Delta_{c}$, such a coupling shifts the cavity and atomic resonances~\cite{gamel2010time}, and as a result it causes an additional detuning $\delta\approx Ng^{2}/\Delta_{c}$ between cavity and atoms. To avoid this undesired effect, the modulating frequency $\omega_{m}$ needs to be modified to compensate $\delta$, such that $\omega_{m}\approx\omega_{c}-\omega_{q}+Ng^{2}/\Delta_{c}$ (see Appendix~\ref{Detuning arising from non-resonant couplings}). With such a modification, we directly calculate the  parameter $\xi_{\rm SWA}^{2}$ and the correlation $\langle bb \rangle$ obtained using the effective and full Hamiltonians under the spin-wave approximation. We find from Fig.~\ref{fig_SWA}(a) that after compensating the detuning $\delta$, the full dynamics are in excellent agreement with the desired effective dynamics. This allows us to investigate stronger spin squeezing, according to such an effective Hamiltonian.

Based on $H_{\rm eff}^{\rm SWA}$, we derive the steady-state $\langle b^{\dag}b\rangle$ and $\langle bb \rangle$, yielding
\begin{equation}
\langle b^{\dag}b\rangle_{\rm ss}=\mathcal{A}\sinh^{2}\left(r\right),
\end{equation}
and
\begin{equation}
\langle bb\rangle_{\rm ss}=-\mathcal{A}\sinh\left(2r\right)/2,
\end{equation} 
where $\mathcal{A}=4G^{2}C/\left[\left(4G^{2}C+1\right)\left(1+\gamma/\kappa\right)\right]$. Here, $C=Ng^{2}/\kappa\gamma$ is the collective cooperativity. Having $r\geq1$ gives $\left(\langle b^{\dag}b\rangle_{\rm ss}-\langle bb\rangle_{\rm ss}\right)\rightarrow-\mathcal{A}/2$, and therefore a strong spin squeezed state is achieved if $\mathcal{A}\rightarrow1$. More specifically, we consider the steady-state $\xi_{\rm SWA}^{2}$ expressed as
\begin{equation}\label{eq_squeezing_parameter_boson}
\left(\xi^{2}_{\rm SWA}\right)_{\rm ss}=1+\mathcal{A}\left[\exp\left(-2r\right)-1\right].
\end{equation}
This demonstrates that if $G_{+}\rightarrow G_{-}$, then the parameter $r$ and, thus, spin squeezing increases.  However, as $G_{+}\rightarrow G_{-}$,  the effective coupling, $G\sqrt{N}g$, between modes $a$ and $\beta$ tends to zero  (i.e., $G\rightarrow0$), which suppresses the cooling of the mode $\beta$. The optimal SSSS therefore results from a tradeoff between these two processes~\cite{dalla2013dissipative,ma2013dissipation,PhysRevA.90.013838}. Furthermore, we find that for a spin-squeezed steady state, the number of excited atoms scales as  $\langle b^{\dagger}b\rangle\propto e^{2r}$, but at the same time, the spin-wave approximation requires $\langle b^{\dag}b\rangle\ll N$. To demonstrate the squeezing scaling, we assume that in the steady state, $\langle b^{\dag}b\rangle\propto N^{\mu}$, where $0<\mu<1$. In this case, $\langle b^{\dag}b\rangle\ll N$, and consequently $\xi^{2}_{\rm SWA}\propto N^{-\mu}$, is justified even for $\mu\rightarrow1$, as long as $N$ is sufficiently large. Hence, our approach can, in principle, enable spin squeezing to be far below the standard quantum limit, and approach the Heisenberg limit in a large ensemble. 

To consider the squeezing time, we adiabatically eliminate the cavity mode (see Appendix~\ref{Adiabatic elimination of the cavity mode}), yielding
\begin{equation}\dot{\rho}_{\rm spin}=\frac{\gamma_{c}}{2}\mathcal{L}\left(\beta\right)\rho_{\rm spin}+\frac{\gamma}{2}\mathcal{L}\left(b\right)\rho_{\rm spin},
\end{equation}
where $\rho_{\rm spin}$ describes the reduced density matrix of the collective spin, and $\gamma_{c}=4G^{2}Ng^{2}/\kappa$ represents the cavity-induced atomic decay. According to this adiabatic master equation, $\langle b^{\dag}b\rangle$ and $\langle bb\rangle$ evolve as 
\begin{equation}
X=\left(X_{\rm ini}-X_{\rm ss}\right)\exp\left[-\left(\gamma_{c}+\gamma\right)t\right]+X_{\rm ss},
\end{equation}
where $X=\langle b^{\dag}b\rangle$, $\langle bb\rangle$, and $X_{\rm ini}$ refers to the initial $X$.
We therefore find that the atomic ensemble can be driven into a spin-squeezed state from any initial state in the spin-$\frac{N}{2}$ manifold. Under time evolution, $\xi_{\rm SWA}^{2}$ is given by
\begin{equation}\label{sq_xi_SWA}
	\xi_{\rm SWA}^{2}=\left(\xi_{\rm SWA}^{2}\right)_{\rm ss}-\left[\left(\xi_{\rm SWA}^{2}\right)_{\rm ss}-1\right]\exp\left[-\left(\gamma_{c}+\gamma\right)t\right].
\end{equation}
Here, we have assumed, for simplicity, that $\langle b^{\dag}b\rangle_{\rm ini}=\langle bb\rangle_{\rm ini}=0$. This expression predicts that time evolution leads to an exponential squeezing with a rate $\gamma_{c}+\gamma$, as plotted in Fig.~\ref{fig_SWA}(b). For a realistic setup, e.g., a nitrogen-vacancy (NV) spin ensemble coupled to a superconducting resonator (see below),  a negligibly small spin decay rate $\gamma\rightarrow0$ and a typical collective coupling $\sqrt{N}g\approx2\pi\times10$~MHz could result in a spin-squeezed steady state of $\approx-20$~dB in a squeezing time $\approx8$~$\mu$s. This allows us to neglect spin decoherence, because the coherence time in ensembles of NV centers can experimentally reach the order of ms~\cite{stanwix2010coherence} or even $\sim 1$~s~\cite{bar2013solid}. 

\section{Proposed experimental implementation} 
As an example, we now consider a hybrid quantum system~\cite{xiang2013hybrid,RevModPhys.85.623,PhysRevLett.117.015502}, where a superconducting transmission line (STL), terminated by a superconducting quantum interference device (SQUID), is magnetically coupled to an NV spin ensemble in diamond (see Appendix~\ref{Possible implementations with hybrid quantum systems} for details). The coherent coupling of an STL cavity to an NV spin ensemble has already been widely implemented in experiments~\cite{kubo2010strong, amsuss2011cavity, kubo2011hybrid, PhysRevA.85.012333, putz2014protecting,grezes2014multimode, astner2017coherent}. In particular, Refs.~\cite{kubo2010strong,kubo2011hybrid,PhysRevA.85.012333} used a SQUID to control the cavity frequency. Therefore to achieve a parametrically driven Floquet cavity, we connect a SQUID to one end of the STL. We then assume the driving phase $f\left(t\right)$ across the SQUID loop to be 
\begin{equation}
f\left(t\right)=f_{0}+\left[f_{1}+f_{2}\left(t\right)\right]\cos\left(\omega_{L}t+\theta_{L}\right)+f_{3}\sin\left(\omega_{m}t\right).
\end{equation}
Here, the components $f_{1}$ and $f_{2}\left(t\right)$ result in the drivings $\Omega$ and $\Omega_{1}\left(t\right)$, respectively, while the component $f_{3}$  is to modulate the cavity frequency $\omega_{c}$. Moreover, the electronic ground state of NV centers is a spin triplet, whose $m_{s}=0$ and $m_{s}=\pm1$ sublevels are labeled by $|0\rangle$ and $|\pm1\rangle$. There exists a zero-field splitting $\approx2.87$~GHz between state $|0\rangle$ and states $|\pm1\rangle$. In the presence of an external magnetic field, the states $|\pm 1\rangle$ are further split through the Zeeman effect, which enables a two-level atom with $|0\rangle$ as the ground state and $|-1\rangle$ (or $|+1\rangle$) as the excited state. When the diamond containing an NV spin ensemble is placed on top of the STL, the cavity photon can drive the transition $|0\rangle\rightarrow|-1\rangle$ (or $\rightarrow|+1\rangle$) via a magnetic coupling.

\section{Conclusions}
We have introduced an experimentally feasible method for how to implement Heisenberg-limited SSSS of atomic ensembles in a weakly and parametrically driven Floquet cavity. This method demonstrates a counterintuitive phenomenon: the weak squeezing of light can induce strong spin squeezing. This approach does not require multiple actions on atoms, thus greatly reducing the experimental complexity. We have also shown an anti-Stokes scattering process, induced by an atom, of a correlated photon pair, where one photon of the photon pair is scattered into a higher-energy mode by absorbing a fraction of the energy of the other photon, and the remaining energy of the absorbed photon excites the atom. If the scattered photon is further absorbed by another atom before being lost, then such a scattering process can also generate an atom-pair excitation and, as a consequence, can enable TAT spin squeezing. The two distinct atomic transitions demonstrated are functionally similar to, but experimentally simpler than, the double off-resonant Raman transition in multi-level atoms widely used for generating spin squeezing~\cite{sorensen2002entangling,dalla2013dissipative}. Thus, we could expect that our method can provide a universal building block for implementing spin squeezed states, and simulating ultrastrong light-matter interaction~\cite{kockum2019ultrastrong,forn2019ultrastrong} and quantum many-body phase transition~\cite{kirton2019introduction}.

\begin{acknowledgments}
We thank Fabrizio Minganti, Nathan Shammah, and Vincenzo Macr\`{i} for their valuable discussions.
	Y.-H.C. is supported by the Japan Society for the Promotion of Science (JSPS) Foreign Postdoctoral Fellowship No. P19028. A.M. is supported by the Polish National Science Centre (NCN)
under the Maestro Grant No. DEC-2019/34/A/ST2/00081.
	F.N. is supported in part by: NTT Research,
Army Research Office (ARO) (Grant No. W911NF-18-1-0358),
Japan Science and Technology Agency (JST)
(via the Q-LEAP program and the CREST Grant No. JPMJCR1676),
Japan Society for the Promotion of Science (JSPS)
(via the KAKENHI Grant No. JP20H00134, and the JSPS-RFBR Grant No. JPJSBP120194828), and
the Grant No. FQXi-IAF19-06 from the Foundational Questions Institute Fund (FQXi),
a donor advised fund of the Silicon Valley Community Foundation.
\end{acknowledgments}

\appendix 

\setcounter{equation}{0} \setcounter{figure}{0}
\setcounter{table}{0} \makeatletter
\renewcommand{\thefigure}{A\arabic{figure}}

\section{Effective Hamiltonian and decay of the collective spin}
\label{Effective Hamiltonian and decay of the collective spin}

Let us first derive the effective Hamiltonian $H_{\rm eff}$. We begin with the full Hamiltonian in a rotating frame,
\begin{equation}\label{seq:full Hamiltonian 00}
H\left(t\right)=H_{0}+H_{1}\left(t\right), 
\end{equation}
where
\begin{align}
H_{0}=\;&\Delta_{c}a^{\dag}a+\Delta_{q}S_{z}\nonumber\\
&+g\left(aS_{+}+{\rm H.c.}\right)+\frac{1}{2}\Omega\left[\exp\left(i\theta_{L}\right)a^{2}+{\rm H.c.}\right],\\\
H_{1}\left(t\right)=\;&A_{m}\sin\left(\omega_{m}t\right)a^{\dag}a\nonumber\\
&+\frac{1}{2}\Omega_{1}\left(t\right)\left[\exp\left(i\theta_{L}\right)a^{2}+{\rm H.c.}\right].
\end{align}
Here, $\Delta_{c/q}=\omega_{c/q}-\omega_{L}/2$, where $\omega_{c}$ is the cavity frequency, $\omega_{q}$ is the atomic transition frequency, and $\omega_{L}$ is the frequency of the two-photon driving. The cavity mode $a$ is dressed by the detuned two-photon driving $\Omega$, and becomes a squeezed mode $\alpha$. This squeezing operation can be described by the Bogoliubov transformation,
\begin{equation}\label{seq:Bogoliubov transformation}
\alpha=\cosh\left(r_{c}\right)a+\exp\left(-i\theta_{L}\right)\sinh\left(r_{c}\right)a^{\dag},
\end{equation} 
where 
\begin{equation}
r_{c}=\frac{1}{4}\ln\frac{\Delta_{c}+\Omega}{\Delta_{c}-\Omega}
\end{equation}
determines the degree of squeezing of the cavity field. It then follows that
\begin{equation}\label{seq:squeezed mode}
\Delta_{c}a^{\dag}a+\frac{1}{2}\Omega\left[\exp\left(i\theta_{L}\right)a^{2}+{\rm H.c.}\right]=\omega_{s}\alpha^{\dag}\alpha,
\end{equation}
where $\omega_{s}=\sqrt{\Delta_{c}^{2}-\Omega^{2}}$ is the squeezed-mode frequency. It is seen from Eqs.~(\ref{seq:Bogoliubov transformation}) and~(\ref{seq:squeezed mode}) that, inside the cavity, there exist an upper squeezing sideband at $\left(\omega_{L}/2+\omega_{s}\right)$ and a lower squeezing sideband at $\left(\omega_{L}/2-\omega_{s}\right)$.
The Hamiltonian $H\left(t\right)$, when expressed in terms of the mode $\alpha$, is transformed to
\begin{align}\label{seq:full Hamiltonian 02}
H\left(t\right)=&\left[\omega_{s}+A_{m}^{\prime}\sin\left(\omega_{m}t\right)\right]\alpha^{\dag}\alpha+\Delta_{q}J_{z}\nonumber\\
&+g\cosh\left(r_{c}\right)\left(\alpha S_{+}+{\rm H.c.}\right)\nonumber\\
&-g\sinh\left(r_{c}\right)\left(e^{i\theta_{L}}\alpha S_{-}+{\rm H.c.}\right),
\end{align}
where $A_{m}^{\prime}=A_{m}\cosh\left(2r_{c}\right)[1-\tanh^{2}\left(2r_{c}\right)]$. In Eq.~(\ref{seq:full Hamiltonian 02}), we have assumed that $\Omega_{1}\left(t\right)=A_{m}\tanh\left(2r_{c}\right)\sin\left(\omega_{m}t\right)$, such that an undesired parametric driving of the mode $\alpha$ can be eliminated. The last two terms of Eq.~(\ref{seq:full Hamiltonian 02}) describe two distinct spin-cavity couplings, which are associated with the upper and lower squeezing sidebands, respectively.

We now focus our discussion on the limit $\Omega\ll \Delta_{c}$, where light squeezing inside the cavity is very weak. Such weak squeezing can avoid two-photon correlation noise and thermal noise, which are generally considered detrimental in strong-squeezing processes~\cite{milburn1981production,lu2015squeezed}. In this limit, we have
\begin{equation}
r_{c}\approx\frac{\Omega}{2\Delta_{c}}\ll1,
\end{equation}
which, in turn, gives
\begin{equation}
\cosh\left(r_{c}\right)\approx1\gg\sinh\left(r_{c}\right)\approx\frac{\Omega}{2\Delta_{c}}.
\end{equation} 
Consequently, the squeezed mode $\alpha$ can, according to the Bogoliubov transformation in Eq.~(\ref{seq:Bogoliubov transformation}), be approximated by the bare mode $a$, i.e., 
\begin{equation}
\alpha\approx a.
\end{equation}
The Hamiltonian $H\left(t\right)$ is therefore approximated by
\begin{align}\label{seq:full Hamiltonian 03}
H\left(t\right)\approx H^{\prime}\left(t\right)=&\left[\omega_{s}+A_{m}^{\prime}\sin\left(\omega_{m}t\right)\right]a^{\dag}a+\Delta_{q}J_{z}\nonumber\\
&+g\cosh\left(r_{c}\right)\left(a S_{+}+{\rm H.c.}\right)\nonumber\\
&-g\sinh\left(r_{c}\right)\left(e^{i\theta_{L}}a S_{-}+{\rm H.c.}\right).
\end{align}
Note that, in the limit of $\Omega\ll\Delta_{c}$, the upper squeezing sideband becomes the cavity resonance due to $\omega_{L}/2+\omega_{s}\approx\omega_{c}$, and the lower squeezing sideband is likewise shifted to $\omega_{L}-\omega_{c}$  (i.e., $\omega_{L}/2-\omega_{s}\approx\omega_{L}-\omega_{c}$).

\begin{figure*}[t]
	\centering
	\includegraphics[width=10.0cm]{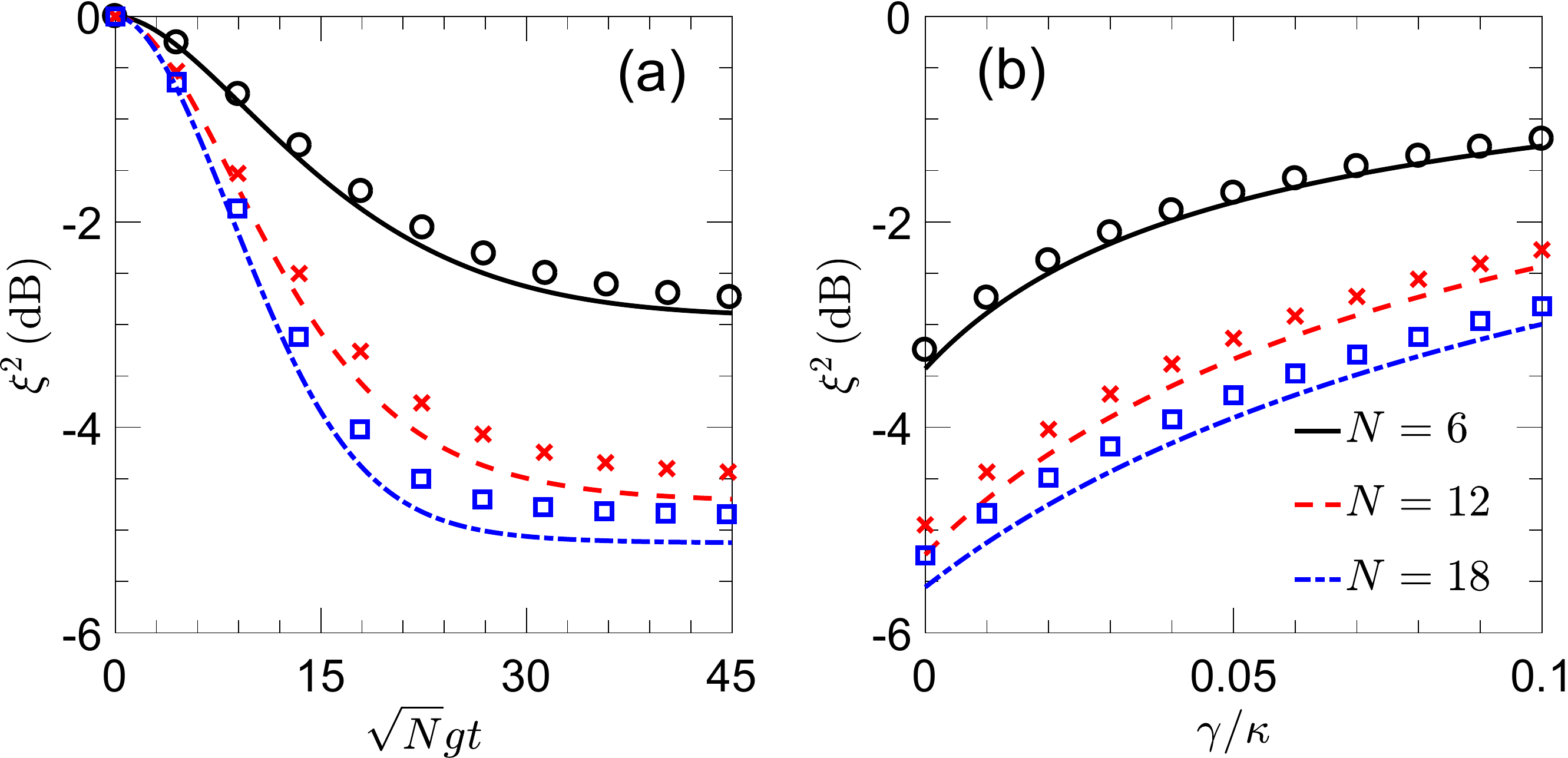}
	\caption{Spin squeezing parameter $\xi^{2}$. (a) shows the time evolution for $\gamma=0.01\kappa$, and in (b) the ratio $\gamma/\kappa$ is varied at a fixed time $\sqrt{N}gt=45$, for $N=6$, $12$, and $18$. In both plots, curves and symbols are results obtained using the effective ($H_{\rm eff}$) and full [$H\left(t\right)$] Hamiltonians, respectively. We have assumed that $g=0.5\kappa$, $\Delta_{c}=200\kappa$, $\Omega=0.2\Delta_{c}$, $A_{m}=0.34\omega_{m}$, $\gamma=0.01\kappa$, and also that all atoms are initialized in the ground state and the cavity is in the vacuum.}\label{sfig-spin}
\end{figure*}

Upon introducing a unitary transformation
\begin{equation}
U\left(t\right)=\exp\left\{i\left[\omega_{s}t-\eta_{m}\cos\left(\omega_{m}t\right)\right]a^{\dag}a+i\Delta_{q}S_{z}t\right\},
\end{equation}
with $\eta_{m}=A_{m}^{\prime}/\omega_{m}$,
$H^{\prime}\left(t\right)$ in Eq.~(\ref{seq:full Hamiltonian 03}) is then transformed to
\begin{widetext}
\begin{align}\label{seq:full Hamiltonian 04}
H^{\prime}\left(t\right)=\;&g\cosh\left(r_{c}\right)\sum_{n=-\infty}^{+\infty}\left\{i^{n}J_{n}\left(\eta_{m}\right)aS_{+}\exp\left[-i\left(\omega_{s}-\Delta_{q}-n\omega_{m}t\right)t\right]+{\rm H.c.}\right\}\nonumber\\
&-g\sinh\left(r_{c}\right)\sum_{n=-\infty}^{+\infty}\left\{e^{i\theta_{L}}i^{n}J_{n}\left(\eta_{m}\right)aS_{-}\exp\left[-i\left(\omega_{s}+\Delta_{q}-n\omega_{m}t\right)t\right]+{\rm H.c.}\right\},
\end{align}
\end{widetext}
where we have used the Jacobi-Anger identity
\begin{equation}
\exp\left[i\eta_{m}\cos\left(\omega_{m}t\right)\right]=\sum_{n=-\infty}^{+\infty}i^nJ_{n}\left(\eta_{m}\right)\exp\left(in\omega_{m}t\right),
\end{equation}
with $J_{n}\left(\eta_{m}\right)$ being the $n$th-order Bessel function of the first kind.

We find that, when $\omega_{s}+\Delta_{q}=0$ (i.e., $\omega_{q}\approx\omega_{L}-\omega_{c}$), the last sum in Eq.~(\ref{seq:full Hamiltonian 04}) contains a resonant coupling of the form 
\begin{equation}
\exp\left(i\theta_{L}\right)aS_{-}+\exp\left(-i\theta_{L}\right)a^{\dag}S_{+},
\end{equation}
with strength $g\sinh\left(r_{c}\right)J_{0}\left(\eta_{m}\right)\approx g\Omega/2\Delta_{c}$. Such a coupling, which originates from the lower squeezing sideband at $\left(\omega_{L}-\omega_{c}\right)$, describes the anti-Stokes scattering process of a driving photon pair interacting with an atom. Specifically, one photon of the photon pair is scattered into the cavity resonance by absorbing part of the energy of the other photon, and simultaneously the remaining energy of the absorbed photon excites the atom. When we further choose $2\omega_{s}=\omega_{m}$ (i.e., $\omega_{q}\approx\omega_{c}-\omega_{m}$), the first sum in Eq.~(\ref{seq:full Hamiltonian 04}) also contains a resonant coupling of the form 
\begin{equation}
aS_{+}+a^{\dag}S_{-},
\end{equation}
with strength $g\cosh\left(r_{c}\right)J_{1}\left(\eta_{m}\right)\approx gA_{m}/2\omega_{m}$. This coupling, which is mediated via a first-order Floquet sideband at $\left(\omega_{c}-\omega_{m}\right)$, describes that a detuned atom can emit a photon into the cavity resonance.  Under the assumptions, $g\ll\Delta_{c}$ and $A_{m}\ll\omega_{m}$ (i.e., $\eta_{m}\ll 1$), off-resonant couplings can be neglected, and thus the system dynamics is determined by the following effective Hamiltonian 
\begin{equation}\label{seq:effective Hamiltonian}
H_{\rm eff}=ga^{\dag}\left(G_{-}S_{-}+G_{+}S_{+}\right)+{\rm H.c.},
\end{equation}
where $G_{-}=A_{m}/2\omega_{m}$ and $G_{+}=\Omega/2\Delta_{c}$. Here, we have set $\theta_{L}=-\pi/2$ and a phase factor $i$ has been absorbed into $a$.

\begin{figure*}[t]
	\centering
	\includegraphics[width=16.0cm]{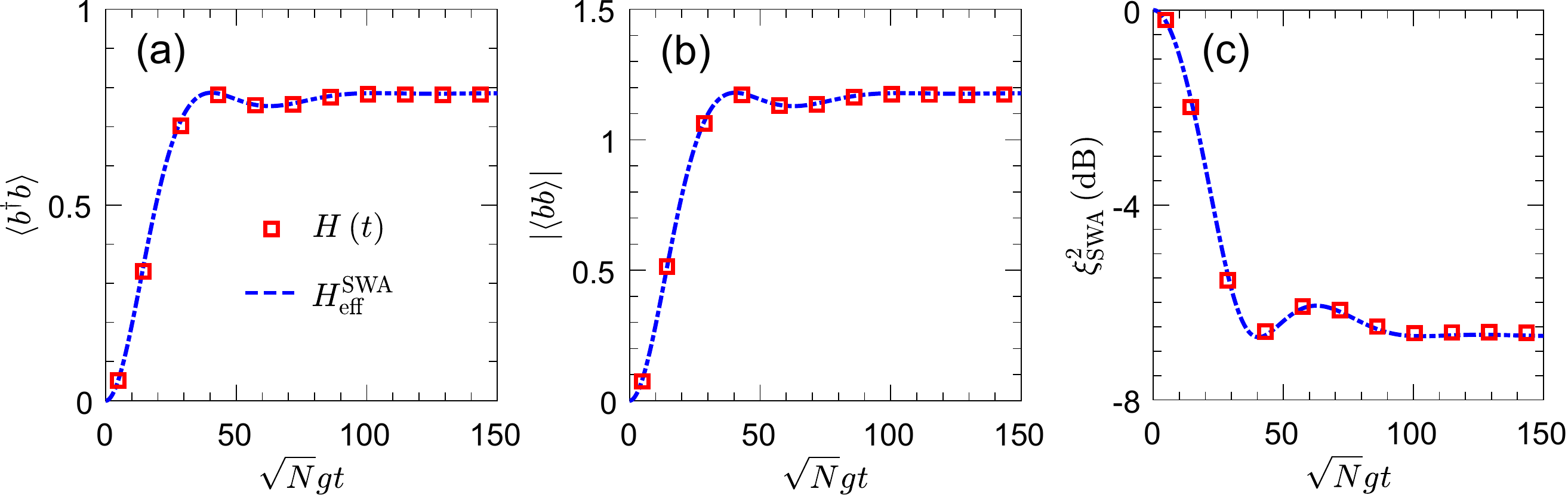}
	\caption{Evolution of (a) the excited-atom number $\langle b^{\dag}b\rangle$, (b) the two-atom correlation $\langle bb\rangle$, and (c) the spin squeezing parameter $\xi^{2}_{\rm SWA}$. In all plots, squares are obtained from the full Hamiltonian $H\left(t\right)$ by compensating the detuning $\delta$, and dashed curves are given by the effective Hamiltonian $H_{\rm eff}^{\rm SWA}$. Here, we have made the spin-wave approximation for $H\left(t\right)$. We have assumed that  $\Delta_{c}=200\kappa$, $\Omega=0.1\Delta_{c}$, $A_{m}=0.15\omega_{m}$, $\gamma=0.01\kappa$, $\sqrt{N}g=10\kappa$,  and also that all atoms are initialized in the ground state and the cavity is in the vacuum.}\label{sfig-SWA}
\end{figure*}

We now consider the dissipative dynamics of the system. The dissipative dynamics can be described with the Lindblad operator 
\begin{equation}
\mathcal{L}\left(o\right)\rho=2o\rho o^{\dag}-o^{\dag}o\rho-\rho o^{\dag}o, 
\end{equation}
such that $\frac{\kappa}{2}\mathcal{L}\left(a\right)\rho$ corresponds to cavity loss, and $\frac{\gamma}{2}\sum_{j=1}^{N}\mathcal{L}\left(\sigma_{j}^{-}\right)\rho$ to atomic spontaneous emission. It is, in general, very difficult to perform numerical simulations for a large ensemble, because the Hilbert space of the ensemble grows as $2^{N}$. In order to reduce the dimension of this Hilbert space, we follow the method in Refs.~\cite{gelhausen2017many,shammah2018open,macri2020spin}, and perform a Fourier transformation,
\begin{equation}
\widetilde{\sigma}_{k}^{-}=\frac{1}{\sqrt{N}}\sum_{j}\exp\left(-ikj\right)\sigma_{j}^{-}.
\end{equation}
It then follows, using $\sqrt{N}\widetilde{\sigma}_{k=0}^{\pm}=S_{\pm}$, that
\begin{equation}\label{seq:atomic emission in momentum space}
\sum_{j}\mathcal{L}\left(\sigma_{j}^{-}\right)\rho=\frac{1}{N}\mathcal{L}\left(S_{-}\right)\rho+\sum_{k\neq0}\mathcal{L}\left(\widetilde{\sigma}_{k}^{-}\right)\rho,
\end{equation}
where the first and second terms on the right-hand side describe the dissipative processes of the zero and nonzero momentum modes, respectively. It is seen, from the full Hamiltonian $H\left(t\right)$ in Eq.~(\ref{seq:full Hamiltonian 00}) or the effective Hamiltonian $H_{\rm eff}$ in Eq.~(\ref{seq:effective Hamiltonian}), that the coherent dynamics only involves the zero ($k=0$) momentum mode. This implies that we can only focus on the zero momentum mode; that is,
\begin{equation}\label{seq:zero momentum mode dissipation}
\sum_{j}\mathcal{L}\left(\sigma_{j}^{-}\right)\rho=\frac{1}{N}\mathcal{L}\left(S_{-}\right)\rho.
\end{equation}
This is valid in the steady-state limit or the long-time limit, because the nonzero momentum modes in Eq.~(\ref{seq:atomic emission in momentum space}) only decay. In particular, such a reduction can exactly describe the dissipative dynamics of an atomic ensemble initially in the ground state. Therefore, 
the dynamics of the system is driven by the following master equation 
\begin{equation}\label{seq:master equation 01}
\dot{\rho}=i\left[\rho,\mathcal{H}\right]+\frac{\kappa}{2}\mathcal{L}\left(a\right)\rho+\frac{\gamma}{2N}\sum_{j=1}^{N}\mathcal{L}\left(S_{-}\right)\rho,
\end{equation}
where $\mathcal{H}$ can be taken to be $H\left(t\right)$ for the full dynamics or to be $H_{\rm eff}$ for the effective dynamics. 

In Fig.~\ref{sfig-spin}, we numerically integrated the master equation in Eq.~(\ref{seq:master equation 01}), with the full Hamiltonian $H\left(t\right)$ and the effective Hamiltonian $H_{\rm eff}$. Specifically, we plot the spin squeezing parameter $\xi^{2}$ versus the scaled evolution time $\sqrt{N}gt$ in Fig.~\ref{sfig-spin}(a) and versus the ratio $\gamma/\kappa$ in Fig.~\ref{sfig-spin}(b). The result in this figure reveals that $H_{\rm eff}$ can describe well the dynamics of the system. The divergence between them mainly arises from neglecting an off-resonant coupling to the zero-order Floquet sideband. In the next section, we discuss how to remove the detrimental effect induced by such an off-resonant coupling under the spin-wave approximation.

\section{Detuning arising from non-resonant couplings}	
\label{Detuning arising from non-resonant couplings}
Under the spin-wave approximation (i.e., $S_{-}\approx\sqrt{N}b$), the Hamiltonian $H^{\prime}\left(t\right)$ in Eq.~(\ref{seq:full Hamiltonian 04}) becomes
\begin{widetext}
\begin{align}\label{seq:full Hamiltonian SWA}
H^{\prime}_{\rm SWA}\left(t\right)=\;&g_{\rm col}\cosh\left(r_{c}\right)\sum_{n=-\infty}^{+\infty}\left\{i^{n}J_{n}\left(\eta_{m}\right)ab^{\dag}\exp\left[-i\left(\omega_{s}-\Delta_{q}-n\omega_{m}t\right)t\right]+{\rm H.c.}\right\}\nonumber\\
&-g_{\rm col}\sinh\left(r_{c}\right)\sum_{n=-\infty}^{+\infty}\left\{e^{i\theta_{L}}i^{n}J_{n}\left(\eta_{m}\right)ab\exp\left[-i\left(\omega_{s}+\Delta_{q}-n\omega_{m}t\right)t\right]+{\rm H.c.}\right\},
\end{align}
\end{widetext}
where $g_{\rm col}=\sqrt{N}g$ represents a collective coupling. It is seen that, when $\omega_{s}+\Delta_{q}=0$ and $2\omega_{s}-\omega_{m}=0$, the off-resonant coupling to the zero-order ($n=0$) Floquet sideband, given by
\begin{equation}
\mathcal{V}_{0}\left(t\right)=g_{0}\left[ab^{\dag}\exp\left(-i2\omega_{s}t\right)+{\rm H.c.}\right]
\end{equation}
with $g_{0}=g_{\rm col}\cosh\left(r_{c}\right)J_{0}\left(\eta_{m}\right)$, dominates other off-resonant couplings, 
due to the property that $J_{0}\left(\eta_{m}\right)\gg \left|J_{n\neq0}\left(\eta_{m}\right)\right|$ for $\eta_{m}\ll1$. Therefore, we may drop these counter-rotating terms for $n\neq0$. 

\begin{figure*}[t]
	\centering
	\includegraphics[width=16.0cm]{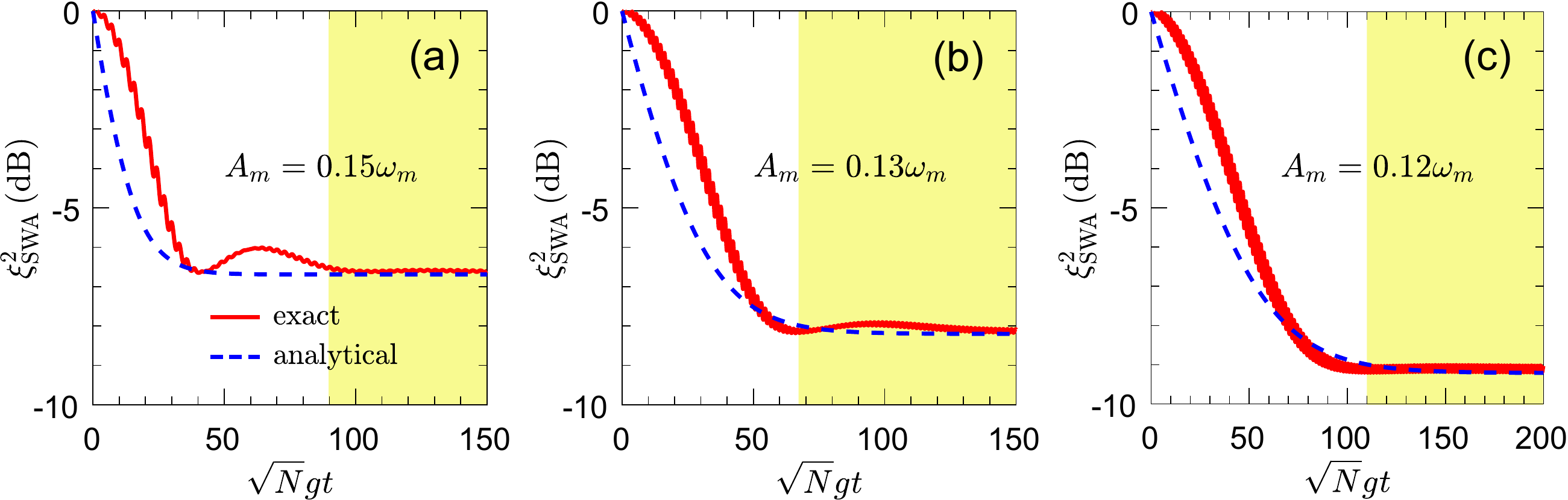}
	\caption{Evolution of the spin squeezing parameter $\xi^{2}_{\rm SWA}$ for (a) $A_{m}/\omega_{m}=0.15$, (b) $0.13$, and (c) $0.12$. Solid curves are obtained from the full Hamiltonian $H\left(t\right)$ in Eq.~(\ref{seq:full Hamiltonian 00}), while dashed curves are analytical predictions given by Eq.~(\ref{seq:analy-xi}). The analytical expression can predict well the squeezing of the collective spin, in particular, for the steady-state behavior (yellow regions). Here, we have made the spin-wave approximation for $H\left(t\right)$. In all plots, we have assumed that  $\Delta_{c}=200\kappa$, $\Omega=0.1\Delta_{c}$, $\gamma=0.01\kappa$, $\sqrt{N}g=10\kappa$, and also that all atoms are initialized in the ground state and the cavity is in the vacuum.}\label{sfig-adiab}
\end{figure*}

As demonstrated above, two resonant couplings in $H_{\rm SWA}^{\prime}\left(t\right)$ lead to the effective Hamiltonian 
\begin{align}	\label{seq:H_eff_SWA}
H_{\rm eff}^{\rm SWA}=\;&g_{\rm col}\,a^{\dag}\left(G_{-}b+G_{+}b^{\dag}\right)+{\rm H.c.},\nonumber\\
=\;&G\,g_{\rm col}\left(a^{\dag}\beta+{\rm H.c.}\right).
\end{align}
Here, we have defined a squeezed mode, $\beta=\cosh\left(r\right)b+\sinh\left(r\right)b^{\dag}$, of the collective spin, with $G^{2}=G_{-}^{2}-G_{+}^{2}$ and $\tanh\left(r\right)=G_{+}/G_{-}$. 

Furthermore, after time averaging~\cite{gamel2010time}, the effective dynamics of the coupling $\mathcal{V}_{0}\left(t\right)$ is determined by  
\begin{equation}
\overline{\mathcal{V}}_{0}\left(t\right)=\frac{g^{2}_{0}}{2\omega_{s}}\left(a^{\dag}a-b^{\dag}b\right).
\end{equation}
This implies that the coupling $\mathcal{V}_{0}\left(t\right)$ shifts the cavity resonance frequency and the atomic transition frequency by $+g_{0}^{2}/2\omega_{s}$ and $-g_{0}^{2}/2\omega_{s}$, respectively. This, in turn, enables an additional detuning of  $\delta=g_{0}^{2}/\omega_{s}\approx g_{\rm col}^{2}/\Delta_{c}$ between cavity and atoms. For the effective Hamiltonian $H_{\rm eff}^{\rm SWA}$, the detuning $\delta$ has no effect on the coupling of the form $\left(ab+a^{\dag}b^{\dag}\right)$, but it causes the coupling $\left(a^{\dag}b+ab^{\dag}\right)$ to become far off-resonant if $g_{\rm col}$ is comparable to $\Omega$. As a result, the degree of spin squeezing decreases, and even the desired dynamics is destroyed. To remove such a detrimental effect, we need to modify the resonant condition $2\omega_{s}=\omega_{m}$ (i.e., $\omega_{q}\approx\omega_{c}-\omega_{m}$) to be 
\begin{equation}
2\omega_{s}=\omega_{m}-\delta, \; {\rm or} \; \; \omega_{q}\approx\omega_{c}-\omega_{m}+g_{\rm col}^{2}/\Delta_{c},
\end{equation}
which compensates the detuning $\delta$. In Fig.~\ref{sfig-SWA}, we use the full Hamiltonian $H\left(t\right)$ by compensating the detuning $\delta$ to numerically calculate the excited-atom number $\langle b^{\dag}b\rangle$, the two-atom correlation $\langle bb\rangle$, and the spin squeezing parameter $\xi^{2}_{\rm SWA}$. We then compare them with the predictions of the effective Hamiltonian $H_{\rm eff}^{\rm SWA}$. Note that the full Hamiltonian $H\left(t\right)$ has been obtained under the spin-wave approximation. We see from Fig.~\ref{sfig-SWA} that, when the detuning $\delta$ is compensated, the full dynamics is in excellent agreement with the desired effective dynamics.

\begin{figure*}[t]
	\centering
	\includegraphics[width=15.0cm]{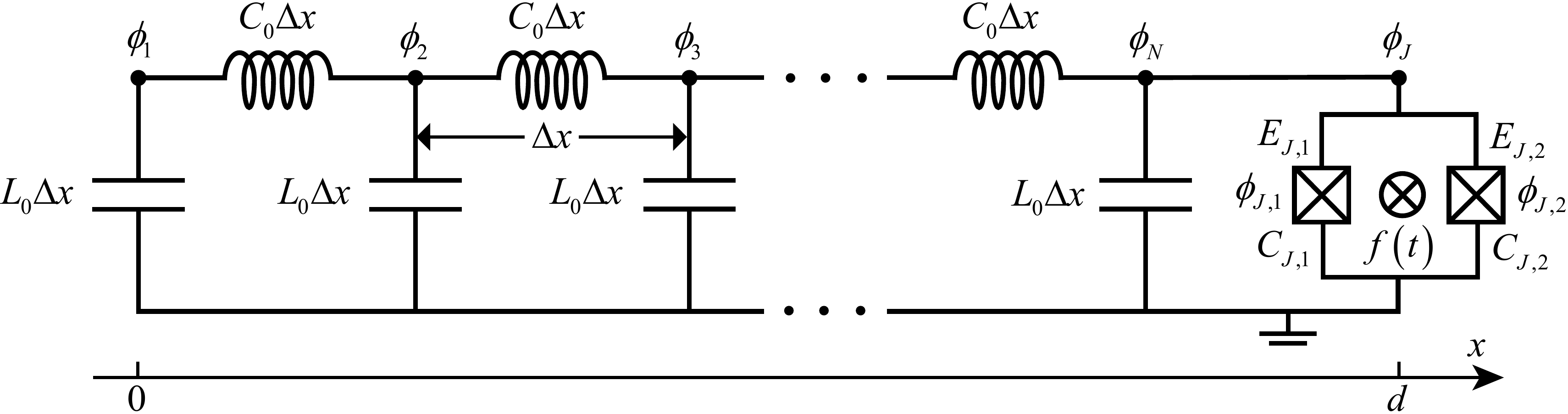}
	\caption{Equivalent circuits for an STL terminated by a SQUID. We assume that the left end, at $x=0$, of the STL is open, and its right end, at $x=d$, is connected to the SQUID. The STL of length $d$ has a characteristic capacitance $C_{0}$ and inductance $L_{0}$ per unit length. The STL is modeled as a series of $LC$ circuits each with a capacitance $C_{0}\Delta x$ and a inductance $L_{0}\Delta x$. Here, $\Delta x$ is a small distance. We assume $\phi_{i}$ ($i=1,2,3, \cdots, N$) to be the node phases between these $LC$ circuits. The SQUID consists of two Josephson junctions, and we use $E_{J,i}$, $C_{J,i}$, and $\phi_{J,i}$ ($i=1,2$) to label the Josephson energy, capacitance, and phase of the $i$th junction, respectively. The phases $\phi_{J,i}$ are determined by a driving phase $f\left(t\right)$ across the SQUID, such that $f\left(t\right)=\left(\phi_{J, 1}-\phi_{J, 2}\right)/2$. The effective phase $\phi_{J}$ of the SQUID is given by $\phi_{J}=\left(\phi_{J, 1}+\phi_{J, 2}\right)/2$. In the continuum limit $N\rightarrow\infty$, we have $\Delta x\rightarrow dx$ and $\phi_{i}\rightarrow\phi\left(x,t\right)$}\label{sfig_superconducting_circuits}
\end{figure*}

\section{Adiabatic elimination of the cavity mode}	
\label{Adiabatic elimination of the cavity mode}
We now discuss how to adiabatically eliminate the cavity mode. To begin, we consider the master equation with the effective Hamiltonian $H_{\rm eff}^{\rm SWA}$,
\begin{equation}\label{seq:master equation_SWA}
\dot{\rho}=i\left[\rho,H_{\rm eff}^{\rm SWA}\right]+\frac{\kappa}{2}\mathcal{L}\left(a\right)\rho+\frac{\gamma}{2}\sum_{j=1}^{N}\mathcal{L}\left(b\right)\rho.
\end{equation}
As mentioned already, we work within the limit $\Omega\ll\Delta_{c}$, and the squeezing of the cavity field is very weak. In this case, the occupation of the cavity mode is very low, such that we can only consider the vacuum state $\ket{0}$ and the single-photon state $\ket{1}$ of the cavity mode. The density matrix, $\rho$, of the system can therefore be expanded as 
\begin{equation}
\rho=\rho_{00}\ket{0}\bra{0}+\rho_{11}\ket{1}\bra{1}+\rho_{01}\ket{0}\bra{1}+\rho_{10}\ket{1}\bra{0}.
\end{equation}
Upon substituting this expression into the master equation in Eq.~(\ref{seq:master equation_SWA}), we obtain
\begin{align}
\label{seq:rho00}
\dot{\rho}_{00}=&iGg_{\rm col}\left(\rho_{01}\beta-\beta^{\dag}\rho_{10}\right)+\kappa\rho_{11}+\frac{\gamma}{2}\mathcal{L}\left(b\right)\rho_{00},\\
\label{seq:rho11}
\dot{\rho}_{11}=&iGg_{\rm col}\left(\rho_{10}\beta^{\dag}-\beta\rho_{01}\right)-\kappa\rho_{11}+\frac{\gamma}{2}\mathcal{L}\left(b\right)\rho_{11},\\
\dot{\rho}_{01}=&iGg_{\rm col}\left(\rho_{00}\beta^{\dag}-\beta^{\dag}\rho_{11}\right)-\frac{\kappa}{2}\rho_{01}+\frac{\gamma}{2}\mathcal{L}\left(b\right)\rho_{01},
\end{align}
and $\rho_{10}=\rho^{\dag}_{01}$. It then follows, on setting $\dot{\rho}_{01}=0$, that 
\begin{align}\label{seq:rho01}
\rho_{01}=\frac{i2Gg_{\rm col}}{\kappa}\left(\rho_{00}\beta^{\dag}-\beta^{\dag}\rho_{11}\right).
\end{align}
Here, we have assumed $\gamma\ll\kappa$. This assumption is generally valid because, for a typical atomic ensemble, e.g., a nitrogen-vacancy (NV) spin ensemble, the atomic decay rate $\gamma$ is negligible compared to the cavity loss rate $\kappa$.  Then, substituting Eq.~(\ref{seq:rho01}) into Eqs.~(\ref{seq:rho00}) and (\ref{seq:rho11}) leads to the following adiabatic master equation
\begin{equation}\label{seq:adiabatic master equation}
\dot{\rho}_{\rm spin}=\frac{\gamma_{c}}{2}\mathcal{L}\left(\beta\right)\rho_{\rm spin}+\frac{\gamma}{2}\mathcal{L}\left(b\right)\rho_{\rm spin},
\end{equation}
where $\rho_{\rm spin}$ is the reduced density matrix of the collective spin, and $\gamma_{c}=4G^{2}g_{\rm col}^{2}/\kappa$ represents the cavity-induced atomic decay.
We analytically find, according to Eq.~(\ref{seq:adiabatic master equation}), that
\begin{align}
\langle b^{\dag}b\rangle\left(t\right)=&\left[\langle b^{\dag}b\rangle_{\rm ini}-\langle b^{\dag}b\rangle_{\rm ss}\right]\exp\left[-\left(\gamma_{c}+\gamma\right)t\right]+\langle  b^{\dag}b\rangle_{\rm ss},\\
\langle bb\rangle\left(t\right)=&\left[\langle bb\rangle_{\rm ini}-\langle bb\rangle_{\rm ss}\right]\exp\left[-\left(\gamma_{c}+\gamma\right)t\right]+\langle  bb\rangle_{\rm ss}.
\end{align}
Here, $\langle b^{\dag}b\rangle_{\rm ini}$ is the initial excited-atom number, $\langle bb\rangle_{\rm ini}$ is the initial two-atom correlation, and  the corresponding steady-state values are
\begin{align}\label{seq:analy-bdagb}
\langle b^{\dag}b\rangle_{\rm ss}=\mathcal{A}\sinh^{2}\left(r\right),\; \; \;\langle bb\rangle_{\rm ss}=-\frac{1}{2}\mathcal{A}\sinh\left(2r\right),
\end{align}
where $\mathcal{A}=\left(\gamma_{c}/\gamma\right)/\left[\left(\gamma_{c}/\gamma+1\right)\left(1+\gamma/\kappa\right)\right]$. It follows, using $\xi^{2}_{\rm SWA}=1+2\left(\langle b^{\dag}b\rangle-\left|\langle bb\rangle\right|\right)$, that
\begin{equation}\label{seq:analy-xi}
\xi^{2}_{\rm SWA}=\left(\xi^{2}_{\rm SWA}\right)_{\rm ss}-\left[\left(\xi^{2}_{\rm SWA}\right)_{\rm ss}-1\right]\exp\left[-\left(\gamma_{c}+\gamma\right)t\right],
\end{equation}
where, for simplicity, we have assumed $\langle b^{\dag}b\rangle_{\rm ini}=\langle bb\rangle_{\rm ini}=0$.

In Fig.~\ref{sfig-adiab}, we compare the analytical $\xi_{\rm SWA}^{2}$ in Eq.~(\ref{seq:analy-xi}) with the exact numerical simulations of the full Hamiltonian $H\left(t\right)$ in Eq.~(\ref{seq:full Hamiltonian 00}). This figure shows a good agreement, in particular, for the steady-state behavior (yellow regions). The oscillation of red solid curves results from the reversible energy exchange between cavity and atoms (i.e., Rabi oscillation). However, this Rabi oscillation vanishes in the limit $G_{+}\rightarrow G_{-}$, as shown in Fig.~\ref{sfig-adiab}. This is because the coupling, $Gg_{\rm col}$, in the effective Hamiltonian $H_{\rm eff}^{\rm SWA}$ becomes smaller when $G_{+}$ approaches $G_{-}$. Thus, Eqs.~(\ref{seq:analy-bdagb}) and~(\ref{seq:analy-xi}) may be used to analytically predict stronger steady-state spin squeezing.

\section{Proposed experimental implementation with hybrid quantum systems and its feasibility}
\label{Possible implementations with hybrid quantum systems}

In this section, we consider a hybrid system, where a superconducting transmission line (STL) is terminated by a superconducting quantum interference device (SQUID) and is magnetically coupled to an NV spin ensemble in diamond. The strong coupling between the STL cavity and the NV spin ensemble has already been widely implemented experimentally~\cite{kubo2010strong, amsuss2011cavity, kubo2011hybrid, PhysRevA.85.012333, putz2014protecting,grezes2014multimode, astner2017coherent}. In particular, in Refs.~\cite{kubo2010strong,kubo2011hybrid,PhysRevA.85.012333}, a SQUID has already been used to tune the cavity frequency.

\subsection{Proposed experimental implementation}

We first show how to use an STL terminated by a SQUID to implement a parametrically driven Floquet cavity. The equivalent circuit for this setup is  schematically illustrated in Fig.~\ref{sfig_superconducting_circuits}. The STL of length $d$ can be divided into $N$ segments of equal length $\Delta x$, and then this can be modeled as a series of $LC$ circuits each with a capacitance $C_{0}\Delta x$ and an inductance $L_{0}\Delta x$. Here, $C_{0}$ and $L_{0}$ are the characteristic capacitance and inductance per unit length, respectively. The Lagrangian for the STL is therefore given by~\cite{PhysRevB.74.224506,johansson2010dynamical,PhysRevB.87.184501}:
\begin{equation}
\mathcal{L}_{\rm STL}=\left(\frac{\hbar}{2e}\right)^{2}\frac{C_{0}}{2}\sum_{i=1}^{N-1}\left[ \dot{\phi}_{i}^{2}\Delta x-v^{2}\frac{\left(\phi_{i+1}-\phi_{i}\right)^{2}}{\Delta x} \right],
\end{equation}
where $\phi_{i}$ is the node phase, and $v=1/\sqrt{L_{0}C_{0}}$ is the speed of light in the STL. In the continuum limit $N\rightarrow \infty$, we have $\Delta x\rightarrow dx$, and $\phi_{i}\rightarrow\phi\left(x,t\right)$. As a result, $\mathcal{L}_{\rm STL}$ becomes
\begin{equation}
\mathcal{L}_{\rm STL}=\left(\frac{\hbar}{2e}\right)^{2}\frac{C_{0}}{2}\int_{0}^{d}dx\left(\dot{\phi}^{2}-v^{2}\phi^{\prime 2}\right).
\end{equation}
The Lagrangian for the SQUID is
\begin{equation}
\mathcal{L}_{\rm SQUID}=\sum_{i=1,2}\left[\left(\frac{\hbar}{2e}\right)^{2}\frac{C_{J,i}}{2}\dot{\phi}_{J,i}^{2}+E_{J,i}\cos\left(\phi_{J,i}\right)\right].
\end{equation}
Here, $E_{J,i}$, $C_{J,i}$, and $\phi_{J,i}$ are, respectively, the Josephson energy, capacitance, and phase of the $i$th component Josephson junction in the SQUID loop. The phases $\phi_{J,i}$ of the Josephson junctions depend on the external magnetic flux, such that $\left(\phi_{J,1}-\phi_{J,2}\right)$ is determined by a driving phase $f\left(t\right)$ across the SQUID, yielding $\phi_{J,1}-\phi_{J,2}=2f\left(t\right)$.  We assume that the SQUID is symmetric, i.e., $C_{J,1}=C_{J,2}=C_{J}$ and $E_{J,1}=E_{J,2}=E_{J}$. The Lagrangian $\mathcal{L}_{\rm SQUID}$ is reduced to
\begin{equation}\label{seq_SQUID_Lagrangian}
\mathcal{L}_{\rm SQUID}=\left(\frac{\hbar}{2e}\right)^{2}\frac{2C_{J}}{2}\dot{\phi_{d}}+2E_{J}\cos \left[f\left(t\right)\right]\cos\left(\phi_{d}\right),
\end{equation}
where we have assumed that an effective phase of the SQUID, $\phi_{J}=\left(\phi_{J, 1}+\phi_{J, 2}\right)/2$, is equal to the boundary phase of the STL, $\phi_{d}=\phi\left(d,t\right)$. The cavity Lagrangian, including the STL and SQUID Lagrangians, is
\begin{equation}
\mathcal{L}_{\rm cavity}=\mathcal{L}_{\rm STL}+\mathcal{L}_{\rm SQUID}.
\end{equation}

We now discuss how to quantize the system. We begin with the massless scalar Klein-Gordon equation~\cite{PhysRevLett.123.054301},
\begin{equation}\label{seq_wave_equation}
\ddot{\phi}-v^{2}\phi''=0,
\end{equation}
which results from the Lagrangian $\mathcal{L}_{\rm STL}$. This wave equation is complemented 
with two boundary conditions $\phi'_{0}=0$ at the open end of the STL, and
\begin{multline}\label{seq_boundary_condition_second01}
2C_{J}\left(\frac{\hbar}{2e}\right)^{2}\ddot{\phi}_{d}+2E_{J}\cos \left[f\left(t\right)\right]\sin\left(\phi_{d}\right)\\+\frac{1}{L_{0}}\left(\frac{\hbar}{2e}\right)^{2}\phi'_{d}=0,
\end{multline}
at the end connected to the SQUID. We tune the driving phase $f\left(t\right)$ to be
\begin{align}
f\left(t\right)=&f_{0}+f_{1}\cos\left(\omega_{L1}t+\theta_{L1}\right)\nonumber\\
&+f_{2}\left(t\right)\cos\left(\omega_{L2}t+\theta_{L2}\right)
+f_{3}\cos\left(\omega_{L3}t+\theta_{L3}\right),
\end{align}
where $f_{0}$, $f_{1}$ and $f_{3}$ are time-independent, but $f_{2}\left(t\right)$ is time-dependent. We restrict our discussion to the case where $f_{1}$, $f_{2}\left(t\right)$, and $f_{3}$ are much weaker than $f_{0}$. As we demonstrate below, $f_{1}$ corresponds to the two-photon driving with a time-independent amplitude, $f_{2}\left(t\right)$ to another two-photon driving with a time-dependent amplitude, and $f_{3}$ to the cavity-frequency modulation. Following the procedure in Ref.~\cite{PhysRevB.87.184501}, the solution of the wave function in Eq.~(\ref{seq_wave_equation}) is given by
\begin{equation}\label{seq_solution}
\phi\left(x,t\right)=\frac{2e}{\hbar}\sqrt{\frac{2}{C_{0}d}}\sum_{n}q_{n}\left(t\right)\cos\left(k_{n}x\right),
\end{equation}
and the cavity Lagrangian $\mathcal{L}_{\rm cavity}$, accordingly, becomes
\begin{equation}
\mathcal{L}_{\rm cavity}=\frac{1}{2}\sum_{n}\left(M_{n}\dot{q}_{n}^{2}-M_{n}\omega_{n}^{2}q_{n}^{2}\right)-V.
\end{equation}
Here, $M_{n}$ is an effective mass, defined as  
\begin{equation}
M_{n}=1+\frac{\sin\left(2k_{n}d\right)}{2k_{n}d}+\frac{4C_{J}}{C_{0}d}\cos^{2}\left(k_{n}d\right),
\end{equation}
and $V$ is a nonlinear potential, defined as
\begin{equation}
V=-2E_{J}\left\{\cos\left[f\left(t\right)\right]\cos\left(\phi_{d}\right)+\frac{\phi_{d}^{2}}{2}\cos\left(f_{0}\right)\right\}.
\end{equation}
Consequently, the canonical-conjugate variable of $q_{n}$ is 
\begin{equation}
p_{n}=\frac{\partial \mathcal{L}_{\rm cavity}}{\partial \dot{q}_{n}}=M_{n}\dot{q}_{n}.
\end{equation}
thereby resulting in the cavity Hamiltonian
\begin{equation}
H_{\rm cavity}=H_{0}+V,
\end{equation}
with a free Hamiltonian
\begin{equation}
H_{0}=\frac{1}{2}\sum_{n}\left(\frac{p_{n}^{2}}{M_{n}}+M_{n}\omega_{n}^{2}q_{n}^{2}\right).
\end{equation}
We find that $H_{0}$ describes a collection of independent harmonic oscillators, but $V$ can provide either linear or nonlinear interactions between them.

Following the standard quantization procedure, we replace the c-numbers $q_{n}$ and $p_{n}$ by operators, which obey the canonical commutation relation $\left[q_{n},p_{m}\right]=i\hbar\delta_{nm}$. We then introduce the annihilation and creation operators $a_{n}$ and $a_{n}^{\dag}$
\begin{align}
q_{n}&=q_{{\rm zpf},n}\left(a_{n}+a_{n}^{\dag}\right),\\
p_{n}&=\frac{-i\hbar}{2q_{{\rm zpf},n}}\left(a_{n}-a_{n}^{\dag}\right),
\end{align}
where $q_{{\rm zpf},n}=\sqrt{\hbar/\left(2M_{n}\omega_{n}\right)}$ is the zero-point fluctuation of the variable $q_{n}$. Here, $a_{n}$ and $a_{n}^{\dag}$ obey the canonical  commutation relation $\left[a_{n},a_{m}^{\dag}\right]=\delta_{nm}$. With these definitions, the free Hamiltonian $H_{0}$ is transformed to
\begin{equation}
H_{0}=\sum_{n}\hbar
\omega_{n}\left(a_{n}^{\dag}a_{n}+\frac{1}{2}\right).
\end{equation}
We find that the quantized STL contains infinitely many modes, but the existence of the driving phase $f\left(t\right)$ enables us to selectively excite a desired mode, e.g., the fundamental mode $a_{0}$ (see below). The nonlinear potential $V$ can be approximated as
\begin{align}
V=&-E_{J}\sin \left(f_{0}\right)\big[f_{1}\cos\left(\omega_{L1}t+\theta_{L1}\right)\nonumber\\
&+f_{2}\left(t\right)\cos\left(\omega_{L2}t+\theta_{L2}\right)+f_{3}\cos\left(\omega_{L3}t+\theta_{L3}\right)\big]\phi_{d}^{2},
\end{align}
by assuming that $\left\{f_{1},f_{2}\left(t\right),f_{3}\right\}\ll f_{0}$ and $\phi_{d}\ll 1$. According to the solution $\phi\left(x,t\right)$ in Eq.~(\ref{seq_solution}), the quadratic potential $V$ can be expressed, in terms of the modes $a_{n}$, as
\begin{widetext}
\begin{align}
V=&-\left(\frac{2e}{\hbar}\right)^{2}\left(\frac{2}{C_{0}d}\right)E_{J}\sin \left(f_{0}\right)\left[f_{1}\cos\left(\omega_{L1}t+\theta_{L1}\right)+f_{2}\left(t\right)\cos\left(\omega_{L2}t+\theta_{L2}\right)+f_{3}\cos\left(\omega_{L3}t+\theta_{L3}\right)\right]\nonumber\\
&\times\sum_{n,m}q_{{\rm zpf},n}q_{{\rm zpf},m}\left(a_{n}+a_{n}^{\dag}\right)\left(a_{m}+a_{m}^{\dag}\right)\cos\left(k_{n}d\right)\cos\left(k_{m}d\right).
\end{align}
\end{widetext}
This means that the potential can excite or couple different modes. To select the fundamental mode $a_{0}$, we further assume that $\omega_{L1}=\omega_{L2}\approx2\omega_{0}$ and $\omega_{L3}\ll\omega_{0}$. In this case, we can only focus on the $a_{0}$ mode and other modes can be neglected, yielding 
\begin{align}
V=&A_{m}\sin\left(\omega_{m}t\right)a^{\dag}_{0}a_{0}\nonumber\\
&+\frac{1}{2}\left[\Omega+\Omega_{1}\left(t\right)\right]\left\{\exp\left[i\left(\omega_{L}t+\theta_{L}\right)\right]a^{2}_{0}+{\rm H.c.}\right\}.
\end{align}
Here, $\omega_{L}=\omega_{L1}=\omega_{L2}$, $\omega_{m}=\omega_{L3}$,  $\theta_{L}=\theta_{L1}=\theta_{L2}$, and $\theta_{L3}=3\pi/2$. Moreover, we have defined
\begin{align}
A_{m}=&\; \Omega f_{3}/f_{1}, \qquad	\Omega_{1}\left(t\right)=\; \Omega f_{2}\left(t\right)/f_{1},\nonumber\\
\Omega=&-2\left(\frac{2e}{\hbar}\right)^{2}\frac{E_{J}}{C_{0}d}q_{{\rm zpf},0}^{2}f_{1}\sin\left(f_{0}\right)\cos^{2}\left(k_{0}d\right).
\end{align}
In a frame rotating at $\omega_{L}/2$, the cavity Hamiltonian becomes (hereafter, we set $\hbar=1$)
\begin{align}\label{seq_desried_cavity_Hamiltonian}
H_{\rm cavity}=&\Delta_{c}a^{\dag}a+A_{m}\sin\left(\omega_{m}t\right)a^{\dag}a\nonumber\\
&+\frac{1}{2}\left[\Omega+\Omega_{1}\left(t\right)\right]\left[\exp\left(i\theta_{L}\right)a^{2}+{\rm H.c.}\right],
\end{align}
where we have written $a_{0}\equiv a$. The Hamiltonian in Eq.~(\ref{seq_desried_cavity_Hamiltonian}) describes a \emph{parametrically driven Floquet cavity}.

\begin{figure}[b]
	\centering
	\includegraphics[width=6.0cm]{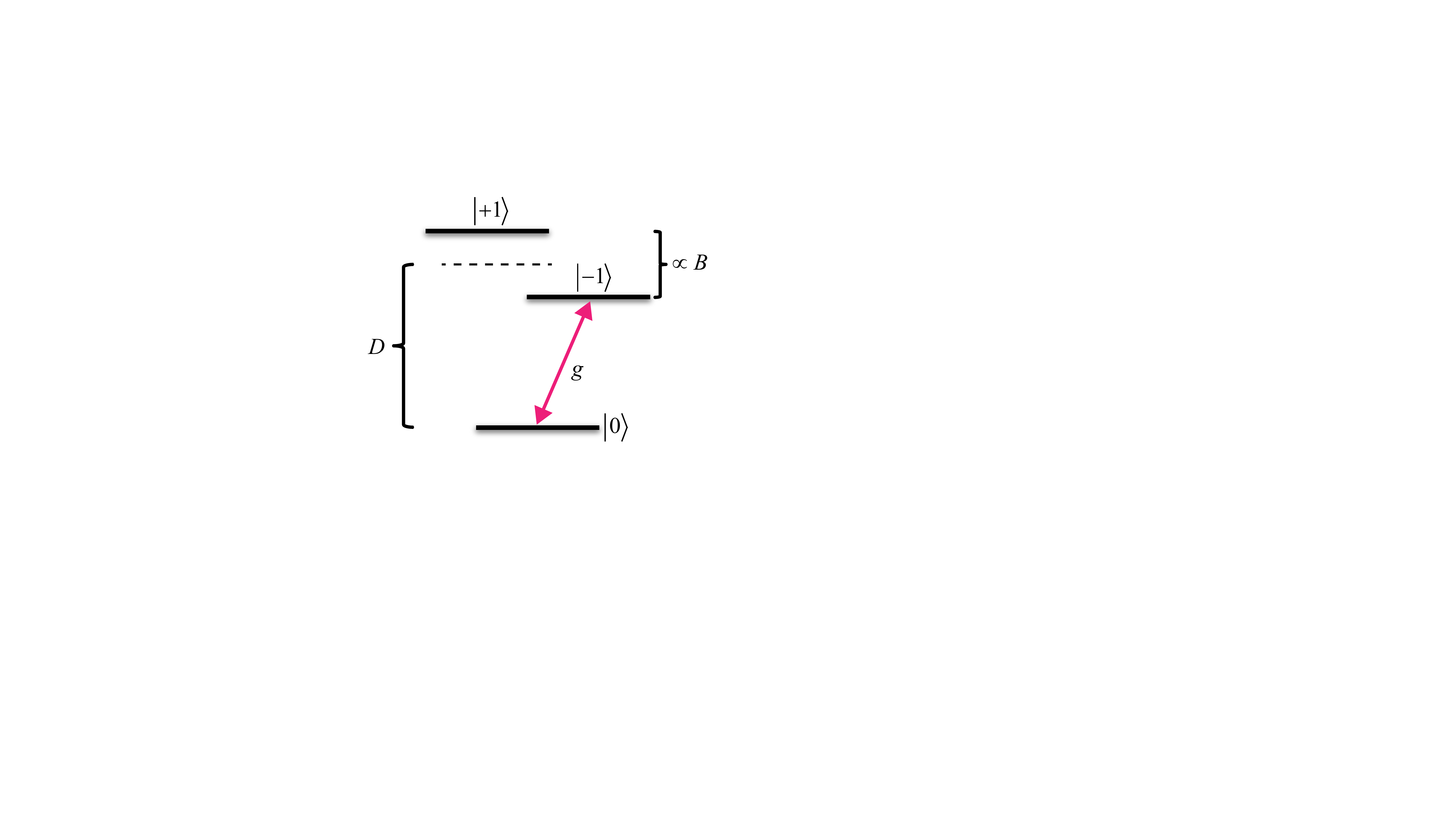}
	\caption{Level structure of a single NV spin in the electronic-ground state. This is a spin triplet consisting of states $|0\rangle$, $|-1\rangle$, and $|+1\rangle$. The zero-field splitting is $D\approx2.87$~GHz, while the Zeeman splitting between the states $|\pm1\rangle$ is proportional to the applied magnetic field $B$. We focus on, e.g., the $|0\rangle\rightarrow|-1\rangle$ transition, and assume that this spin transition is coupled to the cavity mode with a strength $g$.}\label{sfig_NV_center}
\end{figure}

Below let us consider the coupling of such a cavity to an NV spin ensemble in diamond. The electronic ground state of a single NV center is a long-lived spin triplet, whose  $m_{s}=0$ and $m_{s}=\pm1$ sublevels we label by $|0\rangle$ and $|\pm1\rangle$, respectively. The level structure is shown in Fig.~\ref{sfig_NV_center}. If there is no external magnetic field, the states $|\pm 1\rangle$ are degenerate, and due to the spin-spin interaction, are separated from the state $|0\rangle$ by the zero-field splitting $D\approx2.87$~GHz. In the presence of an external magnetic field $B$, the Zeeman splitting, which depends on the magnetic field strength, appears between the states $|\pm1\rangle$. This yields a two-level atom or a qubit, with $|0\rangle$ as the ground state and either $|-1\rangle$ or $|+1\rangle$ as the excited state. Here, we focus on, e.g., the $|0\rangle\rightarrow|-1\rangle$ transition, and the $|0\rangle\rightarrow|+1\rangle$ transition can be neglected due to large detuning. When a diamond containing an NV spin ensemble is placed on top of an STL, the STL mode $a$ can magnetically couple to the $|0\rangle\rightarrow|-1\rangle$ transition. Therefore, the collective spin-cavity coupling can be described by the following Hamiltonian
\begin{table*}[t]
	\caption{Some experimental parameters for recent experiments reporting the coupling between an NV spin ensemble and an STL cavity. Here, $\omega_{c}$ is the cavity frequency, $Q$ is the quality factor of the cavity, $\kappa$ is the loss rate of the cavity, $N$ is the number of NV centers in the ensemble, $g_{\rm col}$ is the collective coupling of the ensemble to the cavity, $\gamma_{\phi}$ is the dephasing rate of the ensemble, and $\gamma$ is the energy relaxation rate of the ensemble. Note that the superscript ``$\star$" indicates that the cavity frequency is tunable via a SQUID.} \label{stab:table}
	\setlength{\tabcolsep}{4.6mm}{
		\begin{tabular}{|c|c|c|c|c|c|c|c|}
			\hline \hline
			Ref. & $\frac{\omega_{c}}{2\pi}$ (GHz) & $Q$ & $\frac{\kappa}{2\pi}$ (MHz) & $N$ & $\frac{g_{\rm col}}{2\pi}$ (MHz) & $\frac{\gamma_{\phi}}{2\pi}$ (MHz) & $\frac{\gamma}{2\pi}$ (Hz) \\
			\hline
			\cite{kubo2010strong} & $2.87^{\star}$ &  $\sim1.9\times10^{3}$& $\sim1.5$ & $\sim10^{12}$ & $\sim11$ & $\sim3$ &--\\ 
			\hline
			\cite{amsuss2011cavity} & $2.701$ & $\sim3.2\times10^{3}$ & $\sim0.8$ & $\sim10^{12}$ & $\sim10$ & -- &$\sim0.004$\\
			\hline
			\cite{kubo2011hybrid} & $3.004^{\star}$ & -- &-- & $\sim10^{11}$ & $\sim 3$ & $\sim0.02$ &--\\
			\hline
			\cite{PhysRevA.85.012333} & $2.88^{\star}$ & $\sim1.8\times10^{3}$ & $\sim 1.6$ & $\sim10^{12}$ & $\sim 11$ & $\sim 5.3$ &--\\
			\hline
			\cite{putz2014protecting} & $2.6899$ & $\sim3.0\times10^{3}$ & $\sim0.8$ & $\sim10^{12}$ & $\sim9$ & $\sim5.2$ &--\\
			\hline
			\cite{grezes2014multimode} & $2.88$ & $\sim80$ & $\sim36$ & -- & $\sim5$ & $\sim0.02$ &$<0.005$\\
			\hline
			\cite{astner2017coherent} & $2.7491$  & $\sim4.3\times10^{3}$ & $\sim0.6$ & -- & $\sim10$ & -- & --\\
			\hline
	\end{tabular}}
\end{table*}
\begin{equation}\label{seq_coupling_Hamiltonian_single}
H_{\rm int}=\sum_{j=1}^{N}g_{j}\left(a^{\dag}\sigma^{-}_{j}+a\sigma_{j}^{+}\right),
\end{equation}
where $\sigma_{j}^{-}=|0\rangle_{j}\langle -1|$ is the lowering operator for the $j$th spin qubit, $\sigma_{j}^{+}=\left(\sigma_{j}^{-}\right)^{\dag}$, $g_{j}$ is the single spin-cavity coupling strength, and $N$ is the total number of spins. Such a spin ensemble can also be described with collective spin operators \begin{align}
S_{z}=\frac{1}{2}\sum_{j=1}^{N}\sigma_{j}^{z},\quad {\rm and} \quad S_{\pm}=\frac{1}{g}\sum_{j=1}^{N}g_{j}\sigma_{j}^{\pm}.
\end{align}
Here, $g^{2}=\frac{1}{N}\sum_{j=1}^{N}g_{j}^{2}$.  The Hamiltonian $H_{\rm int}$ is accordingly transformed into
\begin{equation}\label{seq_coupling_Hamiltonian_collective}
H_{\rm int}=g\left(aS_{+}+a^{\dag}S_{-}\right).
\end{equation}
Furthermore, we assume, for simplicity but without loss of generality, that $g_{j}$ is a constant, such that $g_{j}=g$, yielding $S_{\pm}=\sum_{j=1}^{N}\sigma_{j}^{\pm}$. Combined with the cavity Hamiltonian in Eq.~(\ref{seq_desried_cavity_Hamiltonian}), the full Hamiltonian
for the system becomes
\begin{equation}\label{seq_total_desried_Hamiltonian}
H=H_{0}+H_{1}\left(t\right),
\end{equation}
where 
\begin{align}
H_{0}=&\Delta_{c}a^{\dag}a+\Delta_{q}S_{z}+g\left(aS_{+}+a^{\dag}S_{-}\right)\nonumber\\
&+\frac{1}{2}\Omega\left[\exp\left(i\theta_{L}\right)a^{2}+{\rm H.c.}\right],
\end{align}
and 
\begin{equation}
H_{1}\left(t\right)=A_{m}\sin\left(\omega_{m}t\right)a^{\dag}a+\frac{1}{2}\Omega_{1}\left(t\right)\left[\exp\left(i\theta_{L}\right)a^{2}+{\rm H.c.}\right].
\end{equation}
It is seen that the Hamiltonian $H$ in Eq.~(\ref{seq_total_desried_Hamiltonian}) is exactly the one applied by us in the main article.

\subsection{Experimental feasibility}
In Table.~\ref{stab:table}, we list some relevant parameters reported in recent experiments demonstrating the coupling between an NV spin ensemble and an STL cavity.
In addition to these parameters listed in Table~\ref{stab:table}, the coherence time of NV spin ensembles, with spin-echo sequences, has experimentally reached the order of ms (i.e., $\gamma_{\phi}/2\pi\sim0.16$~kHz)~\cite{stanwix2010coherence}, and harnessing dynamical-decoupling sequences can further make this coherence time close to one second (i.e., $\gamma_{\phi}/2\pi\sim0.16$~Hz)~\cite{bar2013solid}.

Note that Refs.~\cite{kubo2010strong,kubo2011hybrid,PhysRevA.85.012333} used a SQUID to tune the resonance frequency of an STL cavity coupled to an NV spin ensemble. This setup is similar to the one we have already proposed for a possible implementation of our proposal. 

The analytical $\xi^{2}_{\rm SWA}$ in Eq.~(\ref{seq:analy-xi}) predicts that, for typical parameters $g_{\rm col}/2\pi=10$~MHz, $\kappa/2\pi=1.0$~MHz, and $\gamma=0$ in Table~\ref{stab:table}, a spin-squeezed steady state of $\approx-12$~dB can be achieved for a squeezing time $\approx0.8$~$\mu$s, or $\approx-20$~dB for $\approx8$~$\mu$s. This justifies neglecting spin decoherence, which, as described above, could be made much slower. We also find, according to an exponential squeezing given in Eq.~(\ref{seq:analy-xi}), that by properly increasing $\gamma_{c}$, we can achieve a shorter squeezing time.

Moreover, in addition to the NV spin ensembles, ion spin ensembles~\cite{PhysRevLett.105.140501,PhysRevLett.110.157001,wisby2014coupling} and P1 center ensembles~\cite{ranjan2013probing} can also couple to an STL cavity.  In a recent experiment~\cite{hattermann2017coupling}, the coupling of an ensemble of $^{87}$Rb atoms to an STL cavity has already been reported.

Hence, we  expect that our proposal could be realized with current technologies.


\begin{thebibliography}{79}%
	\makeatletter
	\providecommand \@ifxundefined [1]{%
		\@ifx{#1\undefined}
	}%
	\providecommand \@ifnum [1]{%
		\ifnum #1\expandafter \@firstoftwo
		\else \expandafter \@secondoftwo
		\fi
	}%
	\providecommand \@ifx [1]{%
		\ifx #1\expandafter \@firstoftwo
		\else \expandafter \@secondoftwo
		\fi
	}%
	\providecommand \natexlab [1]{#1}%
	\providecommand \enquote  [1]{``#1''}%
	\providecommand \bibnamefont  [1]{#1}%
	\providecommand \bibfnamefont [1]{#1}%
	\providecommand \citenamefont [1]{#1}%
	\providecommand \href@noop [0]{\@secondoftwo}%
	\providecommand \href [0]{\begingroup \@sanitize@url \@href}%
	\providecommand \@href[1]{\@@startlink{#1}\@@href}%
	\providecommand \@@href[1]{\endgroup#1\@@endlink}%
	\providecommand \@sanitize@url [0]{\catcode `\\12\catcode `\$12\catcode
		`\&12\catcode `\#12\catcode `\^12\catcode `\_12\catcode `\%12\relax}%
	\providecommand \@@startlink[1]{}%
	\providecommand \@@endlink[0]{}%
	\providecommand \url  [0]{\begingroup\@sanitize@url \@url }%
	\providecommand \@url [1]{\endgroup\@href {#1}{\urlprefix }}%
	\providecommand \urlprefix  [0]{URL }%
	\providecommand \Eprint [0]{\href }%
	\providecommand \doibase [0]{http://dx.doi.org/}%
	\providecommand \selectlanguage [0]{\@gobble}%
	\providecommand \bibinfo  [0]{\@secondoftwo}%
	\providecommand \bibfield  [0]{\@secondoftwo}%
	\providecommand \translation [1]{[#1]}%
	\providecommand \BibitemOpen [0]{}%
	\providecommand \bibitemStop [0]{}%
	\providecommand \bibitemNoStop [0]{.\EOS\space}%
	\providecommand \EOS [0]{\spacefactor3000\relax}%
	\providecommand \BibitemShut  [1]{\csname bibitem#1\endcsname}%
	\let\auto@bib@innerbib\@empty
	\bibitem [{\citenamefont {Kitagawa}\ and\ \citenamefont
		{Ueda}(1993)}]{kitagawa1993squeezed}%
	\BibitemOpen
	\bibfield  {author} {\bibinfo {author} {\bibfnamefont {M.}~\bibnamefont
			{Kitagawa}}\ and\ \bibinfo {author} {\bibfnamefont {M.}~\bibnamefont
			{Ueda}},\ }\bibfield  {title} {\enquote {\bibinfo {title} {Squeezed spin
				states},}\ }\href {https://link.aps.org/doi/10.1103/PhysRevA.47.5138}
	{\bibfield  {journal} {\bibinfo  {journal} {Phys. Rev. A}\ }\textbf {\bibinfo
			{volume} {47}},\ \bibinfo {pages} {5138} (\bibinfo {year}
		{1993})}\BibitemShut {NoStop}%
	\bibitem [{\citenamefont {Wineland}\ \emph {et~al.}(1992)\citenamefont
		{Wineland}, \citenamefont {Bollinger}, \citenamefont {Itano}, \citenamefont
		{Moore},\ and\ \citenamefont {Heinzen}}]{wineland1992spin}%
	\BibitemOpen
	\bibfield  {author} {\bibinfo {author} {\bibfnamefont {D.~J.}\ \bibnamefont
			{Wineland}}, \bibinfo {author} {\bibfnamefont {J.~J.}\ \bibnamefont
			{Bollinger}}, \bibinfo {author} {\bibfnamefont {W.~M.}\ \bibnamefont
			{Itano}}, \bibinfo {author} {\bibfnamefont {F.~L.}\ \bibnamefont {Moore}}, \
		and\ \bibinfo {author} {\bibfnamefont {D.~J.}\ \bibnamefont {Heinzen}},\
	}\bibfield  {title} {\enquote {\bibinfo {title} {Spin squeezing and reduced
				quantum noise in spectroscopy},}\ }\href
	{https://link.aps.org/doi/10.1103/PhysRevA.46.R6797} {\bibfield  {journal}
		{\bibinfo  {journal} {Phys. Rev. A}\ }\textbf {\bibinfo {volume} {46}},\
		\bibinfo {pages} {R6797} (\bibinfo {year} {1992})}\BibitemShut {NoStop}%
	\bibitem [{\citenamefont {Wineland}\ \emph {et~al.}(1994)\citenamefont
		{Wineland}, \citenamefont {Bollinger}, \citenamefont {Itano},\ and\
		\citenamefont {Heinzen}}]{wineland1994squeezed}%
	\BibitemOpen
	\bibfield  {author} {\bibinfo {author} {\bibfnamefont {D.~J.}\ \bibnamefont
			{Wineland}}, \bibinfo {author} {\bibfnamefont {J.~J.}\ \bibnamefont
			{Bollinger}}, \bibinfo {author} {\bibfnamefont {W.~M.}\ \bibnamefont
			{Itano}}, \ and\ \bibinfo {author} {\bibfnamefont {D.~J.}\ \bibnamefont
			{Heinzen}},\ }\bibfield  {title} {\enquote {\bibinfo {title} {Squeezed atomic
				states and projection noise in spectroscopy},}\ }\href
	{https://link.aps.org/doi/10.1103/PhysRevA.50.67} {\bibfield  {journal}
		{\bibinfo  {journal} {Phys. Rev. A}\ }\textbf {\bibinfo {volume} {50}},\
		\bibinfo {pages} {67} (\bibinfo {year} {1994})}\BibitemShut {NoStop}%
	\bibitem [{\citenamefont {Ma}\ \emph {et~al.}(2011)\citenamefont {Ma},
		\citenamefont {Wang}, \citenamefont {Sun},\ and\ \citenamefont
		{Nori}}]{ma2011quantum}%
	\BibitemOpen
	\bibfield  {author} {\bibinfo {author} {\bibfnamefont {J.}~\bibnamefont
			{Ma}}, \bibinfo {author} {\bibfnamefont {X.}~\bibnamefont {Wang}}, \bibinfo
		{author} {\bibfnamefont {C.-P.}\ \bibnamefont {Sun}}, \ and\ \bibinfo
		{author} {\bibfnamefont {F.}~\bibnamefont {Nori}},\ }\bibfield  {title}
	{\enquote {\bibinfo {title} {Quantum spin squeezing},}\ }\href
	{http://www.sciencedirect.com/science/article/pii/S0370157311002201}
	{\bibfield  {journal} {\bibinfo  {journal} {Phys. Rep.}\ }\textbf {\bibinfo
			{volume} {509}},\ \bibinfo {pages} {89--165} (\bibinfo {year}
		{2011})}\BibitemShut {NoStop}%
	\bibitem [{\citenamefont {Pezz{\`e}}\ \emph {et~al.}(2018)\citenamefont
		{Pezz{\`e}}, \citenamefont {Smerzi}, \citenamefont {Oberthaler},
		\citenamefont {Schmied},\ and\ \citenamefont {Treutlein}}]{pezze2018quantum}%
	\BibitemOpen
	\bibfield  {author} {\bibinfo {author} {\bibfnamefont {L.}~\bibnamefont
			{Pezz{\`e}}}, \bibinfo {author} {\bibfnamefont {A.}~\bibnamefont {Smerzi}},
		\bibinfo {author} {\bibfnamefont {M.~K.}\ \bibnamefont {Oberthaler}},
		\bibinfo {author} {\bibfnamefont {R.}~\bibnamefont {Schmied}}, \ and\
		\bibinfo {author} {\bibfnamefont {P.}~\bibnamefont {Treutlein}},\ }\bibfield
	{title} {\enquote {\bibinfo {title} {Quantum metrology with nonclassical
				states of atomic ensembles},}\ }\href
	{https://link.aps.org/doi/10.1103/RevModPhys.90.035005} {\bibfield  {journal}
		{\bibinfo  {journal} {Rev. Mod. Phys.}\ }\textbf {\bibinfo {volume} {90}},\
		\bibinfo {pages} {035005} (\bibinfo {year} {2018})}\BibitemShut {NoStop}%
	\bibitem [{\citenamefont {S{\o}rensen}\ \emph {et~al.}(2001)\citenamefont
		{S{\o}rensen}, \citenamefont {Duan}, \citenamefont {Cirac},\ and\
		\citenamefont {Zoller}}]{sorensen2001many}%
	\BibitemOpen
	\bibfield  {author} {\bibinfo {author} {\bibfnamefont {A.}~\bibnamefont
			{S{\o}rensen}}, \bibinfo {author} {\bibfnamefont {L.-M.}\ \bibnamefont
			{Duan}}, \bibinfo {author} {\bibfnamefont {J.~I.}\ \bibnamefont {Cirac}}, \
		and\ \bibinfo {author} {\bibfnamefont {P.}~\bibnamefont {Zoller}},\
	}\bibfield  {title} {\enquote {\bibinfo {title} {Many-particle entanglement
				with {B}ose-{E}instein condensates},}\ }\href
	{https://doi.org/10.1038/35051038} {\bibfield  {journal} {\bibinfo  {journal}
			{Nature}\ }\textbf {\bibinfo {volume} {409}},\ \bibinfo {pages} {63}
		(\bibinfo {year} {2001})}\BibitemShut {NoStop}%
	\bibitem [{\citenamefont {Orzel}\ \emph {et~al.}(2001)\citenamefont {Orzel},
		\citenamefont {Tuchman}, \citenamefont {Fenselau}, \citenamefont {Yasuda},\
		and\ \citenamefont {Kasevich}}]{orzel2001squeezed}%
	\BibitemOpen
	\bibfield  {author} {\bibinfo {author} {\bibfnamefont {C.}~\bibnamefont
			{Orzel}}, \bibinfo {author} {\bibfnamefont {A.~K.}\ \bibnamefont {Tuchman}},
		\bibinfo {author} {\bibfnamefont {M.~L.}\ \bibnamefont {Fenselau}}, \bibinfo
		{author} {\bibfnamefont {M.}~\bibnamefont {Yasuda}}, \ and\ \bibinfo {author}
		{\bibfnamefont {M.~A.}\ \bibnamefont {Kasevich}},\ }\bibfield  {title}
	{\enquote {\bibinfo {title} {Squeezed states in a {B}ose-{E}instein
				condensate},}\ }\href {https://science.sciencemag.org/content/291/5512/2386}
	{\bibfield  {journal} {\bibinfo  {journal} {Science}\ }\textbf {\bibinfo
			{volume} {291}},\ \bibinfo {pages} {2386--2389} (\bibinfo {year}
		{2001})}\BibitemShut {NoStop}%
	\bibitem [{\citenamefont {Est\`{e}ve}\ \emph {et~al.}(2008)\citenamefont
		{Est\`{e}ve}, \citenamefont {Gross}, \citenamefont {Weller}, \citenamefont
		{Giovanazzi},\ and\ \citenamefont {Oberthaler}}]{esteve2008squeezing}%
	\BibitemOpen
	\bibfield  {author} {\bibinfo {author} {\bibfnamefont {J.}~\bibnamefont
			{Est\`{e}ve}}, \bibinfo {author} {\bibfnamefont {C.}~\bibnamefont {Gross}},
		\bibinfo {author} {\bibfnamefont {A.}~\bibnamefont {Weller}}, \bibinfo
		{author} {\bibfnamefont {S.}~\bibnamefont {Giovanazzi}}, \ and\ \bibinfo
		{author} {\bibfnamefont {M.~K.}\ \bibnamefont {Oberthaler}},\ }\bibfield
	{title} {\enquote {\bibinfo {title} {Squeezing and entanglement in a
				{B}ose--{E}instein condensate},}\ }\href
	{https://doi.org/10.1038/nature07332} {\bibfield  {journal} {\bibinfo
			{journal} {Nature}\ }\textbf {\bibinfo {volume} {455}},\ \bibinfo {pages}
		{1216} (\bibinfo {year} {2008})}\BibitemShut {NoStop}%
	\bibitem [{\citenamefont {Riedel}\ \emph {et~al.}(2010)\citenamefont {Riedel},
		\citenamefont {B{\"o}hi}, \citenamefont {Li}, \citenamefont {H{\"a}nsch},
		\citenamefont {Sinatra},\ and\ \citenamefont {Treutlein}}]{riedel2010atom}%
	\BibitemOpen
	\bibfield  {author} {\bibinfo {author} {\bibfnamefont {M.~F.}\ \bibnamefont
			{Riedel}}, \bibinfo {author} {\bibfnamefont {P.}~\bibnamefont {B{\"o}hi}},
		\bibinfo {author} {\bibfnamefont {Y.}~\bibnamefont {Li}}, \bibinfo {author}
		{\bibfnamefont {T.~W.}\ \bibnamefont {H{\"a}nsch}}, \bibinfo {author}
		{\bibfnamefont {A.}~\bibnamefont {Sinatra}}, \ and\ \bibinfo {author}
		{\bibfnamefont {P.}~\bibnamefont {Treutlein}},\ }\bibfield  {title} {\enquote
		{\bibinfo {title} {Atom-chip-based generation of entanglement for quantum
				metrology},}\ }\href {https://doi.org/10.1038/nature08988} {\bibfield
		{journal} {\bibinfo  {journal} {Nature}\ }\textbf {\bibinfo {volume} {464}},\
		\bibinfo {pages} {1170} (\bibinfo {year} {2010})}\BibitemShut {NoStop}%
	\bibitem [{\citenamefont {Gross}\ \emph {et~al.}(2010)\citenamefont {Gross},
		\citenamefont {Zibold}, \citenamefont {Nicklas}, \citenamefont {Est\`{e}ve},\
		and\ \citenamefont {Oberthaler}}]{gross2010nonlinear}%
	\BibitemOpen
	\bibfield  {author} {\bibinfo {author} {\bibfnamefont {C.}~\bibnamefont
			{Gross}}, \bibinfo {author} {\bibfnamefont {T.}~\bibnamefont {Zibold}},
		\bibinfo {author} {\bibfnamefont {E.}~\bibnamefont {Nicklas}}, \bibinfo
		{author} {\bibfnamefont {J.}~\bibnamefont {Est\`{e}ve}}, \ and\ \bibinfo
		{author} {\bibfnamefont {M.~K.}\ \bibnamefont {Oberthaler}},\ }\bibfield
	{title} {\enquote {\bibinfo {title} {Nonlinear atom interferometer surpasses
				classical precision limit},}\ }\href {https://doi.org/10.1038/nature08919}
	{\bibfield  {journal} {\bibinfo  {journal} {Nature}\ }\textbf {\bibinfo
			{volume} {464}},\ \bibinfo {pages} {1165} (\bibinfo {year}
		{2010})}\BibitemShut {NoStop}%
	\bibitem [{\citenamefont {L{\"u}cke}\ \emph {et~al.}(2011)\citenamefont
		{L{\"u}cke}, \citenamefont {Scherer}, \citenamefont {Kruse}, \citenamefont
		{Pezz{\'e}}, \citenamefont {Deuretzbacher}, \citenamefont {Hyllus},
		\citenamefont {Peise}, \citenamefont {Ertmer}, \citenamefont {Arlt},
		\citenamefont {Santos}, \citenamefont {Smerzi},\ and\ \citenamefont
		{Klempt}}]{lucke2011twin}%
	\BibitemOpen
	\bibfield  {author} {\bibinfo {author} {\bibfnamefont {B.}~\bibnamefont
			{L{\"u}cke}}, \bibinfo {author} {\bibfnamefont {M.}~\bibnamefont {Scherer}},
		\bibinfo {author} {\bibfnamefont {J.}~\bibnamefont {Kruse}}, \bibinfo
		{author} {\bibfnamefont {L.}~\bibnamefont {Pezz{\'e}}}, \bibinfo {author}
		{\bibfnamefont {F.}~\bibnamefont {Deuretzbacher}}, \bibinfo {author}
		{\bibfnamefont {P.}~\bibnamefont {Hyllus}}, \bibinfo {author} {\bibfnamefont
			{J.}~\bibnamefont {Peise}}, \bibinfo {author} {\bibfnamefont
			{W.}~\bibnamefont {Ertmer}}, \bibinfo {author} {\bibfnamefont
			{J.}~\bibnamefont {Arlt}}, \bibinfo {author} {\bibfnamefont {L.}~\bibnamefont
			{Santos}}, \bibinfo {author} {\bibfnamefont {A.}~\bibnamefont {Smerzi}}, \
		and\ \bibinfo {author} {\bibfnamefont {C.}~\bibnamefont {Klempt}},\
	}\bibfield  {title} {\enquote {\bibinfo {title} {Twin matter waves for
				interferometry beyond the classical limit},}\ }\href
	{https://science.sciencemag.org/content/334/6057/773} {\bibfield  {journal}
		{\bibinfo  {journal} {Science}\ }\textbf {\bibinfo {volume} {334}},\ \bibinfo
		{pages} {773--776} (\bibinfo {year} {2011})}\BibitemShut {NoStop}%
	\bibitem [{\citenamefont {Yu}\ \emph {et~al.}(2014)\citenamefont {Yu},
		\citenamefont {Fan}, \citenamefont {Zhu}, \citenamefont {Chen}, \citenamefont
		{Jia},\ and\ \citenamefont {Nori}}]{yu2014creating}%
	\BibitemOpen
	\bibfield  {author} {\bibinfo {author} {\bibfnamefont {L.}~\bibnamefont
			{Yu}}, \bibinfo {author} {\bibfnamefont {J.}~\bibnamefont {Fan}}, \bibinfo
		{author} {\bibfnamefont {S.}~\bibnamefont {Zhu}}, \bibinfo {author}
		{\bibfnamefont {G.}~\bibnamefont {Chen}}, \bibinfo {author} {\bibfnamefont
			{S.}~\bibnamefont {Jia}}, \ and\ \bibinfo {author} {\bibfnamefont
			{F.}~\bibnamefont {Nori}},\ }\bibfield  {title} {\enquote {\bibinfo {title}
			{Creating a tunable spin squeezing via a time-dependent collective
				atom-photon coupling},}\ }\href
	{https://link.aps.org/doi/10.1103/PhysRevA.89.023838} {\bibfield  {journal}
		{\bibinfo  {journal} {Phys. Rev. A}\ }\textbf {\bibinfo {volume} {89}},\
		\bibinfo {pages} {023838} (\bibinfo {year} {2014})}\BibitemShut {NoStop}%
	\bibitem [{\citenamefont {Luo}\ \emph {et~al.}(2017)\citenamefont {Luo},
		\citenamefont {Zou}, \citenamefont {Wu}, \citenamefont {Liu}, \citenamefont
		{Han}, \citenamefont {Tey},\ and\ \citenamefont
		{You}}]{luo2017deterministic}%
	\BibitemOpen
	\bibfield  {author} {\bibinfo {author} {\bibfnamefont {X.-Y.}\ \bibnamefont
			{Luo}}, \bibinfo {author} {\bibfnamefont {Y.-Q.}\ \bibnamefont {Zou}},
		\bibinfo {author} {\bibfnamefont {L.-N.}\ \bibnamefont {Wu}}, \bibinfo
		{author} {\bibfnamefont {Q.}~\bibnamefont {Liu}}, \bibinfo {author}
		{\bibfnamefont {M.-F.}\ \bibnamefont {Han}}, \bibinfo {author} {\bibfnamefont
			{M.~K.}\ \bibnamefont {Tey}}, \ and\ \bibinfo {author} {\bibfnamefont
			{L.}~\bibnamefont {You}},\ }\bibfield  {title} {\enquote {\bibinfo {title}
			{Deterministic entanglement generation from driving through quantum phase
				transitions},}\ }\href {https://science.sciencemag.org/content/355/6325/620}
	{\bibfield  {journal} {\bibinfo  {journal} {Science}\ }\textbf {\bibinfo
			{volume} {355}},\ \bibinfo {pages} {620--623} (\bibinfo {year}
		{2017})}\BibitemShut {NoStop}%
	\bibitem [{\citenamefont {Fadel}\ \emph {et~al.}(2018)\citenamefont {Fadel},
		\citenamefont {Zibold}, \citenamefont {D{\'e}camps},\ and\ \citenamefont
		{Treutlein}}]{fadel2018spatial}%
	\BibitemOpen
	\bibfield  {author} {\bibinfo {author} {\bibfnamefont {M.}~\bibnamefont
			{Fadel}}, \bibinfo {author} {\bibfnamefont {T.}~\bibnamefont {Zibold}},
		\bibinfo {author} {\bibfnamefont {B.}~\bibnamefont {D{\'e}camps}}, \ and\
		\bibinfo {author} {\bibfnamefont {P.}~\bibnamefont {Treutlein}},\ }\bibfield
	{title} {\enquote {\bibinfo {title} {Spatial entanglement patterns and
				{E}instein-{P}odolsky-{R}osen steering in {B}ose-{E}instein condensates},}\
	}\href {https://science.sciencemag.org/content/360/6387/409} {\bibfield
		{journal} {\bibinfo  {journal} {Science}\ }\textbf {\bibinfo {volume}
			{360}},\ \bibinfo {pages} {409--413} (\bibinfo {year} {2018})}\BibitemShut
	{NoStop}%
	\bibitem [{\citenamefont {Kuzmich}\ \emph {et~al.}(1997)\citenamefont
		{Kuzmich}, \citenamefont {M{\o}lmer},\ and\ \citenamefont
		{Polzik}}]{kuzmich1997spin}%
	\BibitemOpen
	\bibfield  {author} {\bibinfo {author} {\bibfnamefont {A.}~\bibnamefont
			{Kuzmich}}, \bibinfo {author} {\bibfnamefont {K.}~\bibnamefont {M{\o}lmer}},
		\ and\ \bibinfo {author} {\bibfnamefont {E.~S.}\ \bibnamefont {Polzik}},\
	}\bibfield  {title} {\enquote {\bibinfo {title} {Spin {S}queezing in an
				{E}nsemble of {A}toms {I}lluminated with {S}queezed {L}ight},}\ }\href
	{https://link.aps.org/doi/10.1103/PhysRevLett.79.4782} {\bibfield  {journal}
		{\bibinfo  {journal} {Phys. Rev. Lett.}\ }\textbf {\bibinfo {volume} {79}},\
		\bibinfo {pages} {4782} (\bibinfo {year} {1997})}\BibitemShut {NoStop}%
	\bibitem [{\citenamefont {Hald}\ \emph {et~al.}(1999)\citenamefont {Hald},
		\citenamefont {S{\o}rensen}, \citenamefont {Schori},\ and\ \citenamefont
		{Polzik}}]{hald1999spin}%
	\BibitemOpen
	\bibfield  {author} {\bibinfo {author} {\bibfnamefont {J.}~\bibnamefont
			{Hald}}, \bibinfo {author} {\bibfnamefont {J.~L.}\ \bibnamefont
			{S{\o}rensen}}, \bibinfo {author} {\bibfnamefont {C.}~\bibnamefont {Schori}},
		\ and\ \bibinfo {author} {\bibfnamefont {E.~S.}\ \bibnamefont {Polzik}},\
	}\bibfield  {title} {\enquote {\bibinfo {title} {Spin {S}queezed {A}toms: {A}
				{M}acroscopic {E}ntangled {E}nsemble {C}reated by {L}ight},}\ }\href
	{https://link.aps.org/doi/10.1103/PhysRevLett.83.1319} {\bibfield  {journal}
		{\bibinfo  {journal} {Phys. Rev. Lett.}\ }\textbf {\bibinfo {volume} {83}},\
		\bibinfo {pages} {1319} (\bibinfo {year} {1999})}\BibitemShut {NoStop}%
	\bibitem [{\citenamefont {Julsgaard}\ \emph {et~al.}(2001)\citenamefont
		{Julsgaard}, \citenamefont {Kozhekin},\ and\ \citenamefont
		{Polzik}}]{julsgaard2001experimental}%
	\BibitemOpen
	\bibfield  {author} {\bibinfo {author} {\bibfnamefont {B.}~\bibnamefont
			{Julsgaard}}, \bibinfo {author} {\bibfnamefont {A.}~\bibnamefont {Kozhekin}},
		\ and\ \bibinfo {author} {\bibfnamefont {E.~S.}\ \bibnamefont {Polzik}},\
	}\bibfield  {title} {\enquote {\bibinfo {title} {Experimental long-lived
				entanglement of two macroscopic objects},}\ }\href
	{https://doi.org/10.1038/35096524} {\bibfield  {journal} {\bibinfo  {journal}
			{Nature}\ }\textbf {\bibinfo {volume} {413}},\ \bibinfo {pages} {400}
		(\bibinfo {year} {2001})}\BibitemShut {NoStop}%
	\bibitem [{\citenamefont {Kuzmich}\ \emph {et~al.}(2000)\citenamefont
		{Kuzmich}, \citenamefont {Mandel},\ and\ \citenamefont
		{Bigelow}}]{kuzmich2000generation}%
	\BibitemOpen
	\bibfield  {author} {\bibinfo {author} {\bibfnamefont {A.}~\bibnamefont
			{Kuzmich}}, \bibinfo {author} {\bibfnamefont {L.}~\bibnamefont {Mandel}}, \
		and\ \bibinfo {author} {\bibfnamefont {N.~P.}\ \bibnamefont {Bigelow}},\
	}\bibfield  {title} {\enquote {\bibinfo {title} {Generation of {S}pin
				{S}queezing via {C}ontinuous {Q}uantum {N}ondemolition {M}easurement},}\
	}\href {https://link.aps.org/doi/10.1103/PhysRevLett.85.1594} {\bibfield
		{journal} {\bibinfo  {journal} {Phys. Rev. Lett.}\ }\textbf {\bibinfo
			{volume} {85}},\ \bibinfo {pages} {1594} (\bibinfo {year}
		{2000})}\BibitemShut {NoStop}%
	\bibitem [{\citenamefont {Koschorreck}\ \emph {et~al.}(2010)\citenamefont
		{Koschorreck}, \citenamefont {Napolitano}, \citenamefont {Dubost},\ and\
		\citenamefont {Mitchell}}]{koschorreck2010quantum}%
	\BibitemOpen
	\bibfield  {author} {\bibinfo {author} {\bibfnamefont {M.}~\bibnamefont
			{Koschorreck}}, \bibinfo {author} {\bibfnamefont {M.}~\bibnamefont
			{Napolitano}}, \bibinfo {author} {\bibfnamefont {B.}~\bibnamefont {Dubost}},
		\ and\ \bibinfo {author} {\bibfnamefont {M.~W.}\ \bibnamefont {Mitchell}},\
	}\bibfield  {title} {\enquote {\bibinfo {title} {Quantum {N}ondemolition
				{M}easurement of {L}arge-{S}pin {E}nsembles by {D}ynamical {D}ecoupling},}\
	}\href {https://link.aps.org/doi/10.1103/PhysRevLett.105.093602} {\bibfield
		{journal} {\bibinfo  {journal} {Phys. Rev. Lett.}\ }\textbf {\bibinfo
			{volume} {105}},\ \bibinfo {pages} {093602} (\bibinfo {year}
		{2010})}\BibitemShut {NoStop}%
	\bibitem [{\citenamefont {Chalopin}\ \emph {et~al.}(2018)\citenamefont
		{Chalopin}, \citenamefont {Bouazza}, \citenamefont {Evrard}, \citenamefont
		{Makhalov}, \citenamefont {Dreon}, \citenamefont {Dalibard}, \citenamefont
		{Sidorenkov},\ and\ \citenamefont {Nascimbene}}]{chalopin2018quantum}%
	\BibitemOpen
	\bibfield  {author} {\bibinfo {author} {\bibfnamefont {T.}~\bibnamefont
			{Chalopin}}, \bibinfo {author} {\bibfnamefont {C.}~\bibnamefont {Bouazza}},
		\bibinfo {author} {\bibfnamefont {A.}~\bibnamefont {Evrard}}, \bibinfo
		{author} {\bibfnamefont {V.}~\bibnamefont {Makhalov}}, \bibinfo {author}
		{\bibfnamefont {D.}~\bibnamefont {Dreon}}, \bibinfo {author} {\bibfnamefont
			{J.}~\bibnamefont {Dalibard}}, \bibinfo {author} {\bibfnamefont {L.~A.}\
			\bibnamefont {Sidorenkov}}, \ and\ \bibinfo {author} {\bibfnamefont
			{S.}~\bibnamefont {Nascimbene}},\ }\bibfield  {title} {\enquote {\bibinfo
			{title} {Quantum-enhanced sensing using non-classical spin states of a highly
				magnetic atom},}\ }\href {https://doi.org/10.1038/s41467-018-07433-1}
	{\bibfield  {journal} {\bibinfo  {journal} {Nat. Commun.}\ }\textbf {\bibinfo
			{volume} {9}},\ \bibinfo {pages} {4955} (\bibinfo {year} {2018})}\BibitemShut
	{NoStop}%
	\bibitem [{\citenamefont {Evrard}\ \emph {et~al.}(2019)\citenamefont {Evrard},
		\citenamefont {Makhalov}, \citenamefont {Chalopin}, \citenamefont
		{Sidorenkov}, \citenamefont {Dalibard}, \citenamefont {Lopes},\ and\
		\citenamefont {Nascimbene}}]{evrard2019enhanced}%
	\BibitemOpen
	\bibfield  {author} {\bibinfo {author} {\bibfnamefont {A.}~\bibnamefont
			{Evrard}}, \bibinfo {author} {\bibfnamefont {V.}~\bibnamefont {Makhalov}},
		\bibinfo {author} {\bibfnamefont {T.}~\bibnamefont {Chalopin}}, \bibinfo
		{author} {\bibfnamefont {L.~A.}\ \bibnamefont {Sidorenkov}}, \bibinfo
		{author} {\bibfnamefont {J.}~\bibnamefont {Dalibard}}, \bibinfo {author}
		{\bibfnamefont {R.}~\bibnamefont {Lopes}}, \ and\ \bibinfo {author}
		{\bibfnamefont {S.}~\bibnamefont {Nascimbene}},\ }\bibfield  {title}
	{\enquote {\bibinfo {title} {Enhanced {M}agnetic {S}ensitivity with
				{N}on-{G}aussian {Q}uantum {F}luctuations},}\ }\href
	{https://link.aps.org/doi/10.1103/PhysRevLett.122.173601} {\bibfield
		{journal} {\bibinfo  {journal} {Phys. Rev. Lett.}\ }\textbf {\bibinfo
			{volume} {122}},\ \bibinfo {pages} {173601} (\bibinfo {year}
		{2019})}\BibitemShut {NoStop}%
	\bibitem [{\citenamefont {You}\ and\ \citenamefont
		{Nori}(2011)}]{you2011atomic}%
	\BibitemOpen
	\bibfield  {author} {\bibinfo {author} {\bibfnamefont {J.~Q.}\ \bibnamefont
			{You}}\ and\ \bibinfo {author} {\bibfnamefont {F.}~\bibnamefont {Nori}},\
	}\bibfield  {title} {\enquote {\bibinfo {title} {Atomic physics and quantum
				optics using superconducting circuits},}\ }\href
	{http://dx.doi.org/10.1038/nature10122} {\bibfield  {journal} {\bibinfo
			{journal} {Nature}\ }\textbf {\bibinfo {volume} {474}},\ \bibinfo {pages}
		{589} (\bibinfo {year} {2011})}\BibitemShut {NoStop}%
	\bibitem [{\citenamefont {Gu}\ \emph {et~al.}(2017)\citenamefont {Gu},
		\citenamefont {Kockum}, \citenamefont {Miranowicz}, \citenamefont {Liu},\
		and\ \citenamefont {Nori}}]{gu2017microwave}%
	\BibitemOpen
	\bibfield  {author} {\bibinfo {author} {\bibfnamefont {X.}~\bibnamefont
			{Gu}}, \bibinfo {author} {\bibfnamefont {A.~F.}\ \bibnamefont {Kockum}},
		\bibinfo {author} {\bibfnamefont {A.}~\bibnamefont {Miranowicz}}, \bibinfo
		{author} {\bibfnamefont {Y.-x.}\ \bibnamefont {Liu}}, \ and\ \bibinfo
		{author} {\bibfnamefont {F.}~\bibnamefont {Nori}},\ }\bibfield  {title}
	{\enquote {\bibinfo {title} {Microwave photonics with superconducting quantum
				circuits},}\ }\href
	{http://www.sciencedirect.com/science/article/pii/S0370157317303290}
	{\bibfield  {journal} {\bibinfo  {journal} {Phys. Rep.}\ }\textbf {\bibinfo
			{volume} {718-719}},\ \bibinfo {pages} {1--102} (\bibinfo {year}
		{2017})}\BibitemShut {NoStop}%
	\bibitem [{\citenamefont {Banerjee}(1996)}]{banerjee1996generation}%
	\BibitemOpen
	\bibfield  {author} {\bibinfo {author} {\bibfnamefont {A.}~\bibnamefont
			{Banerjee}},\ }\bibfield  {title} {\enquote {\bibinfo {title} {Generation of
				atomic-squeezed states in an optical cavity with an injected squeezed
				vacuum},}\ }\href {https://link.aps.org/doi/10.1103/PhysRevA.54.5327}
	{\bibfield  {journal} {\bibinfo  {journal} {Phys. Rev. A}\ }\textbf {\bibinfo
			{volume} {54}},\ \bibinfo {pages} {5327} (\bibinfo {year}
		{1996})}\BibitemShut {NoStop}%
	\bibitem [{\citenamefont {S{\o}rensen}\ and\ \citenamefont
		{M{\o}lmer}(2002)}]{sorensen2002entangling}%
	\BibitemOpen
	\bibfield  {author} {\bibinfo {author} {\bibfnamefont {A.~S.}\ \bibnamefont
			{S{\o}rensen}}\ and\ \bibinfo {author} {\bibfnamefont {K.}~\bibnamefont
			{M{\o}lmer}},\ }\bibfield  {title} {\enquote {\bibinfo {title} {Entangling
				atoms in bad cavities},}\ }\href
	{https://link.aps.org/doi/10.1103/PhysRevA.66.022314} {\bibfield  {journal}
		{\bibinfo  {journal} {Phys. Rev. A}\ }\textbf {\bibinfo {volume} {66}},\
		\bibinfo {pages} {022314} (\bibinfo {year} {2002})}\BibitemShut {NoStop}%
	\bibitem [{\citenamefont {Leroux}\ \emph {et~al.}(2010)\citenamefont {Leroux},
		\citenamefont {Schleier-Smith},\ and\ \citenamefont
		{Vuleti{\'c}}}]{leroux2010implementation}%
	\BibitemOpen
	\bibfield  {author} {\bibinfo {author} {\bibfnamefont {I.~D.}\ \bibnamefont
			{Leroux}}, \bibinfo {author} {\bibfnamefont {M.~H.}\ \bibnamefont
			{Schleier-Smith}}, \ and\ \bibinfo {author} {\bibfnamefont {V.}~\bibnamefont
			{Vuleti{\'c}}},\ }\bibfield  {title} {\enquote {\bibinfo {title}
			{Implementation of {C}avity {S}queezing of a {C}ollective {A}tomic {S}pin},}\
	}\href {https://link.aps.org/doi/10.1103/PhysRevLett.104.073602} {\bibfield
		{journal} {\bibinfo  {journal} {Phys. Rev. Lett.}\ }\textbf {\bibinfo
			{volume} {104}},\ \bibinfo {pages} {073602} (\bibinfo {year}
		{2010})}\BibitemShut {NoStop}%
	\bibitem [{\citenamefont {Schleier-Smith}\ \emph {et~al.}(2010)\citenamefont
		{Schleier-Smith}, \citenamefont {Leroux},\ and\ \citenamefont
		{Vuleti{\'c}}}]{schleier2010states}%
	\BibitemOpen
	\bibfield  {author} {\bibinfo {author} {\bibfnamefont {M.~H.}\ \bibnamefont
			{Schleier-Smith}}, \bibinfo {author} {\bibfnamefont {I.~D.}\ \bibnamefont
			{Leroux}}, \ and\ \bibinfo {author} {\bibfnamefont {V.}~\bibnamefont
			{Vuleti{\'c}}},\ }\bibfield  {title} {\enquote {\bibinfo {title} {States of
				an {E}nsemble of {T}wo-{L}evel {A}toms with {R}educed {Q}uantum
				{U}ncertainty},}\ }\href
	{https://link.aps.org/doi/10.1103/PhysRevLett.104.073604} {\bibfield
		{journal} {\bibinfo  {journal} {Phys. Rev. Lett.}\ }\textbf {\bibinfo
			{volume} {104}},\ \bibinfo {pages} {073604} (\bibinfo {year}
		{2010})}\BibitemShut {NoStop}%
	\bibitem [{\citenamefont {Bohnet}\ \emph {et~al.}(2014)\citenamefont {Bohnet},
		\citenamefont {Cox}, \citenamefont {Norcia}, \citenamefont {Weiner},
		\citenamefont {Chen},\ and\ \citenamefont {Thompson}}]{bohnet2014reduced}%
	\BibitemOpen
	\bibfield  {author} {\bibinfo {author} {\bibfnamefont {J.~G.}\ \bibnamefont
			{Bohnet}}, \bibinfo {author} {\bibfnamefont {Ke.~C.}\ \bibnamefont {Cox}},
		\bibinfo {author} {\bibfnamefont {M.~A.}\ \bibnamefont {Norcia}}, \bibinfo
		{author} {\bibfnamefont {J.~M.}\ \bibnamefont {Weiner}}, \bibinfo {author}
		{\bibfnamefont {Z.}~\bibnamefont {Chen}}, \ and\ \bibinfo {author}
		{\bibfnamefont {J.~K.}\ \bibnamefont {Thompson}},\ }\bibfield  {title}
	{\enquote {\bibinfo {title} {Reduced spin measurement back-action for a phase
				sensitivity ten times beyond the standard quantum limit},}\ }\href
	{https://doi.org/10.1038/nphoton.2014.151} {\bibfield  {journal} {\bibinfo
			{journal} {Nat. Photon.}\ }\textbf {\bibinfo {volume} {8}},\ \bibinfo {pages}
		{731} (\bibinfo {year} {2014})}\BibitemShut {NoStop}%
	\bibitem [{\citenamefont {Hosten}\ \emph {et~al.}(2016)\citenamefont {Hosten},
		\citenamefont {Engelsen}, \citenamefont {Krishnakumar},\ and\ \citenamefont
		{Kasevich}}]{hosten2016measurement}%
	\BibitemOpen
	\bibfield  {author} {\bibinfo {author} {\bibfnamefont {O.}~\bibnamefont
			{Hosten}}, \bibinfo {author} {\bibfnamefont {N.~J.}\ \bibnamefont
			{Engelsen}}, \bibinfo {author} {\bibfnamefont {R.}~\bibnamefont
			{Krishnakumar}}, \ and\ \bibinfo {author} {\bibfnamefont {M.~A.}\
			\bibnamefont {Kasevich}},\ }\bibfield  {title} {\enquote {\bibinfo {title}
			{Measurement noise 100 times lower than the quantum-projection limit using
				entangled atoms},}\ }\href {https://doi.org/10.1038/nature16176} {\bibfield
		{journal} {\bibinfo  {journal} {Nature}\ }\textbf {\bibinfo {volume} {529}},\
		\bibinfo {pages} {505} (\bibinfo {year} {2016})}\BibitemShut {NoStop}%
	\bibitem [{\citenamefont {Cox}\ \emph {et~al.}(2016)\citenamefont {Cox},
		\citenamefont {Greve}, \citenamefont {Weiner},\ and\ \citenamefont
		{Thompson}}]{cox2016deterministic}%
	\BibitemOpen
	\bibfield  {author} {\bibinfo {author} {\bibfnamefont {K.~C.}\ \bibnamefont
			{Cox}}, \bibinfo {author} {\bibfnamefont {G.~P.}\ \bibnamefont {Greve}},
		\bibinfo {author} {\bibfnamefont {J.~M.}\ \bibnamefont {Weiner}}, \ and\
		\bibinfo {author} {\bibfnamefont {J.~K.}\ \bibnamefont {Thompson}},\
	}\bibfield  {title} {\enquote {\bibinfo {title} {Deterministic {S}queezed
				{S}tates with {C}ollective {M}easurements and {F}eedback},}\ }\href
	{https://link.aps.org/doi/10.1103/PhysRevLett.116.093602} {\bibfield
		{journal} {\bibinfo  {journal} {Phys. Rev. Lett.}\ }\textbf {\bibinfo
			{volume} {116}},\ \bibinfo {pages} {093602} (\bibinfo {year}
		{2016})}\BibitemShut {NoStop}%
	\bibitem [{\citenamefont {Zhang}\ \emph {et~al.}(2017)\citenamefont {Zhang},
		\citenamefont {Zhou}, \citenamefont {Zhou}, \citenamefont {Guo},\ and\
		\citenamefont {Zhou}}]{zhang2017cavity}%
	\BibitemOpen
	\bibfield  {author} {\bibinfo {author} {\bibfnamefont {Y.-C.}\ \bibnamefont
			{Zhang}}, \bibinfo {author} {\bibfnamefont {X.-F.}\ \bibnamefont {Zhou}},
		\bibinfo {author} {\bibfnamefont {X.}~\bibnamefont {Zhou}}, \bibinfo {author}
		{\bibfnamefont {G.-C.}\ \bibnamefont {Guo}}, \ and\ \bibinfo {author}
		{\bibfnamefont {Z.-W.}\ \bibnamefont {Zhou}},\ }\bibfield  {title} {\enquote
		{\bibinfo {title} {Cavity-{A}ssisted {S}ingle-{M}ode and {T}wo-{M}ode
				{S}pin-{S}queezed {S}tates via {P}hase-{L}ocked {A}tom-{P}hoton
				{C}oupling},}\ }\href
	{https://link.aps.org/doi/10.1103/PhysRevLett.118.083604} {\bibfield
		{journal} {\bibinfo  {journal} {Phys. Rev. Lett.}\ }\textbf {\bibinfo
			{volume} {118}},\ \bibinfo {pages} {083604} (\bibinfo {year}
		{2017})}\BibitemShut {NoStop}%
	\bibitem [{\citenamefont {Lewis-Swan}\ \emph {et~al.}(2018)\citenamefont
		{Lewis-Swan}, \citenamefont {Norcia}, \citenamefont {Cline}, \citenamefont
		{Thompson},\ and\ \citenamefont {Rey}}]{lewis2018robust}%
	\BibitemOpen
	\bibfield  {author} {\bibinfo {author} {\bibfnamefont {R.~J.}\ \bibnamefont
			{Lewis-Swan}}, \bibinfo {author} {\bibfnamefont {M.~A.}\ \bibnamefont
			{Norcia}}, \bibinfo {author} {\bibfnamefont {J.~R.~K.}\ \bibnamefont
			{Cline}}, \bibinfo {author} {\bibfnamefont {J.~K.}\ \bibnamefont {Thompson}},
		\ and\ \bibinfo {author} {\bibfnamefont {A.~M.}\ \bibnamefont {Rey}},\
	}\bibfield  {title} {\enquote {\bibinfo {title} {Robust {S}pin {S}queezing
				via {P}hoton-{M}ediated {I}nteractions on an {O}ptical {C}lock
				{T}ransition},}\ }\href
	{https://link.aps.org/doi/10.1103/PhysRevLett.121.070403} {\bibfield
		{journal} {\bibinfo  {journal} {Phys. Rev. Lett.}\ }\textbf {\bibinfo
			{volume} {121}},\ \bibinfo {pages} {070403} (\bibinfo {year}
		{2018})}\BibitemShut {NoStop}%
	\bibitem [{\citenamefont {Braverman}\ \emph {et~al.}(2019)\citenamefont
		{Braverman}, \citenamefont {Kawasaki}, \citenamefont {Pedrozo-Pe{\~n}afiel},
		\citenamefont {Colombo}, \citenamefont {Shu}, \citenamefont {Li},
		\citenamefont {Mendez}, \citenamefont {Yamoah}, \citenamefont {Salvi},
		\citenamefont {Akamatsu}, \citenamefont {Xiao},\ and\ \citenamefont
		{Vuleti\'{c}}}]{braverman2019near}%
	\BibitemOpen
	\bibfield  {author} {\bibinfo {author} {\bibfnamefont {B.}~\bibnamefont
			{Braverman}}, \bibinfo {author} {\bibfnamefont {A.}~\bibnamefont {Kawasaki}},
		\bibinfo {author} {\bibfnamefont {E.}~\bibnamefont {Pedrozo-Pe{\~n}afiel}},
		\bibinfo {author} {\bibfnamefont {S.}~\bibnamefont {Colombo}}, \bibinfo
		{author} {\bibfnamefont {C.}~\bibnamefont {Shu}}, \bibinfo {author}
		{\bibfnamefont {Z.}~\bibnamefont {Li}}, \bibinfo {author} {\bibfnamefont
			{E.}~\bibnamefont {Mendez}}, \bibinfo {author} {\bibfnamefont
			{M.}~\bibnamefont {Yamoah}}, \bibinfo {author} {\bibfnamefont
			{L.}~\bibnamefont {Salvi}}, \bibinfo {author} {\bibfnamefont
			{D.}~\bibnamefont {Akamatsu}}, \bibinfo {author} {\bibfnamefont
			{Y.}~\bibnamefont {Xiao}}, \ and\ \bibinfo {author} {\bibfnamefont
			{V.}~\bibnamefont {Vuleti\'{c}}},\ }\bibfield  {title} {\enquote {\bibinfo
			{title} {Near-{U}nitary {S}pin {S}queezing in $^{171}\mathrm{Yb}$},}\ }\href
	{https://link.aps.org/doi/10.1103/PhysRevLett.122.223203} {\bibfield
		{journal} {\bibinfo  {journal} {Phys. Rev. Lett.}\ }\textbf {\bibinfo
			{volume} {122}},\ \bibinfo {pages} {223203} (\bibinfo {year}
		{2019})}\BibitemShut {NoStop}%
	\bibitem [{\citenamefont {Song}\ \emph {et~al.}(2019)\citenamefont {Song},
		\citenamefont {Xu}, \citenamefont {Li}, \citenamefont {Zhang}, \citenamefont
		{Zhang}, \citenamefont {Liu}, \citenamefont {Guo}, \citenamefont {Wang},
		\citenamefont {Ren}, \citenamefont {Hao}, \citenamefont {Feng}, \citenamefont
		{Fan}, \citenamefont {Zheng}, \citenamefont {Wang}, \citenamefont {Wang},\
		and\ \citenamefont {Zhu}}]{song2019generation}%
	\BibitemOpen
	\bibfield  {author} {\bibinfo {author} {\bibfnamefont {C.}~\bibnamefont
			{Song}}, \bibinfo {author} {\bibfnamefont {K.}~\bibnamefont {Xu}}, \bibinfo
		{author} {\bibfnamefont {H.}~\bibnamefont {Li}}, \bibinfo {author}
		{\bibfnamefont {Y.-R.}\ \bibnamefont {Zhang}}, \bibinfo {author}
		{\bibfnamefont {X.}~\bibnamefont {Zhang}}, \bibinfo {author} {\bibfnamefont
			{W.}~\bibnamefont {Liu}}, \bibinfo {author} {\bibfnamefont {Q.}~\bibnamefont
			{Guo}}, \bibinfo {author} {\bibfnamefont {Z.}~\bibnamefont {Wang}}, \bibinfo
		{author} {\bibfnamefont {W.}~\bibnamefont {Ren}}, \bibinfo {author}
		{\bibfnamefont {J.}~\bibnamefont {Hao}}, \bibinfo {author} {\bibfnamefont
			{H.}~\bibnamefont {Feng}}, \bibinfo {author} {\bibfnamefont {H.}~\bibnamefont
			{Fan}}, \bibinfo {author} {\bibfnamefont {D.}~\bibnamefont {Zheng}}, \bibinfo
		{author} {\bibfnamefont {D.-W.}\ \bibnamefont {Wang}}, \bibinfo {author}
		{\bibfnamefont {H.}~\bibnamefont {Wang}}, \ and\ \bibinfo {author}
		{\bibfnamefont {S.-Y.}\ \bibnamefont {Zhu}},\ }\bibfield  {title} {\enquote
		{\bibinfo {title} {Generation of multicomponent atomic {S}chr{\"o}dinger cat
				states of up to 20 qubits},}\ }\href {\doibase 10.1126/science.aay0600}
	{\bibfield  {journal} {\bibinfo  {journal} {Science}\ }\textbf {\bibinfo
			{volume} {365}},\ \bibinfo {pages} {574--577} (\bibinfo {year}
		{2019})}\BibitemShut {NoStop}%
	\bibitem [{\citenamefont {Verstraete}\ \emph {et~al.}(2009)\citenamefont
		{Verstraete}, \citenamefont {Wolf},\ and\ \citenamefont
		{Cirac}}]{verstraete2009quantum}%
	\BibitemOpen
	\bibfield  {author} {\bibinfo {author} {\bibfnamefont {F.}~\bibnamefont
			{Verstraete}}, \bibinfo {author} {\bibfnamefont {M.~M.}\ \bibnamefont
			{Wolf}}, \ and\ \bibinfo {author} {\bibfnamefont {J.~I.}\ \bibnamefont
			{Cirac}},\ }\bibfield  {title} {\enquote {\bibinfo {title} {Quantum
				computation and quantum-state engineering driven by dissipation},}\ }\href
	{https://doi.org/10.1038/nphys1342} {\bibfield  {journal} {\bibinfo
			{journal} {Nat. Phys.}\ }\textbf {\bibinfo {volume} {5}},\ \bibinfo {pages}
		{633--636} (\bibinfo {year} {2009})}\BibitemShut {NoStop}%
	\bibitem [{\citenamefont {Krauter}\ \emph {et~al.}(2011)\citenamefont
		{Krauter}, \citenamefont {Muschik}, \citenamefont {Jensen}, \citenamefont
		{Wasilewski}, \citenamefont {Petersen}, \citenamefont {Cirac},\ and\
		\citenamefont {Polzik}}]{krauter2011entanglement}%
	\BibitemOpen
	\bibfield  {author} {\bibinfo {author} {\bibfnamefont {H.}~\bibnamefont
			{Krauter}}, \bibinfo {author} {\bibfnamefont {C.~A.}\ \bibnamefont
			{Muschik}}, \bibinfo {author} {\bibfnamefont {K.}~\bibnamefont {Jensen}},
		\bibinfo {author} {\bibfnamefont {W.}~\bibnamefont {Wasilewski}}, \bibinfo
		{author} {\bibfnamefont {J.~M.}\ \bibnamefont {Petersen}}, \bibinfo {author}
		{\bibfnamefont {J.~I.}\ \bibnamefont {Cirac}}, \ and\ \bibinfo {author}
		{\bibfnamefont {E.~S.}\ \bibnamefont {Polzik}},\ }\bibfield  {title}
	{\enquote {\bibinfo {title} {Entanglement {G}nerated by {D}issipation and
				{S}teady {S}tate {E}ntanglement of {T}wo {M}acroscopic {O}bjects},}\ }\href
	{https://link.aps.org/doi/10.1103/PhysRevLett.107.080503} {\bibfield
		{journal} {\bibinfo  {journal} {Phys. Rev. Lett.}\ }\textbf {\bibinfo
			{volume} {107}},\ \bibinfo {pages} {080503} (\bibinfo {year}
		{2011})}\BibitemShut {NoStop}%
	\bibitem [{\citenamefont {Lin}\ \emph {et~al.}(2013)\citenamefont {Lin},
		\citenamefont {Gaebler}, \citenamefont {Reiter}, \citenamefont {Tan},
		\citenamefont {Bowler}, \citenamefont {S{\o}rensen}, \citenamefont
		{Leibfried},\ and\ \citenamefont {Wineland}}]{lin2013dissipative}%
	\BibitemOpen
	\bibfield  {author} {\bibinfo {author} {\bibfnamefont {Y.}~\bibnamefont
			{Lin}}, \bibinfo {author} {\bibfnamefont {J.~P.}\ \bibnamefont {Gaebler}},
		\bibinfo {author} {\bibfnamefont {F.}~\bibnamefont {Reiter}}, \bibinfo
		{author} {\bibfnamefont {T.~R.}\ \bibnamefont {Tan}}, \bibinfo {author}
		{\bibfnamefont {R.}~\bibnamefont {Bowler}}, \bibinfo {author} {\bibfnamefont
			{A.~S.}\ \bibnamefont {S{\o}rensen}}, \bibinfo {author} {\bibfnamefont
			{D.}~\bibnamefont {Leibfried}}, \ and\ \bibinfo {author} {\bibfnamefont
			{D.~J.}\ \bibnamefont {Wineland}},\ }\bibfield  {title} {\enquote {\bibinfo
			{title} {Dissipative production of a maximally entangled steady state of two
				quantum bits},}\ }\href {http://dx.doi.org/10.1038/nature12801} {\bibfield
		{journal} {\bibinfo  {journal} {Nature}\ }\textbf {\bibinfo {volume} {504}},\
		\bibinfo {pages} {415} (\bibinfo {year} {2013})}\BibitemShut {NoStop}%
	\bibitem [{\citenamefont {Qin}\ \emph {et~al.}(2017)\citenamefont {Qin},
		\citenamefont {Wang}, \citenamefont {Miranowicz}, \citenamefont {Zhong},\
		and\ \citenamefont {Nori}}]{qin2017heralded}%
	\BibitemOpen
	\bibfield  {author} {\bibinfo {author} {\bibfnamefont {W.}~\bibnamefont
			{Qin}}, \bibinfo {author} {\bibfnamefont {X.}~\bibnamefont {Wang}}, \bibinfo
		{author} {\bibfnamefont {A.}~\bibnamefont {Miranowicz}}, \bibinfo {author}
		{\bibfnamefont {Z.}~\bibnamefont {Zhong}}, \ and\ \bibinfo {author}
		{\bibfnamefont {F.}~\bibnamefont {Nori}},\ }\bibfield  {title} {\enquote
		{\bibinfo {title} {Heralded quantum controlled-phase gates with dissipative
				dynamics in macroscopically distant resonators},}\ }\href
	{https://link.aps.org/doi/10.1103/PhysRevA.96.012315} {\bibfield  {journal}
		{\bibinfo  {journal} {Phys. Rev. A}\ }\textbf {\bibinfo {volume} {96}},\
		\bibinfo {pages} {012315} (\bibinfo {year} {2017})}\BibitemShut {NoStop}%
	\bibitem [{\citenamefont {Qin}\ \emph {et~al.}(2018)\citenamefont {Qin},
		\citenamefont {Miranowicz}, \citenamefont {Li}, \citenamefont {L{\"u}},
		\citenamefont {You},\ and\ \citenamefont {Nori}}]{qin2018exponentially}%
	\BibitemOpen
	\bibfield  {author} {\bibinfo {author} {\bibfnamefont {W.}~\bibnamefont
			{Qin}}, \bibinfo {author} {\bibfnamefont {A.}~\bibnamefont {Miranowicz}},
		\bibinfo {author} {\bibfnamefont {P.-B.}\ \bibnamefont {Li}}, \bibinfo
		{author} {\bibfnamefont {X.-Y.}\ \bibnamefont {L{\"u}}}, \bibinfo {author}
		{\bibfnamefont {J.~Q.}\ \bibnamefont {You}}, \ and\ \bibinfo {author}
		{\bibfnamefont {F.}~\bibnamefont {Nori}},\ }\bibfield  {title} {\enquote
		{\bibinfo {title} {Exponentially {E}nhanced {L}ight-{M}atter {I}nteraction,
				{C}ooperativities, and {S}teady-{S}tate {E}ntanglement {U}sing {P}arametric
				{A}mplification},}\ }\href
	{https://link.aps.org/doi/10.1103/PhysRevLett.120.093601} {\bibfield
		{journal} {\bibinfo  {journal} {Phys. Rev. Lett.}\ }\textbf {\bibinfo
			{volume} {120}},\ \bibinfo {pages} {093601} (\bibinfo {year}
		{2018})}\BibitemShut {NoStop}%
	\bibitem [{\citenamefont {Parkins}\ \emph {et~al.}(2006)\citenamefont
		{Parkins}, \citenamefont {Solano},\ and\ \citenamefont
		{Cirac}}]{parkins2006unconditional}%
	\BibitemOpen
	\bibfield  {author} {\bibinfo {author} {\bibfnamefont {A.~S.}\ \bibnamefont
			{Parkins}}, \bibinfo {author} {\bibfnamefont {E.}~\bibnamefont {Solano}}, \
		and\ \bibinfo {author} {\bibfnamefont {J.~I.}\ \bibnamefont {Cirac}},\
	}\bibfield  {title} {\enquote {\bibinfo {title} {Unconditional {T}wo-{M}ode
				{S}queezing of {S}eparated {A}tomic {E}nsembles},}\ }\href
	{https://link.aps.org/doi/10.1103/PhysRevLett.96.053602} {\bibfield
		{journal} {\bibinfo  {journal} {Phys. Rev. Lett.}\ }\textbf {\bibinfo
			{volume} {96}},\ \bibinfo {pages} {053602} (\bibinfo {year}
		{2006})}\BibitemShut {NoStop}%
	\bibitem [{\citenamefont {Zheng}(2012)}]{zheng2012generation}%
	\BibitemOpen
	\bibfield  {author} {\bibinfo {author} {\bibfnamefont {S.-B.}\ \bibnamefont
			{Zheng}},\ }\bibfield  {title} {\enquote {\bibinfo {title} {Generation of
				atomic and field squeezing by adiabatic passage and symmetry breaking},}\
	}\href {https://link.aps.org/doi/10.1103/PhysRevA.86.013828} {\bibfield
		{journal} {\bibinfo  {journal} {Phys. Rev. A}\ }\textbf {\bibinfo {volume}
			{86}},\ \bibinfo {pages} {013828} (\bibinfo {year} {2012})}\BibitemShut
	{NoStop}%
	\bibitem [{\citenamefont {Dalla~Torre}\ \emph {et~al.}(2013)\citenamefont
		{Dalla~Torre}, \citenamefont {Otterbach}, \citenamefont {Demler},
		\citenamefont {Vuletic},\ and\ \citenamefont {Lukin}}]{dalla2013dissipative}%
	\BibitemOpen
	\bibfield  {author} {\bibinfo {author} {\bibfnamefont {E.~G.}\ \bibnamefont
			{Dalla~Torre}}, \bibinfo {author} {\bibfnamefont {J.}~\bibnamefont
			{Otterbach}}, \bibinfo {author} {\bibfnamefont {E.}~\bibnamefont {Demler}},
		\bibinfo {author} {\bibfnamefont {V.}~\bibnamefont {Vuletic}}, \ and\
		\bibinfo {author} {\bibfnamefont {M.~D.}\ \bibnamefont {Lukin}},\ }\bibfield
	{title} {\enquote {\bibinfo {title} {Dissipative {P}reparation of {S}pin
				{S}queezed {A}tomic {E}nsembles in a {S}teady {S}tate},}\ }\href
	{https://link.aps.org/doi/10.1103/PhysRevLett.110.120402} {\bibfield
		{journal} {\bibinfo  {journal} {Phys. Rev. Lett.}\ }\textbf {\bibinfo
			{volume} {110}},\ \bibinfo {pages} {120402} (\bibinfo {year}
		{2013})}\BibitemShut {NoStop}%
	\bibitem [{\citenamefont {Ma}\ \emph {et~al.}(2013)\citenamefont {Ma},
		\citenamefont {Li}, \citenamefont {Fang}, \citenamefont {Gao},\ and\
		\citenamefont {Li}}]{ma2013dissipation}%
	\BibitemOpen
	\bibfield  {author} {\bibinfo {author} {\bibfnamefont {S.-L.}\ \bibnamefont
			{Ma}}, \bibinfo {author} {\bibfnamefont {P.-B.}\ \bibnamefont {Li}}, \bibinfo
		{author} {\bibfnamefont {A.-P.}\ \bibnamefont {Fang}}, \bibinfo {author}
		{\bibfnamefont {S.-Y.}\ \bibnamefont {Gao}}, \ and\ \bibinfo {author}
		{\bibfnamefont {F.-L.}\ \bibnamefont {Li}},\ }\bibfield  {title} {\enquote
		{\bibinfo {title} {Dissipation-assisted generation of steady-state
				single-mode squeezing of collective excitations in a solid-state spin
				ensemble},}\ }\href {https://link.aps.org/doi/10.1103/PhysRevA.88.013837}
	{\bibfield  {journal} {\bibinfo  {journal} {Phys. Rev. A}\ }\textbf {\bibinfo
			{volume} {88}},\ \bibinfo {pages} {013837} (\bibinfo {year}
		{2013})}\BibitemShut {NoStop}%
	\bibitem [{\citenamefont {Borregaard}\ \emph {et~al.}(2017)\citenamefont
		{Borregaard}, \citenamefont {Davis}, \citenamefont {Bentsen}, \citenamefont
		{Schleier-Smith},\ and\ \citenamefont {S{\o}rensen}}]{borregaard2017one}%
	\BibitemOpen
	\bibfield  {author} {\bibinfo {author} {\bibfnamefont {J.}~\bibnamefont
			{Borregaard}}, \bibinfo {author} {\bibfnamefont {E.~J.}\ \bibnamefont
			{Davis}}, \bibinfo {author} {\bibfnamefont {G.~S.}\ \bibnamefont {Bentsen}},
		\bibinfo {author} {\bibfnamefont {M.~H.}\ \bibnamefont {Schleier-Smith}}, \
		and\ \bibinfo {author} {\bibfnamefont {A.~S.}\ \bibnamefont {S{\o}rensen}},\
	}\bibfield  {title} {\enquote {\bibinfo {title} {One-and two-axis squeezing
				of atomic ensembles in optical cavities},}\ }\href
	{https://doi.org/10.1088%2F1367-2630%2Faa8438} {\bibfield  {journal}
		{\bibinfo  {journal} {New J. Phys.}\ }\textbf {\bibinfo {volume} {19}},\
		\bibinfo {pages} {093021} (\bibinfo {year} {2017})}\BibitemShut {NoStop}%
	\bibitem [{\citenamefont {Liu}\ \emph {et~al.}(2019)\citenamefont {Liu},
		\citenamefont {Wang}, \citenamefont {Yan}, \citenamefont {Jiang},
		\citenamefont {Xiong},\ and\ \citenamefont {Wang}}]{liu2019spin}%
	\BibitemOpen
	\bibfield  {author} {\bibinfo {author} {\bibfnamefont {G.}~\bibnamefont
			{Liu}}, \bibinfo {author} {\bibfnamefont {Y.-N.}\ \bibnamefont {Wang}},
		\bibinfo {author} {\bibfnamefont {L.-F.}\ \bibnamefont {Yan}}, \bibinfo
		{author} {\bibfnamefont {N.-Q.}\ \bibnamefont {Jiang}}, \bibinfo {author}
		{\bibfnamefont {W.}~\bibnamefont {Xiong}}, \ and\ \bibinfo {author}
		{\bibfnamefont {M.-F.}\ \bibnamefont {Wang}},\ }\bibfield  {title} {\enquote
		{\bibinfo {title} {Spin squeezing via one-and two-axis twisting induced by a
				single off-resonance stimulated {R}aman scattering in a cavity},}\ }\href
	{https://link.aps.org/doi/10.1103/PhysRevA.99.043840} {\bibfield  {journal}
		{\bibinfo  {journal} {Phys. Rev. A}\ }\textbf {\bibinfo {volume} {99}},\
		\bibinfo {pages} {043840} (\bibinfo {year} {2019})}\BibitemShut {NoStop}%
	\bibitem [{\citenamefont {Jensen}\ \emph {et~al.}(2011)\citenamefont {Jensen},
		\citenamefont {Wasilewski}, \citenamefont {Krauter}, \citenamefont
		{Fernholz}, \citenamefont {Nielsen}, \citenamefont {Owari}, \citenamefont
		{Plenio}, \citenamefont {Serafini}, \citenamefont {Wolf},\ and\ \citenamefont
		{Polzik}}]{jensen2011quantum}%
	\BibitemOpen
	\bibfield  {author} {\bibinfo {author} {\bibfnamefont {K.}~\bibnamefont
			{Jensen}}, \bibinfo {author} {\bibfnamefont {W.}~\bibnamefont {Wasilewski}},
		\bibinfo {author} {\bibfnamefont {H.}~\bibnamefont {Krauter}}, \bibinfo
		{author} {\bibfnamefont {T.}~\bibnamefont {Fernholz}}, \bibinfo {author}
		{\bibfnamefont {B.~M.}\ \bibnamefont {Nielsen}}, \bibinfo {author}
		{\bibfnamefont {M.}~\bibnamefont {Owari}}, \bibinfo {author} {\bibfnamefont
			{M.~B.}\ \bibnamefont {Plenio}}, \bibinfo {author} {\bibfnamefont
			{A.}~\bibnamefont {Serafini}}, \bibinfo {author} {\bibfnamefont {M.~M.}\
			\bibnamefont {Wolf}}, \ and\ \bibinfo {author} {\bibfnamefont {E.~S.}\
			\bibnamefont {Polzik}},\ }\bibfield  {title} {\enquote {\bibinfo {title}
			{Quantum memory for entangled continuous-variable states},}\ }\href
	{https://doi.org/10.1038/nphys1819} {\bibfield  {journal} {\bibinfo
			{journal} {Nat. Phys.}\ }\textbf {\bibinfo {volume} {7}},\ \bibinfo {pages}
		{13} (\bibinfo {year} {2011})}\BibitemShut {NoStop}%
	\bibitem [{\citenamefont {Yan}\ \emph {et~al.}(2017)\citenamefont {Yan},
		\citenamefont {Wu}, \citenamefont {Jia}, \citenamefont {Liu}, \citenamefont
		{Deng}, \citenamefont {Li}, \citenamefont {Wang}, \citenamefont {Xie},\ and\
		\citenamefont {Peng}}]{yan2017establishing}%
	\BibitemOpen
	\bibfield  {author} {\bibinfo {author} {\bibfnamefont {Z.}~\bibnamefont
			{Yan}}, \bibinfo {author} {\bibfnamefont {L.}~\bibnamefont {Wu}}, \bibinfo
		{author} {\bibfnamefont {X.}~\bibnamefont {Jia}}, \bibinfo {author}
		{\bibfnamefont {Y.}~\bibnamefont {Liu}}, \bibinfo {author} {\bibfnamefont
			{R.}~\bibnamefont {Deng}}, \bibinfo {author} {\bibfnamefont {S.}~\bibnamefont
			{Li}}, \bibinfo {author} {\bibfnamefont {H.}~\bibnamefont {Wang}}, \bibinfo
		{author} {\bibfnamefont {C.}~\bibnamefont {Xie}}, \ and\ \bibinfo {author}
		{\bibfnamefont {K.}~\bibnamefont {Peng}},\ }\bibfield  {title} {\enquote
		{\bibinfo {title} {Establishing and storing of deterministic quantum
				entanglement among three distant atomic ensembles},}\ }\href
	{https://doi.org/10.1038/s41467-017-00809-9} {\bibfield  {journal} {\bibinfo
			{journal} {Nat. Commun.}\ }\textbf {\bibinfo {volume} {8}},\ \bibinfo {pages}
		{718} (\bibinfo {year} {2017})}\BibitemShut {NoStop}%
	\bibitem [{\citenamefont {Milburn}\ and\ \citenamefont
		{Walls}(1981)}]{milburn1981production}%
	\BibitemOpen
	\bibfield  {author} {\bibinfo {author} {\bibfnamefont {G.}~\bibnamefont
			{Milburn}}\ and\ \bibinfo {author} {\bibfnamefont {D.~F.}\ \bibnamefont
			{Walls}},\ }\bibfield  {title} {\enquote {\bibinfo {title} {Production of
				squeezed states in a degenerate parametric amplifier},}\ }\href
	{http://www.sciencedirect.com/science/article/pii/0030401881902327}
	{\bibfield  {journal} {\bibinfo  {journal} {Opt. Commun.}\ }\textbf {\bibinfo
			{volume} {39}},\ \bibinfo {pages} {401--404} (\bibinfo {year}
		{1981})}\BibitemShut {NoStop}%
	\bibitem [{\citenamefont {Boyd}(2003)}]{boyd2003nonlinear}%
	\BibitemOpen
	\bibfield  {author} {\bibinfo {author} {\bibfnamefont {R.~W.}\ \bibnamefont
			{Boyd}},\ }\href@noop {} {\emph {\bibinfo {title} {Nonlinear {O}ptics}}}\
	(\bibinfo  {publisher} {Elsevier, North Holland},\ \bibinfo {year}
	{2003})\BibitemShut {NoStop}%
	\bibitem [{\citenamefont {Gelhausen}\ \emph {et~al.}(2017)\citenamefont
		{Gelhausen}, \citenamefont {Buchhold},\ and\ \citenamefont
		{Strack}}]{gelhausen2017many}%
	\BibitemOpen
	\bibfield  {author} {\bibinfo {author} {\bibfnamefont {J.}~\bibnamefont
			{Gelhausen}}, \bibinfo {author} {\bibfnamefont {M.}~\bibnamefont {Buchhold}},
		\ and\ \bibinfo {author} {\bibfnamefont {P.}~\bibnamefont {Strack}},\
	}\bibfield  {title} {\enquote {\bibinfo {title} {Many-body quantum optics
				with decaying atomic spin states: ($\gamma$, $\kappa$) {D}icke model},}\
	}\href {https://link.aps.org/doi/10.1103/PhysRevA.95.063824} {\bibfield
		{journal} {\bibinfo  {journal} {Phys. Rev. A}\ }\textbf {\bibinfo {volume}
			{95}},\ \bibinfo {pages} {063824} (\bibinfo {year} {2017})}\BibitemShut
	{NoStop}%
	\bibitem [{\citenamefont {Shammah}\ \emph {et~al.}(2018)\citenamefont
		{Shammah}, \citenamefont {Ahmed}, \citenamefont {Lambert}, \citenamefont
		{De~Liberato},\ and\ \citenamefont {Nori}}]{shammah2018open}%
	\BibitemOpen
	\bibfield  {author} {\bibinfo {author} {\bibfnamefont {N.}~\bibnamefont
			{Shammah}}, \bibinfo {author} {\bibfnamefont {S.}~\bibnamefont {Ahmed}},
		\bibinfo {author} {\bibfnamefont {N.}~\bibnamefont {Lambert}}, \bibinfo
		{author} {\bibfnamefont {S.}~\bibnamefont {De~Liberato}}, \ and\ \bibinfo
		{author} {\bibfnamefont {F.}~\bibnamefont {Nori}},\ }\bibfield  {title}
	{\enquote {\bibinfo {title} {Open quantum systems with local and collective
				incoherent processes: Efficient numerical simulations using permutational
				invariance},}\ }\href {https://link.aps.org/doi/10.1103/PhysRevA.98.063815}
	{\bibfield  {journal} {\bibinfo  {journal} {Phys. Rev. A}\ }\textbf {\bibinfo
			{volume} {98}},\ \bibinfo {pages} {063815} (\bibinfo {year}
		{2018})}\BibitemShut {NoStop}%
	\bibitem [{\citenamefont {Macr\`{\i}}\ \emph {et~al.}(2020)\citenamefont
		{Macr\`{\i}}, \citenamefont {Nori}, \citenamefont {Savasta},\ and\
		\citenamefont {Zueco}}]{macri2020spin}%
	\BibitemOpen
	\bibfield  {author} {\bibinfo {author} {\bibfnamefont {Vincenzo}\
			\bibnamefont {Macr\`{\i}}}, \bibinfo {author} {\bibfnamefont {Franco}\
			\bibnamefont {Nori}}, \bibinfo {author} {\bibfnamefont {Salvatore}\
			\bibnamefont {Savasta}}, \ and\ \bibinfo {author} {\bibfnamefont {David}\
			\bibnamefont {Zueco}},\ }\bibfield  {title} {\enquote {\bibinfo {title} {Spin
				squeezing by one-photon--two-atom excitation processes in atomic
				ensembles},}\ }\href {\doibase 10.1103/PhysRevA.101.053818} {\bibfield
		{journal} {\bibinfo  {journal} {Phys. Rev. A}\ }\textbf {\bibinfo {volume}
			{101}},\ \bibinfo {pages} {053818} (\bibinfo {year} {2020})}\BibitemShut
	{NoStop}%
	\bibitem [{\citenamefont {Gamel}\ and\ \citenamefont
		{James}(2010)}]{gamel2010time}%
	\BibitemOpen
	\bibfield  {author} {\bibinfo {author} {\bibfnamefont {O.}~\bibnamefont
			{Gamel}}\ and\ \bibinfo {author} {\bibfnamefont {D.~F.~V.}\ \bibnamefont
			{James}},\ }\bibfield  {title} {\enquote {\bibinfo {title} {Time-averaged
				quantum dynamics and the validity of the effective {H}amiltonian model},}\
	}\href {https://link.aps.org/doi/10.1103/PhysRevA.82.052106} {\bibfield
		{journal} {\bibinfo  {journal} {Phys. Rev. A}\ }\textbf {\bibinfo {volume}
			{82}},\ \bibinfo {pages} {052106} (\bibinfo {year} {2010})}\BibitemShut
	{NoStop}%
	\bibitem [{\citenamefont {Wang}\ \emph {et~al.}(2014)\citenamefont {Wang},
		\citenamefont {Li}, \citenamefont {Li}, \citenamefont {Jiang}, \citenamefont
		{Gao},\ and\ \citenamefont {Li}}]{PhysRevA.90.013838}%
	\BibitemOpen
	\bibfield  {author} {\bibinfo {author} {\bibfnamefont {X.}~\bibnamefont
			{Wang}}, \bibinfo {author} {\bibfnamefont {H.-R.}\ \bibnamefont {Li}},
		\bibinfo {author} {\bibfnamefont {P.-B.}\ \bibnamefont {Li}}, \bibinfo
		{author} {\bibfnamefont {C.-W.}\ \bibnamefont {Jiang}}, \bibinfo {author}
		{\bibfnamefont {H.}~\bibnamefont {Gao}}, \ and\ \bibinfo {author}
		{\bibfnamefont {F.-L.}\ \bibnamefont {Li}},\ }\bibfield  {title} {\enquote
		{\bibinfo {title} {Preparing ground states and squeezed states of
				nanomechanical cantilevers by fast dissipation},}\ }\href {\doibase
		10.1103/PhysRevA.90.013838} {\bibfield  {journal} {\bibinfo  {journal} {Phys.
				Rev. A}\ }\textbf {\bibinfo {volume} {90}},\ \bibinfo {pages} {013838}
		(\bibinfo {year} {2014})}\BibitemShut {NoStop}%
	\bibitem [{\citenamefont {Stanwix}\ \emph {et~al.}(2010)\citenamefont
		{Stanwix}, \citenamefont {Pham}, \citenamefont {Maze}, \citenamefont
		{Le~Sage}, \citenamefont {Yeung}, \citenamefont {Cappellaro}, \citenamefont
		{Hemmer}, \citenamefont {Yacoby}, \citenamefont {Lukin},\ and\ \citenamefont
		{Walsworth}}]{stanwix2010coherence}%
	\BibitemOpen
	\bibfield  {author} {\bibinfo {author} {\bibfnamefont {P.~L.}\ \bibnamefont
			{Stanwix}}, \bibinfo {author} {\bibfnamefont {L.~M.}\ \bibnamefont {Pham}},
		\bibinfo {author} {\bibfnamefont {J.~R.}\ \bibnamefont {Maze}}, \bibinfo
		{author} {\bibfnamefont {D.}~\bibnamefont {Le~Sage}}, \bibinfo {author}
		{\bibfnamefont {T.~K.}\ \bibnamefont {Yeung}}, \bibinfo {author}
		{\bibfnamefont {P.}~\bibnamefont {Cappellaro}}, \bibinfo {author}
		{\bibfnamefont {P.~R.}\ \bibnamefont {Hemmer}}, \bibinfo {author}
		{\bibfnamefont {A.}~\bibnamefont {Yacoby}}, \bibinfo {author} {\bibfnamefont
			{M.~D.}\ \bibnamefont {Lukin}}, \ and\ \bibinfo {author} {\bibfnamefont
			{R.~L.}\ \bibnamefont {Walsworth}},\ }\bibfield  {title} {\enquote {\bibinfo
			{title} {Coherence of nitrogen-vacancy electronic spin ensembles in
				diamond},}\ }\href {https://link.aps.org/doi/10.1103/PhysRevB.82.201201}
	{\bibfield  {journal} {\bibinfo  {journal} {Phys. Rev. B}\ }\textbf {\bibinfo
			{volume} {82}},\ \bibinfo {pages} {201201(R)} (\bibinfo {year}
		{2010})}\BibitemShut {NoStop}%
	\bibitem [{\citenamefont {Bar-Gill}\ \emph {et~al.}(2013)\citenamefont
		{Bar-Gill}, \citenamefont {Pham}, \citenamefont {Jarmola}, \citenamefont
		{Budker},\ and\ \citenamefont {Walsworth}}]{bar2013solid}%
	\BibitemOpen
	\bibfield  {author} {\bibinfo {author} {\bibfnamefont {N.}~\bibnamefont
			{Bar-Gill}}, \bibinfo {author} {\bibfnamefont {L.~M.}\ \bibnamefont {Pham}},
		\bibinfo {author} {\bibfnamefont {A.}~\bibnamefont {Jarmola}}, \bibinfo
		{author} {\bibfnamefont {D.}~\bibnamefont {Budker}}, \ and\ \bibinfo {author}
		{\bibfnamefont {R.~L.}\ \bibnamefont {Walsworth}},\ }\bibfield  {title}
	{\enquote {\bibinfo {title} {Solid-state electronic spin coherence time
				approaching one second},}\ }\href {http://dx.doi.org/10.1038/ncomms2771}
	{\bibfield  {journal} {\bibinfo  {journal} {Nat. Commun.}\ }\textbf {\bibinfo
			{volume} {4}},\ \bibinfo {pages} {1743} (\bibinfo {year} {2013})}\BibitemShut
	{NoStop}%
	\bibitem [{\citenamefont {Xiang}\ \emph
		{et~al.}(2013{\natexlab{a}})\citenamefont {Xiang}, \citenamefont {L{\"u}},
		\citenamefont {Li}, \citenamefont {You},\ and\ \citenamefont
		{Nori}}]{xiang2013hybrid}%
	\BibitemOpen
	\bibfield  {author} {\bibinfo {author} {\bibfnamefont {Z.-L.}\ \bibnamefont
			{Xiang}}, \bibinfo {author} {\bibfnamefont {X.-Y.}\ \bibnamefont {L{\"u}}},
		\bibinfo {author} {\bibfnamefont {T.-F.}\ \bibnamefont {Li}}, \bibinfo
		{author} {\bibfnamefont {J.~Q.}\ \bibnamefont {You}}, \ and\ \bibinfo
		{author} {\bibfnamefont {F.}~\bibnamefont {Nori}},\ }\bibfield  {title}
	{\enquote {\bibinfo {title} {Hybrid quantum circuit consisting of a
				superconducting flux qubit coupled to a spin ensemble and a transmission-line
				resonator},}\ }\href {https://link.aps.org/doi/10.1103/PhysRevB.87.144516}
	{\bibfield  {journal} {\bibinfo  {journal} {Phys. Rev. B}\ }\textbf {\bibinfo
			{volume} {87}},\ \bibinfo {pages} {144516} (\bibinfo {year}
		{2013}{\natexlab{a}})}\BibitemShut {NoStop}%
	\bibitem [{\citenamefont {Xiang}\ \emph
		{et~al.}(2013{\natexlab{b}})\citenamefont {Xiang}, \citenamefont {Ashhab},
		\citenamefont {You},\ and\ \citenamefont {Nori}}]{RevModPhys.85.623}%
	\BibitemOpen
	\bibfield  {author} {\bibinfo {author} {\bibfnamefont {Ze-Liang}\
			\bibnamefont {Xiang}}, \bibinfo {author} {\bibfnamefont {Sahel}\ \bibnamefont
			{Ashhab}}, \bibinfo {author} {\bibfnamefont {J.~Q.}\ \bibnamefont {You}}, \
		and\ \bibinfo {author} {\bibfnamefont {Franco}\ \bibnamefont {Nori}},\
	}\bibfield  {title} {\enquote {\bibinfo {title} {Hybrid quantum circuits:
				Superconducting circuits interacting with other quantum systems},}\ }\href
	{\doibase 10.1103/RevModPhys.85.623} {\bibfield  {journal} {\bibinfo
			{journal} {Rev. Mod. Phys.}\ }\textbf {\bibinfo {volume} {85}},\ \bibinfo
		{pages} {623--653} (\bibinfo {year} {2013}{\natexlab{b}})}\BibitemShut
	{NoStop}%
	\bibitem [{\citenamefont {Li}\ \emph {et~al.}(2016)\citenamefont {Li},
		\citenamefont {Xiang}, \citenamefont {Rabl},\ and\ \citenamefont
		{Nori}}]{PhysRevLett.117.015502}%
	\BibitemOpen
	\bibfield  {author} {\bibinfo {author} {\bibfnamefont {Peng-Bo}\ \bibnamefont
			{Li}}, \bibinfo {author} {\bibfnamefont {Ze-Liang}\ \bibnamefont {Xiang}},
		\bibinfo {author} {\bibfnamefont {Peter}\ \bibnamefont {Rabl}}, \ and\
		\bibinfo {author} {\bibfnamefont {Franco}\ \bibnamefont {Nori}},\ }\bibfield
	{title} {\enquote {\bibinfo {title} {Hybrid {Q}uantum {D}evice with
				{N}itrogen-{V}acancy {C}enters in {D}iamond {C}oupled to {C}arbon
				{N}anotubes},}\ }\href {\doibase 10.1103/PhysRevLett.117.015502} {\bibfield
		{journal} {\bibinfo  {journal} {Phys. Rev. Lett.}\ }\textbf {\bibinfo
			{volume} {117}},\ \bibinfo {pages} {015502} (\bibinfo {year}
		{2016})}\BibitemShut {NoStop}%
	\bibitem [{\citenamefont {Kubo}\ \emph {et~al.}(2010)\citenamefont {Kubo},
		\citenamefont {Ong}, \citenamefont {Bertet}, \citenamefont {Vion},
		\citenamefont {Jacques}, \citenamefont {Zheng}, \citenamefont {Dr{\'e}au},
		\citenamefont {Roch}, \citenamefont {Auff{\`e}ves}, \citenamefont {Jelezko},
		\citenamefont {Wrachtrup}, \citenamefont {Barthe}, \citenamefont {Bergonzo},\
		and\ \citenamefont {Esteve}}]{kubo2010strong}%
	\BibitemOpen
	\bibfield  {author} {\bibinfo {author} {\bibfnamefont {Y.}~\bibnamefont
			{Kubo}}, \bibinfo {author} {\bibfnamefont {F.~R.}\ \bibnamefont {Ong}},
		\bibinfo {author} {\bibfnamefont {P.}~\bibnamefont {Bertet}}, \bibinfo
		{author} {\bibfnamefont {D.}~\bibnamefont {Vion}}, \bibinfo {author}
		{\bibfnamefont {V.}~\bibnamefont {Jacques}}, \bibinfo {author} {\bibfnamefont
			{D.}~\bibnamefont {Zheng}}, \bibinfo {author} {\bibfnamefont
			{A.}~\bibnamefont {Dr{\'e}au}}, \bibinfo {author} {\bibfnamefont {J.-F.}\
			\bibnamefont {Roch}}, \bibinfo {author} {\bibfnamefont {A.}~\bibnamefont
			{Auff{\`e}ves}}, \bibinfo {author} {\bibfnamefont {F.}~\bibnamefont
			{Jelezko}}, \bibinfo {author} {\bibfnamefont {J.}~\bibnamefont {Wrachtrup}},
		\bibinfo {author} {\bibfnamefont {M.~F.}\ \bibnamefont {Barthe}}, \bibinfo
		{author} {\bibfnamefont {P.}~\bibnamefont {Bergonzo}}, \ and\ \bibinfo
		{author} {\bibfnamefont {D.}~\bibnamefont {Esteve}},\ }\bibfield  {title}
	{\enquote {\bibinfo {title} {Strong {C}oupling of a {S}pin {E}nsemble to a
				{S}uperconducting {R}esonator},}\ }\href
	{https://link.aps.org/doi/10.1103/PhysRevLett.105.140502} {\bibfield
		{journal} {\bibinfo  {journal} {Phys. Rev. Lett.}\ }\textbf {\bibinfo
			{volume} {105}},\ \bibinfo {pages} {140502} (\bibinfo {year}
		{2010})}\BibitemShut {NoStop}%
	\bibitem [{\citenamefont {Ams{\"u}ss}\ \emph {et~al.}(2011)\citenamefont
		{Ams{\"u}ss}, \citenamefont {Koller}, \citenamefont {N{\"o}bauer},
		\citenamefont {Putz}, \citenamefont {Rotter}, \citenamefont {Sandner},
		\citenamefont {Schneider}, \citenamefont {Schramb{\"o}ck}, \citenamefont
		{Steinhauser}, \citenamefont {Ritsch}, \citenamefont {Schmiedmayer},\ and\
		\citenamefont {Majer}}]{amsuss2011cavity}%
	\BibitemOpen
	\bibfield  {author} {\bibinfo {author} {\bibfnamefont {R.}~\bibnamefont
			{Ams{\"u}ss}}, \bibinfo {author} {\bibfnamefont {Ch.}\ \bibnamefont
			{Koller}}, \bibinfo {author} {\bibfnamefont {T.}~\bibnamefont {N{\"o}bauer}},
		\bibinfo {author} {\bibfnamefont {S.}~\bibnamefont {Putz}}, \bibinfo {author}
		{\bibfnamefont {S.}~\bibnamefont {Rotter}}, \bibinfo {author} {\bibfnamefont
			{K.}~\bibnamefont {Sandner}}, \bibinfo {author} {\bibfnamefont
			{S.}~\bibnamefont {Schneider}}, \bibinfo {author} {\bibfnamefont
			{M.}~\bibnamefont {Schramb{\"o}ck}}, \bibinfo {author} {\bibfnamefont
			{G.}~\bibnamefont {Steinhauser}}, \bibinfo {author} {\bibfnamefont
			{H.}~\bibnamefont {Ritsch}}, \bibinfo {author} {\bibfnamefont
			{J.}~\bibnamefont {Schmiedmayer}}, \ and\ \bibinfo {author} {\bibfnamefont
			{J.}~\bibnamefont {Majer}},\ }\bibfield  {title} {\enquote {\bibinfo {title}
			{Cavity {QED} with {M}agnetically {C}oupled {C}ollective {S}pin {S}tates},}\
	}\href {https://link.aps.org/doi/10.1103/PhysRevLett.107.060502} {\bibfield
		{journal} {\bibinfo  {journal} {Phys. Rev. Lett.}\ }\textbf {\bibinfo
			{volume} {107}},\ \bibinfo {pages} {060502} (\bibinfo {year}
		{2011})}\BibitemShut {NoStop}%
	\bibitem [{\citenamefont {Kubo}\ \emph {et~al.}(2011)\citenamefont {Kubo},
		\citenamefont {Grezes}, \citenamefont {Dewes}, \citenamefont {Umeda},
		\citenamefont {Isoya}, \citenamefont {Sumiya}, \citenamefont {Morishita},
		\citenamefont {Abe}, \citenamefont {Onoda}, \citenamefont {Ohshima},
		\citenamefont {Jacques}, \citenamefont {Dr\'{e}au}, \citenamefont {Roch},
		\citenamefont {Diniz}, \citenamefont {Auffeves}, \citenamefont {Vion},
		\citenamefont {Esteve},\ and\ \citenamefont {Bertet}}]{kubo2011hybrid}%
	\BibitemOpen
	\bibfield  {author} {\bibinfo {author} {\bibfnamefont {Y.}~\bibnamefont
			{Kubo}}, \bibinfo {author} {\bibfnamefont {C.}~\bibnamefont {Grezes}},
		\bibinfo {author} {\bibfnamefont {A.}~\bibnamefont {Dewes}}, \bibinfo
		{author} {\bibfnamefont {T.}~\bibnamefont {Umeda}}, \bibinfo {author}
		{\bibfnamefont {J.}~\bibnamefont {Isoya}}, \bibinfo {author} {\bibfnamefont
			{H.}~\bibnamefont {Sumiya}}, \bibinfo {author} {\bibfnamefont
			{N.}~\bibnamefont {Morishita}}, \bibinfo {author} {\bibfnamefont
			{H.}~\bibnamefont {Abe}}, \bibinfo {author} {\bibfnamefont {S.}~\bibnamefont
			{Onoda}}, \bibinfo {author} {\bibfnamefont {T.}~\bibnamefont {Ohshima}},
		\bibinfo {author} {\bibfnamefont {V.}~\bibnamefont {Jacques}}, \bibinfo
		{author} {\bibfnamefont {A.}~\bibnamefont {Dr\'{e}au}}, \bibinfo {author}
		{\bibfnamefont {J.-F.}\ \bibnamefont {Roch}}, \bibinfo {author}
		{\bibfnamefont {I.}~\bibnamefont {Diniz}}, \bibinfo {author} {\bibfnamefont
			{A.}~\bibnamefont {Auffeves}}, \bibinfo {author} {\bibfnamefont
			{D.}~\bibnamefont {Vion}}, \bibinfo {author} {\bibfnamefont {D.}~\bibnamefont
			{Esteve}}, \ and\ \bibinfo {author} {\bibfnamefont {P.}~\bibnamefont
			{Bertet}},\ }\bibfield  {title} {\enquote {\bibinfo {title} {Hybrid {Q}uantum
				{C}ircuit with a {S}uperconducting {Q}ubit {C}oupled to a {S}pin
				{E}nsemble},}\ }\href
	{https://link.aps.org/doi/10.1103/PhysRevLett.107.220501} {\bibfield
		{journal} {\bibinfo  {journal} {Phys. Rev. Lett.}\ }\textbf {\bibinfo
			{volume} {107}},\ \bibinfo {pages} {220501} (\bibinfo {year}
		{2011})}\BibitemShut {NoStop}%
	\bibitem [{\citenamefont {Kubo}\ \emph {et~al.}(2012)\citenamefont {Kubo},
		\citenamefont {Diniz}, \citenamefont {Dewes}, \citenamefont {Jacques},
		\citenamefont {Dr\'eau}, \citenamefont {Roch}, \citenamefont {Auffeves},
		\citenamefont {Vion}, \citenamefont {Esteve},\ and\ \citenamefont
		{Bertet}}]{PhysRevA.85.012333}%
	\BibitemOpen
	\bibfield  {author} {\bibinfo {author} {\bibfnamefont {Y.}~\bibnamefont
			{Kubo}}, \bibinfo {author} {\bibfnamefont {I.}~\bibnamefont {Diniz}},
		\bibinfo {author} {\bibfnamefont {A.}~\bibnamefont {Dewes}}, \bibinfo
		{author} {\bibfnamefont {V.}~\bibnamefont {Jacques}}, \bibinfo {author}
		{\bibfnamefont {A.}~\bibnamefont {Dr\'eau}}, \bibinfo {author} {\bibfnamefont
			{J.-F.}\ \bibnamefont {Roch}}, \bibinfo {author} {\bibfnamefont
			{A.}~\bibnamefont {Auffeves}}, \bibinfo {author} {\bibfnamefont
			{D.}~\bibnamefont {Vion}}, \bibinfo {author} {\bibfnamefont {D.}~\bibnamefont
			{Esteve}}, \ and\ \bibinfo {author} {\bibfnamefont {P.}~\bibnamefont
			{Bertet}},\ }\bibfield  {title} {\enquote {\bibinfo {title} {Storage and
				retrieval of a microwave field in a spin ensemble},}\ }\href {\doibase
		10.1103/PhysRevA.85.012333} {\bibfield  {journal} {\bibinfo  {journal} {Phys.
				Rev. A}\ }\textbf {\bibinfo {volume} {85}},\ \bibinfo {pages} {012333}
		(\bibinfo {year} {2012})}\BibitemShut {NoStop}%
	\bibitem [{\citenamefont {Putz}\ \emph {et~al.}(2014)\citenamefont {Putz},
		\citenamefont {Krimer}, \citenamefont {Amsuess}, \citenamefont {Valookaran},
		\citenamefont {Noebauer}, \citenamefont {Schmiedmayer}, \citenamefont
		{Rotter},\ and\ \citenamefont {Majer}}]{putz2014protecting}%
	\BibitemOpen
	\bibfield  {author} {\bibinfo {author} {\bibfnamefont {S.}~\bibnamefont
			{Putz}}, \bibinfo {author} {\bibfnamefont {D.~O.}\ \bibnamefont {Krimer}},
		\bibinfo {author} {\bibfnamefont {R.}~\bibnamefont {Amsuess}}, \bibinfo
		{author} {\bibfnamefont {A.}~\bibnamefont {Valookaran}}, \bibinfo {author}
		{\bibfnamefont {T.}~\bibnamefont {Noebauer}}, \bibinfo {author}
		{\bibfnamefont {J.}~\bibnamefont {Schmiedmayer}}, \bibinfo {author}
		{\bibfnamefont {S.}~\bibnamefont {Rotter}}, \ and\ \bibinfo {author}
		{\bibfnamefont {J.}~\bibnamefont {Majer}},\ }\bibfield  {title} {\enquote
		{\bibinfo {title} {Protecting a spin ensemble against decoherence in the
				strong-coupling regime of cavity {QED}},}\ }\href
	{https://doi.org/10.1038/nphys3050} {\bibfield  {journal} {\bibinfo
			{journal} {Nat. Phys.}\ }\textbf {\bibinfo {volume} {10}},\ \bibinfo {pages}
		{720} (\bibinfo {year} {2014})}\BibitemShut {NoStop}%
	\bibitem [{\citenamefont {Grezes}\ \emph {et~al.}(2014)\citenamefont {Grezes},
		\citenamefont {Julsgaard}, \citenamefont {Kubo}, \citenamefont {Stern},
		\citenamefont {Umeda}, \citenamefont {Isoya}, \citenamefont {Sumiya},
		\citenamefont {Abe}, \citenamefont {Onoda}, \citenamefont {Ohshima},
		\citenamefont {Jacques}, \citenamefont {Esteve}, \citenamefont {Vion},
		\citenamefont {Esteve}, \citenamefont {M{\o}lmer},\ and\ \citenamefont
		{Bertet}}]{grezes2014multimode}%
	\BibitemOpen
	\bibfield  {author} {\bibinfo {author} {\bibfnamefont {C.}~\bibnamefont
			{Grezes}}, \bibinfo {author} {\bibfnamefont {B.}~\bibnamefont {Julsgaard}},
		\bibinfo {author} {\bibfnamefont {Y.}~\bibnamefont {Kubo}}, \bibinfo {author}
		{\bibfnamefont {M.}~\bibnamefont {Stern}}, \bibinfo {author} {\bibfnamefont
			{T.}~\bibnamefont {Umeda}}, \bibinfo {author} {\bibfnamefont
			{J.}~\bibnamefont {Isoya}}, \bibinfo {author} {\bibfnamefont
			{H.}~\bibnamefont {Sumiya}}, \bibinfo {author} {\bibfnamefont
			{H.}~\bibnamefont {Abe}}, \bibinfo {author} {\bibfnamefont {S.}~\bibnamefont
			{Onoda}}, \bibinfo {author} {\bibfnamefont {T.}~\bibnamefont {Ohshima}},
		\bibinfo {author} {\bibfnamefont {V.}~\bibnamefont {Jacques}}, \bibinfo
		{author} {\bibfnamefont {J.}~\bibnamefont {Esteve}}, \bibinfo {author}
		{\bibfnamefont {D.}~\bibnamefont {Vion}}, \bibinfo {author} {\bibfnamefont
			{D.}~\bibnamefont {Esteve}}, \bibinfo {author} {\bibfnamefont
			{K.}~\bibnamefont {M{\o}lmer}}, \ and\ \bibinfo {author} {\bibfnamefont
			{P.}~\bibnamefont {Bertet}},\ }\bibfield  {title} {\enquote {\bibinfo {title}
			{Multimode {S}torage and {R}etrieval of {M}icrowave {F}ields in a {S}pin
				{E}nsemble},}\ }\href {https://link.aps.org/doi/10.1103/PhysRevX.4.021049}
	{\bibfield  {journal} {\bibinfo  {journal} {Phys. Rev. X}\ }\textbf {\bibinfo
			{volume} {4}},\ \bibinfo {pages} {021049} (\bibinfo {year}
		{2014})}\BibitemShut {NoStop}%
	\bibitem [{\citenamefont {Astner}\ \emph {et~al.}(2017)\citenamefont {Astner},
		\citenamefont {Nevlacsil}, \citenamefont {Peterschofsky}, \citenamefont
		{Angerer}, \citenamefont {Rotter}, \citenamefont {Putz}, \citenamefont
		{Schmiedmayer},\ and\ \citenamefont {Majer}}]{astner2017coherent}%
	\BibitemOpen
	\bibfield  {author} {\bibinfo {author} {\bibfnamefont {T.}~\bibnamefont
			{Astner}}, \bibinfo {author} {\bibfnamefont {S.}~\bibnamefont {Nevlacsil}},
		\bibinfo {author} {\bibfnamefont {N.}~\bibnamefont {Peterschofsky}}, \bibinfo
		{author} {\bibfnamefont {A.}~\bibnamefont {Angerer}}, \bibinfo {author}
		{\bibfnamefont {S.}~\bibnamefont {Rotter}}, \bibinfo {author} {\bibfnamefont
			{S.}~\bibnamefont {Putz}}, \bibinfo {author} {\bibfnamefont {J.}~\bibnamefont
			{Schmiedmayer}}, \ and\ \bibinfo {author} {\bibfnamefont {J.}~\bibnamefont
			{Majer}},\ }\bibfield  {title} {\enquote {\bibinfo {title} {Coherent
				{C}oupling of {R}emote {S}pin {E}nsembles via a {C}avity {B}us},}\ }\href
	{https://link.aps.org/doi/10.1103/PhysRevLett.118.140502} {\bibfield
		{journal} {\bibinfo  {journal} {Phys. Rev. Lett.}\ }\textbf {\bibinfo
			{volume} {118}},\ \bibinfo {pages} {140502} (\bibinfo {year}
		{2017})}\BibitemShut {NoStop}%
	\bibitem [{\citenamefont {Kockum}\ \emph {et~al.}(2019)\citenamefont {Kockum},
		\citenamefont {Miranowicz}, \citenamefont {De~Liberato}, \citenamefont
		{Savasta},\ and\ \citenamefont {Nori}}]{kockum2019ultrastrong}%
	\BibitemOpen
	\bibfield  {author} {\bibinfo {author} {\bibfnamefont {A.~F.}\ \bibnamefont
			{Kockum}}, \bibinfo {author} {\bibfnamefont {A.}~\bibnamefont {Miranowicz}},
		\bibinfo {author} {\bibfnamefont {S.}~\bibnamefont {De~Liberato}}, \bibinfo
		{author} {\bibfnamefont {S.}~\bibnamefont {Savasta}}, \ and\ \bibinfo
		{author} {\bibfnamefont {F.}~\bibnamefont {Nori}},\ }\bibfield  {title}
	{\enquote {\bibinfo {title} {Ultrastrong coupling between light and
				matter},}\ }\href {https://doi.org/10.1038/s42254-018-0006-2} {\bibfield
		{journal} {\bibinfo  {journal} {Nat. Rev. Phys.}\ }\textbf {\bibinfo {volume}
			{1}},\ \bibinfo {pages} {19--40} (\bibinfo {year} {2019})}\BibitemShut
	{NoStop}%
	\bibitem [{\citenamefont {Forn-D{\'\i}az}\ \emph {et~al.}(2019)\citenamefont
		{Forn-D{\'\i}az}, \citenamefont {Lamata}, \citenamefont {Rico}, \citenamefont
		{Kono},\ and\ \citenamefont {Solano}}]{forn2019ultrastrong}%
	\BibitemOpen
	\bibfield  {author} {\bibinfo {author} {\bibfnamefont {P.}~\bibnamefont
			{Forn-D{\'\i}az}}, \bibinfo {author} {\bibfnamefont {L.}~\bibnamefont
			{Lamata}}, \bibinfo {author} {\bibfnamefont {E.}~\bibnamefont {Rico}},
		\bibinfo {author} {\bibfnamefont {J.}~\bibnamefont {Kono}}, \ and\ \bibinfo
		{author} {\bibfnamefont {E.}~\bibnamefont {Solano}},\ }\bibfield  {title}
	{\enquote {\bibinfo {title} {Ultrastrong coupling regimes of light-matter
				interaction},}\ }\href
	{https://link.aps.org/doi/10.1103/RevModPhys.91.025005} {\bibfield  {journal}
		{\bibinfo  {journal} {Rev. Mod. Phys.}\ }\textbf {\bibinfo {volume} {91}},\
		\bibinfo {pages} {025005} (\bibinfo {year} {2019})}\BibitemShut {NoStop}%
	\bibitem [{\citenamefont {Kirton}\ \emph {et~al.}(2019)\citenamefont {Kirton},
		\citenamefont {Roses}, \citenamefont {Keeling},\ and\ \citenamefont
		{Dalla~Torre}}]{kirton2019introduction}%
	\BibitemOpen
	\bibfield  {author} {\bibinfo {author} {\bibfnamefont {P.}~\bibnamefont
			{Kirton}}, \bibinfo {author} {\bibfnamefont {M.~M.}\ \bibnamefont {Roses}},
		\bibinfo {author} {\bibfnamefont {J.}~\bibnamefont {Keeling}}, \ and\
		\bibinfo {author} {\bibfnamefont {E.~G.}\ \bibnamefont {Dalla~Torre}},\
	}\bibfield  {title} {\enquote {\bibinfo {title} {Introduction to the {D}icke
				model: from equilibrium to nonequilibrium, and vice versa},}\ }\href
	{https://onlinelibrary.wiley.com/doi/abs/10.1002/qute.201800043} {\bibfield
		{journal} {\bibinfo  {journal} {Adv. Quantum Technol.}\ }\textbf {\bibinfo
			{volume} {2}},\ \bibinfo {pages} {1800043} (\bibinfo {year}
		{2019})}\BibitemShut {NoStop}%
	\bibitem [{\citenamefont {L{\"u}}\ \emph {et~al.}(2015)\citenamefont {L{\"u}},
		\citenamefont {Wu}, \citenamefont {Johansson}, \citenamefont {Jing},
		\citenamefont {Zhang},\ and\ \citenamefont {Nori}}]{lu2015squeezed}%
	\BibitemOpen
	\bibfield  {author} {\bibinfo {author} {\bibfnamefont {X.-Y.}\ \bibnamefont
			{L{\"u}}}, \bibinfo {author} {\bibfnamefont {Y.}~\bibnamefont {Wu}}, \bibinfo
		{author} {\bibfnamefont {J.~R.}\ \bibnamefont {Johansson}}, \bibinfo {author}
		{\bibfnamefont {H.}~\bibnamefont {Jing}}, \bibinfo {author} {\bibfnamefont
			{J.}~\bibnamefont {Zhang}}, \ and\ \bibinfo {author} {\bibfnamefont
			{F.}~\bibnamefont {Nori}},\ }\bibfield  {title} {\enquote {\bibinfo {title}
			{Squeezed {O}ptomechanics with {P}hase-{M}atched {A}mplification and
				{D}issipation},}\ }\href
	{https://link.aps.org/doi/10.1103/PhysRevLett.114.093602} {\bibfield
		{journal} {\bibinfo  {journal} {Phys. Rev. Lett.}\ }\textbf {\bibinfo
			{volume} {114}},\ \bibinfo {pages} {093602} (\bibinfo {year}
		{2015})}\BibitemShut {NoStop}%
	\bibitem [{\citenamefont {Wallquist}\ \emph {et~al.}(2006)\citenamefont
		{Wallquist}, \citenamefont {Shumeiko},\ and\ \citenamefont
		{Wendin}}]{PhysRevB.74.224506}%
	\BibitemOpen
	\bibfield  {author} {\bibinfo {author} {\bibfnamefont {M.}~\bibnamefont
			{Wallquist}}, \bibinfo {author} {\bibfnamefont {V.~S.}\ \bibnamefont
			{Shumeiko}}, \ and\ \bibinfo {author} {\bibfnamefont {G.}~\bibnamefont
			{Wendin}},\ }\bibfield  {title} {\enquote {\bibinfo {title} {Selective
				coupling of superconducting charge qubits mediated by a tunable stripline
				cavity},}\ }\href {\doibase 10.1103/PhysRevB.74.224506} {\bibfield  {journal}
		{\bibinfo  {journal} {Phys. Rev. B}\ }\textbf {\bibinfo {volume} {74}},\
		\bibinfo {pages} {224506} (\bibinfo {year} {2006})}\BibitemShut {NoStop}%
	\bibitem [{\citenamefont {Johansson}\ \emph {et~al.}(2010)\citenamefont
		{Johansson}, \citenamefont {Johansson}, \citenamefont {Wilson},\ and\
		\citenamefont {Nori}}]{johansson2010dynamical}%
	\BibitemOpen
	\bibfield  {author} {\bibinfo {author} {\bibfnamefont {J.~R.}\ \bibnamefont
			{Johansson}}, \bibinfo {author} {\bibfnamefont {G.}~\bibnamefont
			{Johansson}}, \bibinfo {author} {\bibfnamefont {C.~M.}\ \bibnamefont
			{Wilson}}, \ and\ \bibinfo {author} {\bibfnamefont {F.}~\bibnamefont
			{Nori}},\ }\bibfield  {title} {\enquote {\bibinfo {title} {Dynamical
				{C}asimir effect in superconducting microwave circuits},}\ }\href
	{https://link.aps.org/doi/10.1103/PhysRevA.82.052509} {\bibfield  {journal}
		{\bibinfo  {journal} {Phys. Rev. A}\ }\textbf {\bibinfo {volume} {82}},\
		\bibinfo {pages} {052509} (\bibinfo {year} {2010})}\BibitemShut {NoStop}%
	\bibitem [{\citenamefont {Wustmann}\ and\ \citenamefont
		{Shumeiko}(2013)}]{PhysRevB.87.184501}%
	\BibitemOpen
	\bibfield  {author} {\bibinfo {author} {\bibfnamefont {W.}~\bibnamefont
			{Wustmann}}\ and\ \bibinfo {author} {\bibfnamefont {V.}~\bibnamefont
			{Shumeiko}},\ }\bibfield  {title} {\enquote {\bibinfo {title} {Parametric
				resonance in tunable superconducting cavities},}\ }\href {\doibase
		10.1103/PhysRevB.87.184501} {\bibfield  {journal} {\bibinfo  {journal} {Phys.
				Rev. B}\ }\textbf {\bibinfo {volume} {87}},\ \bibinfo {pages} {184501}
		(\bibinfo {year} {2013})}\BibitemShut {NoStop}%
	\bibitem [{\citenamefont {Bliokh}\ and\ \citenamefont
		{Nori}(2019)}]{PhysRevLett.123.054301}%
	\BibitemOpen
	\bibfield  {author} {\bibinfo {author} {\bibfnamefont {K.~Y.}\ \bibnamefont
			{Bliokh}}\ and\ \bibinfo {author} {\bibfnamefont {F.}~\bibnamefont {Nori}},\
	}\bibfield  {title} {\enquote {\bibinfo {title} {Klein-{G}ordon
				{R}epresentation of {A}coustic {W}aves and {T}opological {O}rigin of
				{S}urface {A}coustic {M}odes},}\ }\href {\doibase
		10.1103/PhysRevLett.123.054301} {\bibfield  {journal} {\bibinfo  {journal}
			{Phys. Rev. Lett.}\ }\textbf {\bibinfo {volume} {123}},\ \bibinfo {pages}
		{054301} (\bibinfo {year} {2019})}\BibitemShut {NoStop}%
	\bibitem [{\citenamefont {Schuster}\ \emph {et~al.}(2010)\citenamefont
		{Schuster}, \citenamefont {Sears}, \citenamefont {Ginossar}, \citenamefont
		{DiCarlo}, \citenamefont {Frunzio}, \citenamefont {Morton}, \citenamefont
		{Wu}, \citenamefont {Briggs}, \citenamefont {Buckley}, \citenamefont
		{Awschalom},\ and\ \citenamefont {Schoelkopf}}]{PhysRevLett.105.140501}%
	\BibitemOpen
	\bibfield  {author} {\bibinfo {author} {\bibfnamefont {D.~I.}\ \bibnamefont
			{Schuster}}, \bibinfo {author} {\bibfnamefont {A.~P.}\ \bibnamefont {Sears}},
		\bibinfo {author} {\bibfnamefont {E.}~\bibnamefont {Ginossar}}, \bibinfo
		{author} {\bibfnamefont {L.}~\bibnamefont {DiCarlo}}, \bibinfo {author}
		{\bibfnamefont {L.}~\bibnamefont {Frunzio}}, \bibinfo {author} {\bibfnamefont
			{J.~J.~L.}\ \bibnamefont {Morton}}, \bibinfo {author} {\bibfnamefont
			{H.}~\bibnamefont {Wu}}, \bibinfo {author} {\bibfnamefont {G.~A.~D.}\
			\bibnamefont {Briggs}}, \bibinfo {author} {\bibfnamefont {B.~B.}\
			\bibnamefont {Buckley}}, \bibinfo {author} {\bibfnamefont {D.~D.}\
			\bibnamefont {Awschalom}}, \ and\ \bibinfo {author} {\bibfnamefont {R.~J.}\
			\bibnamefont {Schoelkopf}},\ }\bibfield  {title} {\enquote {\bibinfo {title}
			{High-{C}ooperativity {C}oupling of {E}lectron-{S}pin {E}nsembles to
				{S}uperconducting {C}avities},}\ }\href {\doibase
		10.1103/PhysRevLett.105.140501} {\bibfield  {journal} {\bibinfo  {journal}
			{Phys. Rev. Lett.}\ }\textbf {\bibinfo {volume} {105}},\ \bibinfo {pages}
		{140501} (\bibinfo {year} {2010})}\BibitemShut {NoStop}%
	\bibitem [{\citenamefont {Probst}\ \emph {et~al.}(2013)\citenamefont {Probst},
		\citenamefont {Rotzinger}, \citenamefont {W\"unsch}, \citenamefont {Jung},
		\citenamefont {Jerger}, \citenamefont {Siegel}, \citenamefont {Ustinov},\
		and\ \citenamefont {Bushev}}]{PhysRevLett.110.157001}%
	\BibitemOpen
	\bibfield  {author} {\bibinfo {author} {\bibfnamefont {S.}~\bibnamefont
			{Probst}}, \bibinfo {author} {\bibfnamefont {H.}~\bibnamefont {Rotzinger}},
		\bibinfo {author} {\bibfnamefont {S.}~\bibnamefont {W\"unsch}}, \bibinfo
		{author} {\bibfnamefont {P.}~\bibnamefont {Jung}}, \bibinfo {author}
		{\bibfnamefont {M.}~\bibnamefont {Jerger}}, \bibinfo {author} {\bibfnamefont
			{M.}~\bibnamefont {Siegel}}, \bibinfo {author} {\bibfnamefont {A.~V.}\
			\bibnamefont {Ustinov}}, \ and\ \bibinfo {author} {\bibfnamefont {P.~A.}\
			\bibnamefont {Bushev}},\ }\bibfield  {title} {\enquote {\bibinfo {title}
			{Anisotropic {R}are-{E}arth {S}pin {E}nsemble {S}trongly {C}oupled to a
				{S}uperconducting {R}esonator},}\ }\href {\doibase
		10.1103/PhysRevLett.110.157001} {\bibfield  {journal} {\bibinfo  {journal}
			{Phys. Rev. Lett.}\ }\textbf {\bibinfo {volume} {110}},\ \bibinfo {pages}
		{157001} (\bibinfo {year} {2013})}\BibitemShut {NoStop}%
	\bibitem [{\citenamefont {Wisby}\ \emph {et~al.}(2014)\citenamefont {Wisby},
		\citenamefont {de~Graaf}, \citenamefont {Gwilliam}, \citenamefont {Adamyan},
		\citenamefont {Kubatkin}, \citenamefont {Meeson}, \citenamefont
		{Tzalenchuk},\ and\ \citenamefont {Lindstr{\"o}m}}]{wisby2014coupling}%
	\BibitemOpen
	\bibfield  {author} {\bibinfo {author} {\bibfnamefont {I.}~\bibnamefont
			{Wisby}}, \bibinfo {author} {\bibfnamefont {S.~E.}\ \bibnamefont {de~Graaf}},
		\bibinfo {author} {\bibfnamefont {R.}~\bibnamefont {Gwilliam}}, \bibinfo
		{author} {\bibfnamefont {A.}~\bibnamefont {Adamyan}}, \bibinfo {author}
		{\bibfnamefont {S.~E.}\ \bibnamefont {Kubatkin}}, \bibinfo {author}
		{\bibfnamefont {P.~J.}\ \bibnamefont {Meeson}}, \bibinfo {author}
		{\bibfnamefont {A.~Y.}\ \bibnamefont {Tzalenchuk}}, \ and\ \bibinfo {author}
		{\bibfnamefont {T.}~\bibnamefont {Lindstr{\"o}m}},\ }\bibfield  {title}
	{\enquote {\bibinfo {title} {Coupling of a locally implanted rare-earth ion
				ensemble to a superconducting micro-resonator},}\ }\href
	{https://doi.org/10.1063/1.4894455} {\bibfield  {journal} {\bibinfo
			{journal} {Appl. Phys. Lett.}\ }\textbf {\bibinfo {volume} {105}},\ \bibinfo
		{pages} {102601} (\bibinfo {year} {2014})}\BibitemShut {NoStop}%
	\bibitem [{\citenamefont {Ranjan}\ \emph {et~al.}(2013)\citenamefont {Ranjan},
		\citenamefont {de~Lange}, \citenamefont {Schutjens}, \citenamefont
		{Debelhoir}, \citenamefont {Groen}, \citenamefont {Szombati}, \citenamefont
		{Thoen}, \citenamefont {Klapwijk}, \citenamefont {Hanson},\ and\
		\citenamefont {DiCarlo}}]{ranjan2013probing}%
	\BibitemOpen
	\bibfield  {author} {\bibinfo {author} {\bibfnamefont {V.}~\bibnamefont
			{Ranjan}}, \bibinfo {author} {\bibfnamefont {G.}~\bibnamefont {de~Lange}},
		\bibinfo {author} {\bibfnamefont {R.}~\bibnamefont {Schutjens}}, \bibinfo
		{author} {\bibfnamefont {T.}~\bibnamefont {Debelhoir}}, \bibinfo {author}
		{\bibfnamefont {J.~P.}\ \bibnamefont {Groen}}, \bibinfo {author}
		{\bibfnamefont {D.}~\bibnamefont {Szombati}}, \bibinfo {author}
		{\bibfnamefont {D.~J.}\ \bibnamefont {Thoen}}, \bibinfo {author}
		{\bibfnamefont {T.~M.}\ \bibnamefont {Klapwijk}}, \bibinfo {author}
		{\bibfnamefont {R.}~\bibnamefont {Hanson}}, \ and\ \bibinfo {author}
		{\bibfnamefont {L.}~\bibnamefont {DiCarlo}},\ }\bibfield  {title} {\enquote
		{\bibinfo {title} {Probing {D}ynamics of an {E}lectron-{S}pin {E}nsemble via
				a {S}uperconducting {R}esonator},}\ }\href {\doibase
		10.1103/PhysRevLett.110.067004} {\bibfield  {journal} {\bibinfo  {journal}
			{Phys. Rev. Lett.}\ }\textbf {\bibinfo {volume} {110}},\ \bibinfo {pages}
		{067004} (\bibinfo {year} {2013})}\BibitemShut {NoStop}%
	\bibitem [{\citenamefont {Hattermann}\ \emph {et~al.}(2017)\citenamefont
		{Hattermann}, \citenamefont {Bothner}, \citenamefont {Ley}, \citenamefont
		{Ferdinand}, \citenamefont {Wiedmaier}, \citenamefont {S{\'a}rk{\'a}ny},
		\citenamefont {Kleiner}, \citenamefont {Koelle},\ and\ \citenamefont
		{Fort{\'a}gh}}]{hattermann2017coupling}%
	\BibitemOpen
	\bibfield  {author} {\bibinfo {author} {\bibfnamefont {H.}~\bibnamefont
			{Hattermann}}, \bibinfo {author} {\bibfnamefont {D.}~\bibnamefont {Bothner}},
		\bibinfo {author} {\bibfnamefont {L.~Y.}\ \bibnamefont {Ley}}, \bibinfo
		{author} {\bibfnamefont {B.}~\bibnamefont {Ferdinand}}, \bibinfo {author}
		{\bibfnamefont {D.}~\bibnamefont {Wiedmaier}}, \bibinfo {author}
		{\bibfnamefont {L.}~\bibnamefont {S{\'a}rk{\'a}ny}}, \bibinfo {author}
		{\bibfnamefont {R.}~\bibnamefont {Kleiner}}, \bibinfo {author} {\bibfnamefont
			{D.}~\bibnamefont {Koelle}}, \ and\ \bibinfo {author} {\bibfnamefont
			{J.}~\bibnamefont {Fort{\'a}gh}},\ }\bibfield  {title} {\enquote {\bibinfo
			{title} {Coupling ultracold atoms to a superconducting coplanar waveguide
				resonator},}\ }\href {https://doi.org/10.1038/s41467-017-02439-7} {\bibfield
		{journal} {\bibinfo  {journal} {Nat. Commun.}\ }\textbf {\bibinfo {volume}
			{8}},\ \bibinfo {pages} {1--7} (\bibinfo {year} {2017})}\BibitemShut
	{NoStop}%
\end{thebibliography}

%

\end{document}